\documentclass[prd,twocolumn,showpacs,amsmath,amssymb,citeautoscript]{revtex4}

\newcommand{\psib}{{\overline{\psi}}}

\usepackage{graphicx,times}

\begin{document}

\title{Massive fermions without fermion bilinear condensates}
\author{Venkitesh Ayyar and Shailesh Chandrasekharan}
\affiliation{Department of Physics, Box 90305, Duke University,
Durham, North Carolina 27708, USA}

\begin{abstract}
We study a lattice field theory model containing two flavors of massless staggered fermions with an onsite four-fermion interaction. The model contains a $SU(4)$ symmetry which forbids non-zero fermion bilinear mass terms, due to which there is a massless fermion phase at weak couplings. However, even at strong couplings fermion bilinear condensates do not appear in our model, although fermions do become massive. While the existence of this exotic strongly coupled massive fermion phase was established long ago, the nature of the transition between the massless and the massive phase has remained unclear. Using Monte Carlo calculations in three space-time dimensions, we find evidence for a direct second order transition between the two phases suggesting that the exotic lattice phase may have a continuum limit at least in three dimensions. A similar exotic second order critical point was found recently in a bilayer system on a honeycomb lattice.
\end{abstract}


\maketitle

\section{Introduction}
\label{sec1}

It is well known that relativistic four-fermion field theories in three dimensions can contain strongly interacting second order fixed points \cite{Rosenstein:1990nm,Hands:1992be}. The search for such fixed points in four-dimensions has been less successful, although efforts to find them continue in the context of Yukawa models \cite{Gerhold:2007gx,Gies:2009sv,Bulava:2011jp,Molgaard:2014mqa}. One of the motivations for their search is to understand new dynamical mechanisms for fermion mass generation that may be realized in nature. Perturbatively, fermion masses arise from local fermion bilinear terms in the action. Since four-fermion interactions are perturbatively irrelevant in three and higher dimensions, we expect a massless fermion phase at small couplings as long as the interactions are invariant under some subgroup of the chiral symmetry group that prevents fermion bilinear condensates. However when these interactions become strong, symmetries that protect the fermions from becoming massive can break spontaneously leading to non-zero fermion bilinear condensates and massive fermions. This traditional mechanism of mass generation is well known. In this paper we explore another more exotic mechanism of mass generation where fermions become massive without fermion bilinear condensates. As we will explain below, such exotic mechanisms of fermion mass generation are known to occur at strong couplings. In this work we provide evidence that these lattice phases can be connected to massless fermion phases by second order phase transitions, suggesting that the exotic mass generation mechanism may be of interest even in continuum quantum field theory.

Anomaly matching severely constrains the chiral symmetries that can be preserved when fermions become massive \cite{'tHooft:1980xb,PhysRevLett.45.100}. It is necessary for the full chiral symmetry group of free fermions to be broken either explicitly or spontaneously for fermions to become massive. However, there are chiral symmetry subgroups that can remain unbroken, which forbid local fermion bilinear condensates, yet allow for fermions to become massive. Such exotic mechanisms of fermion mass generation have appeared in the literature in the context of QCD like theories \cite{Knecht:1994zb,Stern:1997ri,Stern:1998dy}. In these examples the spontaneous breaking of chiral symmetry occurs through the formation of four-fermion condensates which preserve an unbroken chiral symmetry subgroup that forbids fermion bilinear condensates \cite{PhysRevD.59.016001}. What about four-fermion field theories where the interactions naturally generate the necessary four-fermion condensates that can make fermions massive, but still contain symmetries that forbid fermion bilinear condensates? In such theories, there is no need for any further symmetry breaking in order to make fermions massive, since the four-fermion coupling already breaks the full chiral symmetry group to a subgroup that in principle allows for fermions to become massive. On the other hand since four-fermion interactions are irrelevant perturbatively, there will still be a massless fermion phase at weak couplings. However, as couplings become strong, there can be a phase transition to a phase where fermions become massive without any spontaneous symmetry breaking of the remnant chiral symmetry subgroup. In such a transition there is no local order parameter that distinguish the two phases in the strict sense of the word, although the four-fermion condensate could show a dramatic change in the vicinity of the phase transition. In other words, the remnant chiral symmetry subgroup is realized in the Wigner-Weyl mode in both the phases but in different forms: one containing massless fermions while the other containing massive fermions with some form of parity doubling \cite{Glozman:2007ek}.  In this paper we study an explicit example of such an exotic phase transition in a four-fermion lattice field theory in three dimensions. Interestingly, this phase transition seems to be second order.

It is well known that subgroups of the full chiral symmetry group can be preserved on the lattice even in the presence of interactions. A famous example is staggered fermions, where a $U(1)$ subgroup of the full chiral symmetry group prevents fermion mass terms \cite{Sharatchandra:1981si,vandenDoel:1983mf,Golterman:1984cy}. With more flavors this lattice chiral symmetry group is enhanced and it is interesting to explore if there are subgroups of this remnant lattice chiral symmetry group that forbid fermion bilinear expectation values while still allowing staggered fermions to become massive. Interestingly such an exotic fermion mass generation mechanism was discovered long ago in studies of staggered lattice Yukawa models in four-dimensions within a phase called the strong paramagnetic or PMS phase \cite{Hasenfratz:1988vc,PhysRevD.38.3231,Hasenfratz:1989jr,Lee:1989mi,Lee:1989xq}. Many studies with Wilson fermions followed this discovery in an attempt to explore if the PMS phase can be used to formulate the standard model on the lattice \cite{Bock:1989gg,Bock:1990tv,Bock:1990cx,Aoki:1991mm,Lin:1991csa}. While most of these attempts seem to have failed, as far as we know the rich phase structure that was predicted within various models was only partially verified with Monte Carlo calculations. In particular, results in the intermediate coupling region may not have been reliable since computational techniques were still in their infancy at that time. While most of the studies of the PMS phase focused on four-dimensions, there have been studies more recently in three space-time dimensions where similar phase structures were found \cite{Alonso:1999hh}. Analytic predictions using mean field theory also emerged at the same time  \cite{Abada:1990ds,Stephanov:1990sq,Stephanov:1990pc,Stephanov:1991sv,Ebihara:1991mv}. A review of these early results can be found in \cite{Shigemitsu:1991tc}. 

\begin{figure}
\includegraphics[width=0.48\textwidth]{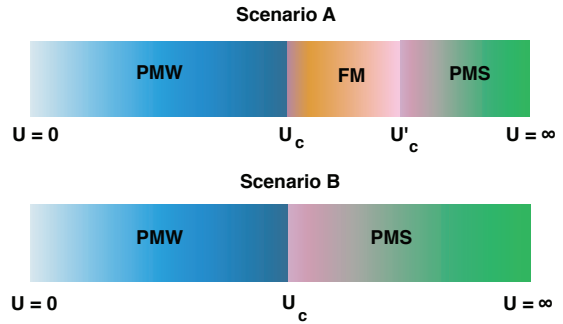}
\caption{\label{pdiag} The two possible phase diagrams for our model based on previous studies. Our work provides strong evidence in favor of scenario B with a second order transition between the PMW phase and the PMS phase.}
\end{figure}

In this work we revisit a simple lattice four-fermion model with two flavors of staggered fermions interacting with an onsite four-fermion coupling. Our model is a limiting case of a lattice Yukawa model studied long ago \cite{Lee:1989mi}. Earlier studies were performed in four-dimensions, where it was established that there is a massless fermion phase at weak couplings and a PMS phase at strong couplings. The weak coupling phase was referred to as the weak paramagnetic or PMW phase. The authors used mean field theory in the intermediate coupling region and found that the two phases are separated from each other by a more conventional massive fermion phase with a non-zero chiral condensate (referred to as the ferromagnetic or FM phase). This phase diagram is shown as scenario A in Fig.~\ref{pdiag}. On the other hand a different mean field theory calculation, which becomes exact in the limit of large dimensions, found a direct first order transition between the massless and the massive phase \cite{Stephanov:1990sq,Stephanov:1990pc,Stephanov:1991sv,Ebihara:1991mv}. This is shown as scenario B in Fig.~\ref{pdiag}. As far as we know, a controlled first principles Monte Carlo calculation has never been performed. In this work we perform such a calculation in three space-time dimensions and find a result consistent with scenario B, but with a second order transition between the PMW and the PMS phase. This second order critical point cannot be described using traditional four-fermion field theory that involves spontaneous symmetry breaking and the formation of a fermion bilinear condensate. Interestingly, a very similar second order transition was recently found in an extended Hubbard model on a bilayer-honeycomb lattice, where it was argued that the exotic critical point is a multi-critical point where three topology driven second order phase transition lines meet \cite{Slagle:2014vma}.

Our paper is organized as follows. In the next section we present our model, its symmetries and the observables we wish to compute. In section \ref{sec3} we discuss how our model can be viewed as a limit of a lattice Yukawa model and argue the presence of the PMW and the PMS phase at weak and strong couplings respectively. We also review results from the mean field theory calculation that predicts a direct first order transition between the two phases. We then discuss the fermion bag approach in \ref{sec4}, which we use to perform Monte Carlo calculations. Section \ref{sec5} contains a discussion of the specific Monte Carlo update procedures we have used in our work. In section \ref{sec6} we present our numerical results and its analysis that provides evidence for a single second order transition between the two phases and in section \ref{sec7} we discuss why we believe there is no order parameter that distinguishes the two phases. Finally, section \ref{sec8} contains our conclusions.

\section{Model and Symmetries} 
\label{sec2}

The model we study contains two flavors of staggered fermions with an onsite four-fermion interaction. The Euclidean action of our model is given by
\begin{equation}
S = S_0 - U \ \sum_{x} \Big\{ \psib_{x,1}\psi_{x,1}\psib_{x,2}\psi_{x,2}\Big\} 
\label{act}
\end{equation}
where $S_0$ is the free massless staggered fermion action
\begin{equation}
S_0 = \sum_{i=1,2}\ \sum_{x,y} {\psib}_{x,i} \ M_{x,y} \ \psi_{y,i}.
\label{freef}
\end{equation}
Here $\psib_{x,i}, \psi_{x,i}, i = 1,2$ are four independent Grassmann valued fields, $M$ is the well known staggered fermion matrix given by
\begin{equation}
M_{x,y} \ =\  \sum_{\hat{\alpha}} \frac{\eta_{x,{\hat{\alpha}}}}{2}\ [\delta_{x,y+\hat{\alpha}} - \delta_{x,y-\hat{\alpha}}],
\label{staggered}
\end{equation}
$x \equiv (x_1,x_2,x_3)$ denotes a lattice site on a $3$ dimensional cubic lattice and $\hat{\alpha} = \hat{1},\hat{2},\hat{3}$  represent unit lattice vectors in the three directions. The staggered fermion phases are defined as usual: $\eta_{x,\hat{1}}=1,, \eta_{x,\hat{2}}=(-1)^{x_1}$, and $\eta_{x,\hat{3}}= (-1)^{x_1+x_2}$. We study cubical lattices of equal size $L$ in each direction with anti-periodic boundary conditions. Since the lattice is cubical we can define a parity for each site using the sign factor $\varepsilon_x = (-1)^{x_1+x_2+x_3}$. If $\varepsilon_x = 1$ we define the site to be even and otherwise it is odd. Our model is just one of the many possible lattice Gross-Neveu models that have been considered in the literature \cite{Hands:1992be,Hands:1992ck,Focht:1995ie,Kogut:1999um,Christofi:2006zt}, however the PMS phase at strong couplings is a peculiarity of our model and is not present in most models. This difference has been pointed out in earlier work \cite{Lee:1989mi}. 

It is easy to verify that the action given in Eq.~(\ref{act}) is symmetric under the usual space-time lattice transformations and internal $SU(4)$ transformations given below \cite{vandenDoel:1983mf,Golterman:1984cy}:
\begin{itemize}
\item[(i)] Space-time translations:
\begin{equation}
 \psi_{x,i} \rightarrow \xi_{x,\hat{\alpha}}\psi_{x+\hat{\hat{\alpha}},i},\   
\overline{\psi}_{x,i} \rightarrow \xi_{x,\hat{\alpha}} \overline{\psi}_{x+\hat{\alpha},i}
\end{equation}
where $\xi_{x,\hat{1}} = ( -1)^{ x_2 + x_3 }$, $\xi_{x,\hat{2}} = ( -1)^{x_3}$, and $\xi_{x,\hat{3}} = 1$.
\item[(ii)] Space-time rotations:
\begin{equation}
 \psi_{x,i} \rightarrow S_R (R^{-1} x ) \psi_{R^{-1} x,i},\  
\overline{\psi}_{x,i} \rightarrow S_R ( R^{-1} x ) \overline{\psi}_{R^{-1} x,i}
\end{equation}
where $R \equiv R ^{( \rho \sigma ) }, \rho \neq \sigma$ is the rotation $x_{\rho} \rightarrow x_{\sigma}$,$x_{\sigma} \rightarrow - x_{\rho}$, and $x_{\tau } \rightarrow x_{\tau}$ when $\tau \neq \rho, \sigma$ and $S_R(x) = \frac {1} {2} 
(1 \pm \eta_{\hat{\rho}} ( x ) \eta_{x,\hat{\sigma}} \mp \xi_{x,\rho} \xi_{x,\sigma} +   \eta_{x,\hat{\rho}}  \eta_{x,\hat{\sigma}} \xi_{x,\rho} \xi_{x,\sigma})$ where the two signs represent the cases $\rho > \sigma$ and $\rho < \sigma$ respectively.
\item[(iii)]Axis reversal:
\begin{eqnarray}
 \psi_{x,i} \rightarrow (-1)^{x_{\rho}} \psi_{(I^\rho x),i}, \overline{\psi}_{x,i} \rightarrow (-1)^{x_{\rho}} \overline{\psi}_{(I^\rho x),i} 
\end{eqnarray}
where $I^\rho( x)$ is the axis reversal operation on $x$ which changes $x_\rho \rightarrow - x_\rho$ and $x_\sigma \rightarrow x_\sigma, \sigma \neq \rho$.
\item[(iv)] Global $SU(4)$ transformations
\begin{subequations}
\begin{equation}
\left(
\begin{array}{c}
\psi_{x_e,1} \cr
\psib_{x_e,1} \cr
\psi_{x_e,2} \cr
\psib_{x_e,2}
\end{array}
\right)
\ \rightarrow \ 
 V \ \left(
\begin{array}{c}
\psi_{x_e,1} \cr
\psib_{x_e,1} \cr
\psi_{x_e,2} \cr
\psib_{x_e,2}
\end{array}
\right)
\end{equation}
\begin{equation}
\left(
\begin{array}{c}
\psi_{x_o,1} \cr
\psib_{x_o,1} \cr
\psi_{x_o,2} \cr
\psib_{x_o,2}
\end{array}
\right)
\ \rightarrow \ 
 V^* \ \left(
\begin{array}{c}
\psi_{x_o,1} \cr
\psib_{x_o,1} \cr
\psi_{x_o,2} \cr
\psib_{x_o,2}
\end{array}
\right)
\end{equation}
\end{subequations}
where $x_e$ and $x_o$ refer to even and odd lattice sites respectively, and $V$ is a $SU(4)$ matrix in the fundamental representation.
\end{itemize}

The free action is invariant under a much bigger symmetry group since it describes four flavors of four component Dirac fermions. While this enhanced symmetry can only be understood in the momentum space formulation, the $SU(4)$ symmetry discussed above and the well known $U_\chi(1)$ symmetry of staggered fermions, implemented through the transformations
\begin{equation}
\psi_{x,i} \rightarrow \mathrm{e}^{i\theta\ \varepsilon_x} \psi_{x,i}, \ \ 
\overline{\psi}_{x,i} \rightarrow \mathrm{e}^{i\theta \varepsilon_x} \overline{\psi}_{x,i},
\end{equation}
are both visible even in position space.  In most staggered four-fermion models, it is the $U_\chi(1)$ symmetry that breaks spontaneously when fermions become massive. In contrast, in our model the interaction term breaks it explicitly by introducing a four-fermion condensate. On the other hand the $SU(4)$ symmetry forbids fermion bilinear condensates. Indeed the six onsite fermion bilinears $\phi_{x,1} = \psib_{x,1}\psi_{x,1}, \ \phi_{x,2} = \psib_{x,2}\psi_{x,2}, \ \phi_{x,3} = \psib_{x,1}\psi_{x,2}, \ \phi_{x,4} = \psib_{x,2}\psi_{x,1}, \ \phi_{x,5} = \psi_{x,1} \psi_{x,2}, \ \phi_{x,6} = \psib_{x,2} \psib_{x,1}$ transform under the sextet representation of $SU(4)$, and cannot acquire an expectation value unless the $SU(4)$ symmetry breaks spontaneously. We believe our model is an example of four-fermion models, discussed in the introduction, where fermions become massive due to four-fermion condensates although fermion bilinear condensates vanish. As discussed in the introduction, since four-fermion interactions are irrelevant there is still a massless fermion phase at weak couplings.  However, as we will see in the next section, at sufficiently strong couplings fermions become massive without fermion bilinear condensates. As far as we can tell, no local order parameters exist that distinguish between the two phases. Thus, fermion mass generation in our model is a question of dynamics rather than symmetry.

Of course it is possible that the $SU(4)$ symmetry still breaks spontaneously at some intermediate couplings. In order to look for such breaking, we can measure correlation functions between the six fermion bilinears. The $SU(4)$ symmetry can be used to relate all of them to two independent correlation functions. In this work we compute the corresponding two independent susceptibilities
\begin{subequations}
\label{susdefs}
\begin{eqnarray}
\chi_{1} &=& \frac{1}{2L^3}\sum_{x,y,x\neq y}\Big\langle \phi_{x,1} \phi_{y,1}\Big\rangle,
\\
\chi_{2} &=& \frac{1}{2L^3}\sum_{x,y,x\neq y}\Big\langle \phi_{x,1} \phi_{y,2}\Big\rangle,
\end{eqnarray}
\end{subequations}
where expectation values are defined as
\begin{equation}
\Big\langle {\cal O} \Big\rangle = \frac{1}{Z}
\int [d\overline{\psi}\ d\psi]\ {\cal O}\ \mathrm{e}^{-S(\overline{\psi},\psi)}
\end{equation}
with $Z$ being the partition function. The presence of a condensate can be inferred when these susceptibilities diverge as $L^3$ for large values of $L$. Another observable that we compute is the local four-point condensate defined by
\begin{equation}
\rho_m = \frac{1}{L^3} \ \sum_x\ \langle \overline{\psi}_{x,1}\psi_{x,1}\overline{\psi}_{x,2}\psi_{x,2} \rangle.
\label{rhom}
\end{equation}
We find that this quantity increases rapidly near the phase transition.

\section{Connection to Yukawa models}
\label{sec3}

Our model can be obtained from many lattice Yukawa models, the simplest being the one in which two flavors of staggered fermions are coupled to an Ising field $\sigma_x=\pm 1$ and whose action is given by
\begin{equation}
S = S_0 \ -\ \kappa \sum_{x,\hat{\alpha}} \sigma_x \sigma_{x+\hat{\alpha}} - 
Y \sum _{i=1,2}\sum_{x} \sigma_x \overline{\psi}_{x,i} \psi_{x,i}.
\label{yukawa}
\end{equation}
Here $\kappa$ is the hopping parameter for the Ising field and $Y$ is the Yukawa coupling. When $\kappa = 0$ it is easy to show that the partition function of the above model is exactly the same as the partition function of our model if we set $U=Y^2$. Note however that the $SU(4)$ symmetry is broken in the Yukawa model for general values of $\kappa$ and is restored (but hidden) when $\kappa = 0$.

The Yukawa model at $\kappa = 0$ can be studied in perturbation theory for both small and large $Y$. At small coupling the fermionic correlation function up to second order in $Y^2$ is given by
\begin{equation}
\langle \psi_{x,i} \overline{\psi}_{y,j} \rangle = \delta_{ij} \Big(M^{-1}_{xy} + Y^4 (M^{-1}\Pi^{(3)} M^{-1})_{xy}\Big)
\end{equation}
where $\Pi^{(n)}_{xy} \equiv (M^{-1}_{xy})^n$ is a matrix in position space. Using this expression and the usual power counting rules of weak coupling perturbation theory that show four-fermion couplings are irrelevant, it is easy to verify that fermions remain massless. Similarly, the bosonic correlation function is given by
\begin{eqnarray}
\langle \overline{\psi}_{x,i}\psi_{x,i} \ \overline{\psi}_{y,j}\psi_{y,j} \rangle &=& \delta_{ij} 
\Big(\Pi^{(2)}_{xy} + Y^4 (\Pi^{(2)}\Pi^{(2)} \Pi^{(2)})_{xy}\Big) 
\nonumber \\
&& + Y^2(1-\delta_{ij}) (\Pi^{(2)}\Pi^{(2)})_{xy},
\end{eqnarray}
which goes to zero when $x$ and $y$ are separated far from each other showing that fermion bilinear condensates vanish. In the leading large coupling limit the fermionic correlation function is given by
\begin{equation}
\langle \psi_{x,i} \overline{\psi}_{y,j} \rangle = \delta_{ij}\ (\frac{1}{Y^2})^{2\ell+2} A_{xy}
\end{equation}
where $2\ell+1$ is the number bonds in the shortest path connecting sites $x$ and $y$. This number is odd since the correlation function is non-zero only if $x$ is an even site and $y$ is an odd site or vice versa. Thus, $\ell=0,1,2,...$ is fixed once $x$ and $y$ are chosen. In general there are many such paths, each of which we can label with the sites along the path as ${\cal P} = (x,z_1,z_2,...z_{2\ell-1},z_{2\ell},y)$. $A_{xy}$ is then given by a sum over amplitudes for each path,
\begin{equation}
A_{xy} = -\sum_{{\cal P}} (M_{x,z_1})^3M_{z_1,z_2}(M_{z_2,z_3})^3...M_{z_{2\ell-1} z_{2\ell}} (M_{z_{2\ell},y})^3.
\end{equation}
Thus, we see that the fermionic correlation function decays as exponentially as $\exp(-(4\ell+4)\ln Y)$ proving that fermions have become massive. Similarly, the bosonic two point correlation function is given by
\begin{eqnarray}
\langle \overline{\psi}_{x,i}\psi_{x,i} \ \overline{\psi}_{y,j}\psi_{y,j} \rangle &=& 
\delta_{ij} (\frac{1}{Y^2})^{2\ell+2} B_{xy}  
\nonumber \\
&& + (1-\delta_{ij}) (\frac{1}{Y^2})^{2\ell+3} C_{xy}.
\end{eqnarray}
If $i = j$ then $x$ and $y$ must have opposite parity like in the fermionic correlation function. This means the total number of bonds in the path is odd ($2\ell+1$) as before. But when $i \neq j$ then $x$ and $y$ must have the same parity for the correlation function to be non-zero. This means the total number of bonds in the path is even and is given by $2\ell+2$. We are excluding the possibility of $x=y$ here. The corresponding amplitudes are given by
\begin{eqnarray}
B_{xy} &=& \sum_{{\cal P}} (M_{x,z_1})^2(M_{z_1,z_2})^2...(M_{z_{2\ell-1} z_{2\ell}})^2(M_{z_{2\ell},y})^2
\nonumber \\
C_{xy} &=& \sum_{{\cal P}} (M_{x,z_1})^2(M_{z_1,z_2})^2...(M_{z_{2\ell-2} z_{2\ell-1}})^2(M_{z_{2\ell+1},y})^2
\nonumber \\
\end{eqnarray}
Thus, bosonic correlations also decay exponentially, which means the fermion bilinear condensates again vanish. This is the proof that there is a PMS phase at strong couplings.

The phase diagram of the Yukawa model was obtained in the mean field approximation by various groups \cite{Lee:1989mi,Stephanov:1990sq,Ebihara:1991mv}.  While each of these calculations yield slightly different results, they qualitatively agree that the generic phase diagram at some $\kappa \neq 0$ is given by scenario A in Fig.~\ref{pdiag}. For $N_f$ flavors of staggered fermions, the calculation at $\kappa=0$ discussed in \cite{Stephanov:1990sq} finds that the critical coupling between the PMW and the FM phase is given by
\begin{equation}
Y^w_c = \frac{d}{2(N_f-1)}
\end{equation}
and between the FM and the PMS phase it is given by
\begin{equation}
Y^s_c = \frac{d (N_f-1)}{2}.
\end{equation}
While $Y^w_c \neq Y_c^s$ for most values of $N_f$, for our model ($N_f=2$) $Y_c^w = Y_c^s = d/2$. This suggests that the FM phase may be absent for all values of $d$ consistent with scenario B of Fig.~\ref{pdiag}. However, the direct transition between the PMW and the PMS phase is found to be first order. A first principles Monte Carlo calculation is clearly necessary to understand what happens at intermediate couplings. In this work we provide evidence for the presence of a direct second order transition between the two phases.

\section{Fermion Bag Approach}
\label{sec4}

Traditional Monte Carlo methods for studying four-fermion field theories are based on introducing an auxiliary field to convert the four-fermion coupling into a fermion bilinear term in the action. In this work we use an alternative Monte Carlo approach introduced a few years ago, called the fermion bag approach \cite{PhysRevD.82.025007}. Interestingly, some sign problems that had remained unsolved with traditional methods, can be solved in the fermion bag approach \cite{Chandrasekharan:2012va, Chandrasekharan:2012fk,PhysRevB.89.111101}. The new approach has also helped in accurately computing the critical exponents with massless fermions \cite{Chandrasekharan:2012va,PhysRevLett.108.140404}. A review of the fermion bag approach can be found in \cite{Chandraepja13}.

\begin{figure}
\includegraphics[width=0.48\textwidth]{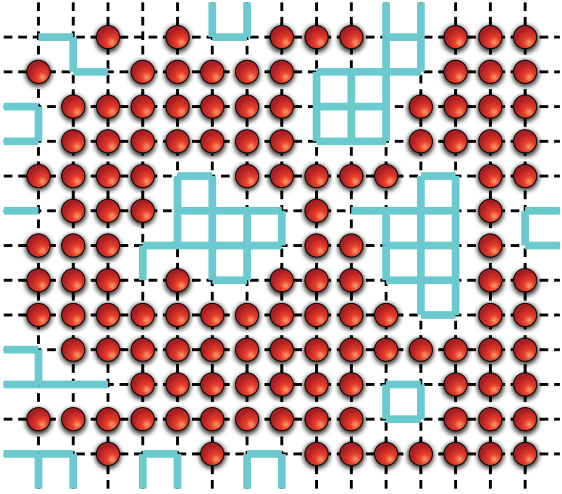}
\caption{\label{monoconf} An example of a monomer configuration $[n]$ showing free fermion bags on a two dimensional lattice.}
\end{figure}

In the fermion bag approach, we rewrite the partition function of our model as a sum over monomer configurations which we denote as $[n]$. Each monomer configuration is defined through a binary lattice field $n_x = 0,1$ which denotes the absence or presence of a monomer at the site $x$ respectively. Figure \ref{monoconf} shows an example of a monomer configuration on a two dimensional lattice. As explained in \cite{Chandrasekharan:2011vy}, there are two dual viewpoints to define fermion bags: (1) A strong coupling viewpoint where lattice sites that do not contain monomers are defined as free fermion bags inside which fermions of both flavors hop freely. Lattice sites with monomers form point-like fermion bags where fermions are pinned; (2) A weak coupling viewpoint where all monomer sites form a fermion bag and fermions of both flavors propagate freely between the monomer sites. Fermion bags of either viewpoint are uniquely defined for every monomer configuration.  An interesting feature of the strong coupling viewpoint is that at sufficiently strong couplings there are many distinct fermion bags, which we label as ${\cal B}=1,2...$, and fermions from one bag cannot hop to a different bag. In contrast in the weak coupling viewpoint there is a single fermion bag containing all monomer sites. Based on these two viewpoints we can write the partition function in two different but equivalent ways:
\begin{subequations}
\begin{eqnarray}
Z \ &=&\  \sum_{[n]}\ U^{N_m} \ \prod_{\cal B}\Big(\mathrm{Det}(W_{\cal B})\Big)^{2}
\\
Z \ &=&\  \Big(\mathrm{Det}(M)\Big) \sum_{[n]}\ U^{N_m} \ \Big(\mathrm{Det}(G)\Big)^{2},
\end{eqnarray}
\end{subequations}
where $N_m$ represents the number of monomers in the configuration, $M$ is the free staggered fermion matrix defined in (\ref{freef}), $W_{\cal B}$ represents the free staggered fermion matrix connecting the sites within the bag ${\cal B}$, and $G$ represents a $N_m \times N_m$ free staggered propagator matrix connecting monomer sites. The elements of $G$ are given by
\begin{equation}
G_{x,y} = \frac{-i}{L^3}
\sum_{k} \mathrm{e}^{ik\cdot (x-y)} 
\frac{\sum_{\alpha'} \eta_{x,\alpha'}\sin k_{\alpha'}}{\sum_\alpha\sin^2k_\alpha}
\end{equation}
where $k \equiv (k_1,k_2,k_3)$ where $k_\alpha= (2n+1)\pi/L, n=0,1..,L-1$ due to anti-periodic boundary conditions. At weak couplings there are very few monomers and the weak coupling viewpoint becomes more useful for calculations and the Boltzmann weight of each monomer configuration is nothing but the sum over all Feynman diagrams. Thus, the weak coupling viewpoint is exactly identical to the well known diagrammatic determinantal Monte Carlo methods \cite{PhysRevLett.81.2514,PhysRevB.76.035116,PhysRevB.87.205102,PhysRevB.79.035320}. On the other hand at strong couplings, when the number of monomers becomes comparable to the volume, the strong coupling view point becomes useful for calculations since free fermion bags become small. As we discuss below, it is also easy to understand some of the strong coupling results of the previous section intuitively.

Expressions for observables can also be derived easily in the fermion bag approach. For example, in the strong coupling viewpoint the two point fermion correlation function is given by
\begin{equation}
\langle \psi_{x,i}\ \psib_{y,i}\rangle \ =\  \frac{1}{Z}\ 
\sum_{[n]}\ U^{N_m} \ \prod_{\cal B}\Big(\mathrm{Det}(W_{\cal B})\Big)^{2}\ W^{-1}_{{\cal B};x,y},
\end{equation}
where $W^{-1}_{{\cal B};x,y}$ is the inverse of the Dirac operator within the free fermion bag ${\cal B}$ that contains the sites $x$ and $y$. It is understood that when either of the sites $x$ or $y$ contains a monomer, that configuration does not contribute to the correlation function. Further, since fermions cannot hop from one fermion bag to another, $x$ and $y$ are also forced to be within the same bag. With this insight it is easy to see why fermion correlations decay exponentially at strong couplings. Since the lattice is filled with monomers, large fermion bags are suppressed exponentially and fermions are confined within small regions. 

The argument that shows that even bosonic correlations decay exponentially is more subtle. In principle, it is possible to have a single insertion of $\overline{\psi}_{i,x}\psi_{i,x}$ within special fermion bags that allow a zero mode in the matrix $W_{\cal B}$.  Clearly, such bags do not contribute to the partition function since without the insertion of $\overline{\psi}_{i,x}\psi_{i,x}$ the determinant $\mathrm{Det}(W_{\cal B})$ vanishes. However, with the insertion of $\overline{\psi}_{i,x}\psi_{i,x}$ one row and one column are removed from the matrix and then the determinant no longer vanishes. This is very similar to the argument of how instantons can contribute to the chiral condensate in single flavor QCD. However, since there are two flavors in our model and $\overline{\psi}_{i,x}\psi_{i,x}$ only involves the flavor $i$, the determinant of the other flavor still vanishes due to the zero mode in $W_{\cal B}$ of the second flavor. Thus, single insertion of a fermion bilinear is forbidden in our model. For this reason bosonic correlation functions also get contribution only when both $x$ and $y$ are within the same bag. For example the expression for one of the correlation functions is given by
\begin{eqnarray}
\langle \psib_x\psi_{x,i}\ \psib_{y,i}\psi_{y,i}\rangle \ &=&\  \frac{1}{Z}\ 
\sum_{[n]}\ U^{N_m} \ \prod_{\cal B}\Big(\mathrm{Det}(W_{\cal B})\Big)^{2}
\nonumber \\
&& \ \ \ \ \ \ \times \Big(W^{-1}_{{\cal B};x,y}\Big)^2.
\end{eqnarray}
Since $x$ and $y$ are within the bag, it too decays exponentially at sufficiently large coupling as we found in the previous section.

\section{Monte Carlo Algorithms}
\label{sec5}

We have constructed three different Monte Carlo algorithms to update the monomer configurations $[n]$. The first is a block algorithm that creates, destroys and moves monomers within blocks. The second is a worm algorithm that creates a pair of {\em half-monomers} near each other (i.e., $\overline{\psi}_{x,i}\psi_{x,i}$ and $\overline{\psi}_{x,j}\psi_{x,j}$) and moves one around until it returns to the vicinity of its pair and detailed balance allows us to destroy the pair. As the half monomer moves around it can create or destroy other monomers. The third is a heat bath sweep algorithm that picks a random site along with every other site on the lattice and performs a heat-bath update on the two sites. Below we provide more details of the three algorithms.

\subsection{Block Algorithm}

In this algorithm a site on the lattice is chosen at random and a local block consisting of $6^3$ sites in its vicinity is chosen to be updated, while the sites outside the block are held fixed. Since much of matrix whose determinant is being calculated does not change during the block update, the computational cost is significantly reduced. Each update within the block involves two steps, adding and removing monomers in pairs followed by moving individual monomers around. Each of these steps is performed roughly $100$ times during the block update.

The first step of the update that involves adding and removing monomers in pairs is performed as follows:
\begin{enumerate}
\item Choose to either add or remove monomers with probability half.
\item If the decision is to add monomers, compute $k_{\rm free}$, the number of pairs of free sites (one even and one odd) within the block in the current configuration where two monomers can be added. Choose one of these pairs at random and add monomers to the sites with probability
\begin{equation}
P = \frac{\Omega_{\rm final} k_{\rm free}}{\Omega_{\rm initial} k_{\rm filled}}.
\end{equation}
With probability $1-P$ keep the old configuration. In the expression above, $k_{\rm filled}$ is the number of pairs of monomer sites (one even and one odd) within the block in the new configuration from where monomers can be removed and $\Omega_{\rm final}$ and $\Omega_{\rm initial}$ are the Boltzmann weights of the final and the initial configurations with and without the two monomers.
\item  If the decision is to remove monomers compute $k_{\rm filled}$, the number of pairs of monomer sites (one even and one odd) within the block in the current configuration from where monomers can be removed. Choose one of these pairs at random and remove monomers from the sites with probability
\begin{equation}
P = \frac{\Omega_{\rm final} k_{\rm filled}}{\Omega_{\rm initial} k_{\rm free}}.
\end{equation}
With probability $1-P$ keep the old configuration. In the above expression $k_{\rm free}$ is the number of free sites (one even and one odd) within the block in the new configuration where monomers can be added and $\Omega_{\rm initial}$ and $\Omega_{\rm final}$ are the Boltzmann weights of the final and initial configurations with and without the two monomers.
\end{enumerate}
The calculation of $\Omega_{\rm final}/\Omega_{\rm initial}$ involves computing a ratio of two determinants with one row and one column added or subtracted and is the most computationally intensive step in the algorithm.

The second step of the update involves moving monomers from one site to another site with the same parity that does not contain a monomer. For high acceptance we move monomers only to an allowed neighboring site but repeat the process many times. The update is as follows:
\begin{enumerate}
\item Pick a monomer site $x$ at random.
\item Pick at random one of the twelve next-to-nearest-neighbor sites of $x$ with the same parity as $x$. We will refer to this site as $y$. Note that $x$ and $y$ belong to diagonally opposite pairs of sites of an elementary square.
\item If $y$ contains a monomer then the update stops. Otherwise the monomer located at $x$ is moved to $y$ with probability
\begin{equation}
P = \frac{\Omega_{\rm final}}{\Omega_{\rm initial}}.
\end{equation}
With probability $1-P$ the monomer at $x$ is left untouched.
\end{enumerate}
Since the move monomer step is repeated many times, monomers diffuse around within the block.

\subsection {Worm Algorithm}

Past experience shows that worm type algorithms are able to reduce autocorrelation times significantly since worm updates are based on correlations within the system \cite{PhysRevLett.87.160601}. A worm type algorithm can be designed for our model as we discuss here. The idea is based on sampling the bosonic correlation function through the worm. Since a monomer is the presence of a four-point vertex $\overline{\psi}_{x,1}\psi_{x,1}\overline{\psi}_{x,2}\psi_{x,2}$ at the site $x$, a half monomer is the presence of a fermion bilinear vertex at the site. Further when the two half monomers are located at the sites with the same parity then they are forced to belong to different flavors and vice versa. In order to understand this algorithm it is useful to define a compatibility condition for two sites $x$ and $y$. Two sites $x$ and $y$ are defined to be compatibile if: (1) $x$ and $y$ have different site-parities, but the same filling i.e either both are free sites or both have monomers, or (2) $x$ and $y$ have the same site-parity, but have opposite filling i.e. one is a free site and the other has a monomer. If $x$ and $y$ are compatible, the head of the worm can in principle move from $x$ to $y$ or vice-versa. Whether it really moves depends of course on probabilities that satisfy detailed balance. On the other hand when $x$ and $y$ are incompatible, the head of the worm cannot move between the two sites. It is also useful to define a set of {\em nearby} sites for a given site $x$. The worm will explore these nearby sites as it proceeds forward. We will define nearby sites to mean: the 6 nearest-neighbor sites, the 12 next-to-nearest neighbor sites and 6 sites that are two lattice spacings away along each direction. Thus, at each step the worm will explore one of 24 nearby sites as it moves ahead. Based on these definitions, the worm update is constructed as follows:
\begin{enumerate}
\item Determine all the possible pairs of nearby sites that satisfy the compatibility conditions described above. Define $k_{\rm pair}$ as the number of such pairs and pick one compatible pair at random. Label the pair of sites randomly as $x$(tail) and $y$(head). The state of the site $x$, whether it is free or contains a monomer, is noted.
\item Create worm: Introduce half-monomers at $x$ and $y$ with probability
\begin{equation}
 P = \frac{\Omega_{\rm final}}{\Omega_{\rm initial}} \frac{e\ k_{pairs}}{12 \ L^3}
\end{equation} 
where $e$ is an enhancement factor to increase the acceptance and $L^3$ is the lattice volume. With probability $1-P$ the update ends, otherwise proceed to the next step.
\item Move worm-head: Pick one of $24$ nearby sites of $y$ at random. Call this site as $z$. If the site $z$ is the first site $x$, proceed to the ``destroy worm'' step. Otherwise try to move the worm-head from $y$ to $z$. There are two possibilities: (1) If $y$ and $z$ are sites with opposite parity, propose a new configuration where $y$ has the opposite filling state of $z$ and move the half-monomer to $z$, (2) If $y$ and $z$ are sites with same parity, propose a new configuration where $y$ has the same filling state of $z$ and move the half-monomer to $z$. Both these proposals are accepted with the Metropolis acceptance probability
\begin{equation}
 P = \frac{\Omega_{\rm final}}{\Omega_{\rm initial}}
\end{equation}
and the worm-head is moved to $z$ from $y$. If the proposal is rejected the worm-head remains at $y$. The ``move worm-head'' step is repeated again.
\item Destroy worm: Propose to remove the half-monomers located at $y$ and $x$ by restoring the site $x$ to the same state as it was in the first step when the update started and restoring $y$ to the unique state that makes it compatible with $x$. Accept the proposal with probability
\begin{equation}
P = \frac{\Omega_{\rm final}}{\Omega_{\rm initial}} \frac{12 \ L^3 }{ e \ k_{pairs} }
\end{equation}
where $k_{pairs}$ is calculated just like the first step but for the final configuration without the half monomers. If the proposal is accepted the update stops. Otherwise the ``move worm-head'' step is repeated.
\end{enumerate}
Since configurations with two half monomers can often have much smaller Boltzmann weights as compared to those without the half monomers, we have introduced an enhancement factor $e = 10$ in the step that creates the worm. However in order to ensure detailed balance we also divide by this factor in the step that destroys the worm.

The worm algorithm can be used to measure $\chi_1$ and $\chi_2$ easily. Let $n_o$ ($n_s$) be the number of $y$ sites generated during the worm update that have the opposite(same) parity as $x$. Then it is easy to argue that
\begin{equation}
\chi_1 = \frac{1}{e} \langle n_o\rangle,\ \ \ \chi_2 = \frac{1}{e} \langle n_s\rangle. 
\end{equation}
The re-weighting factor $e$ is necessary since the half-monomer sector was produced with an enhanced weight. Thus, the total number of steps during the worm update is proportional to $\chi_1 + \chi_2$ on an average. Since in our model we find that the susceptibilities do not grow with volume a single worm update only touches a few lattice sites in the neighborhood of the first site. Hence we have to repeat the worm update sufficient number of times starting with different initial sites in order to ensure that the entire lattice has been updated.

\begin{figure*}
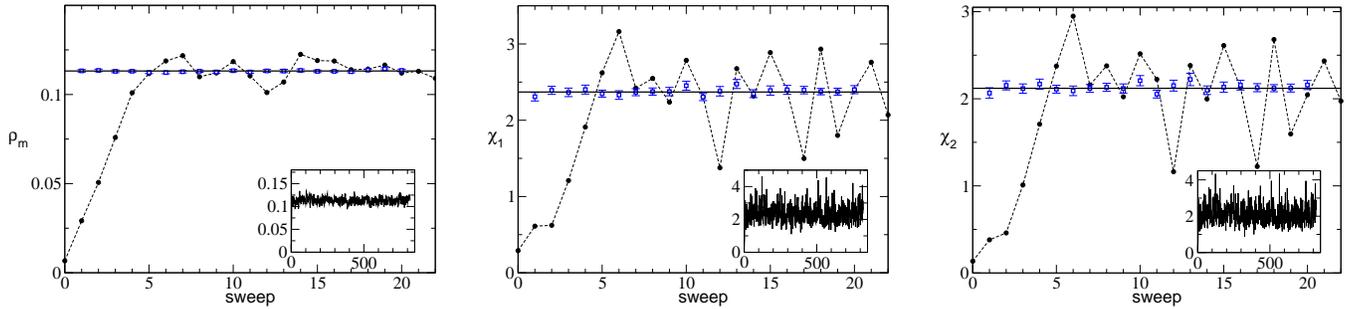

\hbox{
\includegraphics[width=0.32\textwidth]{mono.eps}
\hspace{0.19in}
\includegraphics[width=0.3\textwidth]{uu.eps}
\hspace{0.19in}
\includegraphics[width=0.3\textwidth]{ud.eps}
}
\caption{\label{equil} Plots of equilibration for the three observables $\rho_m$, $\chi_1$ and $\chi_2$, starting from a configuration with zero monomers at $L=20$, $U=0.95$. The insets show the Monte Carlo time history for 900 sweeps using ALG2. The average of the data from the inset is shown as a solid line in the main plots. The open squares are average data from 500 independent runs after a single sweep starting from an equilibrated configuration. The plot demonstrates that instead of running a single computer for many sweeps, one can run many computers for a single sweep and average the data.}
\end{figure*}

\subsection{Heat-Bath Sweep Algorithm}
Although the worm algorithm is normally efficient in computing $\chi_1$ and $\chi_2$, we have noticed some rare but large fluctuations in our data especially for large values of $U$. Worm algorithms can in principle generate rare runaway loops which can cause problems with statistics and computation of errors. 
Hence, in order to check our errors we devised a simple heat-bath sweep algorithm, which guarantees bounded fluctuations. On the other hand the heat-bath sweep is not computationally efficient since the algorithm attempts to add and subtract monomers over long distances. In our work we have only used it as a method to check the accuracy of our worm algorithm results on smaller lattices. The algorithm proceeds as follows:
\begin{enumerate}
\item Pick a site at random (say $x$).
\item Pick every site $y$ on the lattice in a fixed sequence and perform the following heat bath update. If $x$ and $y$ are incompatible sites proceed to the next $y$. Otherwise propose a new configuration where both sites $x$ and $y$ are flipped (i.e., monomer sites are changed to free sites and vice versa). Let $\Omega_{\rm new}$ and $\Omega_{\rm old}$ be the Boltzmann weights of the new and the old configurations. The proposed new configuration is accepted with the heat bath probability
\begin{equation}
P = \frac{\Omega_{\rm new}}{\Omega_{\rm new} + \Omega_{\rm old}}.
\end{equation}
With probability $1-P$ the old configuration is retained and the update moves on to the next site $y$.
\item Once all the sites $y$ are visited the update ends.
\end{enumerate}
This algorithm also allows us to compute the two susceptibilities  $\chi_{1}$ and $ \chi_{2}$ during the heat bath sweep. One can show that
\begin{equation}
\chi_i = \frac{1}{2}\Bigg \langle 
\sum'_y \frac{\sqrt{\Omega_{\rm new}\Omega_{\rm old}}}{\Omega_{\rm new} + \Omega_{\rm old}} \Bigg \rangle
\end{equation}
where prime on the sum indicates that the sites $y$ used in the sum should: (1) have the opposite parity as $x$ for $\chi_1$ and the same parity as $x$ for $\chi_2$ and (2) be compatible with the $x$.

\subsection{Equilibration, Auto-correlation and Parallelization}

We have used the block algorithm (or ALG1), the worm algorithm (or ALG2) and the heat-bath sweep algorithm (or ALG3) as a cross check against each other to make sure they are free of errors. These tests along with comparisons with some exact calculations are discussed in the appendix. In order to study equilibration and autocorrelations we define the concept of a {\em sweep}, as performing the required number of local updates such that all lattice sites are stochastically flipped at least once. For example in the block algorithm we pick roughly $L^3/6^3$ random blocks in a sweep. On the other hand since the worm update involves choosing a site at random and updating a few sites within its neighborhood, a sweep consists of repeating the worm update at least a volume number of times. Each heat bath update on the other hand is exactly one sweep. 

As in previous studies \cite{Chandrasekharan:2011vy} we have observed that worm algorithms based on the fermion bag approach usually produce independent equilibrated configurations within a few sweeps independent of the lattice size. This continues to be true even in our work. We provide some evidence for this in Fig.~\ref{equil} where we show the Monte Carlo time history of our three observables at $L=20$ and $U=0.95$ for 900 sweeps (in the inset) and the first 20 sweeps are shown in the main graph. The solid lines in the main graphs show the average obtained from the whole data set. As one can see, the monomer number reaches the average value in roughly about 5 sweeps and then begins to fluctuate.

If we make the drastic assumption that once equilibration is reached, a single sweep is sufficient to produce another independent configuration, then using several hundred computing cores each starting with an equilibrated configuration but different random number sequences, we should be able to generate an independent configuration after a single sweep from each computer core. We can then average the data from all the cores and propose it as the final average. We can of course continue the runs of each of the cores for several sweeps if necessary and monitor the fluctuations. In Fig.~\ref{equil} the solid squares represent such an average over 500 computer cores for 20 sweeps. It is clear that after each sweep the data from the 500 independent cores produces a number consistent with the average over 900 sweeps on a single core. This feature continues to hold at other lattice sizes and couplings, some of which are shown in the appendix. Based on this result, in our study we use several hundred cores in parallel and run for 5-10 sweeps, where each core starts from an equilibrated configuration. The final answer is obtained as an average over such short runs on hundreds of cores. While we are confident of our errors, in order to be conservative we multiply them by a factor of two uniformly across the board when we analyze our data.

\begin{table*}[!t]
\begin{center}
\begin{tabular}{|r|r|r|r|r||r|r|r|r|r|}
\hline
 $U$ & $L$ & $\rho_m$ & $ \chi_1 $ &  $ \chi_2 $ & $U$ & $L$ & $\rho_m$ & $ \chi_1 $ &  $ \chi_2 $ \\
\hline
0.200 & 8 & 301(7)$\times10^{-5}$ & 440(9)$\times10^{-3}$ & 68(1)$\times10^{-3}$ & 1.050 & 16 & 191(2)$\times10^{-3}$ & 456(9)$\times10^{-2}$ & 433(9)$\times10^{-2}$  \\
0.400 & 8 & 127(1)$\times10^{-4}$ & 476(5)$\times10^{-3}$ & 149(2)$\times10^{-3}$ & 1.080 & 16 & 231(2)$\times10^{-3}$ & 513(6)$\times10^{-2}$ & 490(6)$\times10^{-2}$  \\
0.600 & 8 & 312(2)$\times10^{-4}$ & 552(4)$\times10^{-3}$ & 261(2)$\times10^{-3}$ & 1.100 & 16 & 259(2)$\times10^{-3}$ & 500(6)$\times10^{-2}$ & 478(6)$\times10^{-2}$  \\
0.800 & 8 & 642(4)$\times10^{-4}$ & 717(5)$\times10^{-3}$ & 457(3)$\times10^{-3}$ & 1.120 & 16 & 288(2)$\times10^{-3}$ & 468(6)$\times10^{-2}$ & 447(6)$\times10^{-2}$  \\
1.000 & 8 & 1340(7)$\times10^{-4}$ & 1198(9)$\times10^{-3}$ & 967(8)$\times10^{-3}$ & 1.150 & 16 & 338(1)$\times10^{-3}$ & 393(2)$\times10^{-2}$ & 372(2)$\times10^{-2}$  \\
1.050 & 8 & 1674(10)$\times10^{-4}$ & 145(1)$\times10^{-2}$ & 1228(10)$\times10^{-3}$ & 1.200 & 16 & 415(1)$\times10^{-3}$ & 287(2)$\times10^{-2}$ & 267(2)$\times10^{-2}$  \\
1.080 & 8 & 195(1)$\times10^{-3}$ & 164(1)$\times10^{-2}$ & 142(1)$\times10^{-2}$ & 0.800 & 20 & 643(2)$\times10^{-4}$ & 1026(5)$\times10^{-3}$ & 753(5)$\times10^{-3}$  \\
1.100 & 8 & 217(1)$\times10^{-3}$ & 178(1)$\times10^{-2}$ & 157(1)$\times10^{-2}$ & 0.880 & 20 & 857(2)$\times10^{-4}$ & 1440(9)$\times10^{-3}$ & 1181(8)$\times10^{-3}$  \\
1.120 & 8 & 245(1)$\times10^{-3}$ & 193(1)$\times10^{-2}$ & 172(1)$\times10^{-2}$ & 0.900 & 20 & 922(3)$\times10^{-4}$ & 160(1)$\times10^{-2}$ & 1351(10)$\times10^{-3}$  \\
1.150 & 8 & 290(2)$\times10^{-3}$ & 2093(10)$\times10^{-3}$ & 1885(9)$\times10^{-3}$ & 0.930 & 20 & 1038(3)$\times10^{-4}$ & 199(1)$\times10^{-2}$ & 174(1)$\times10^{-2}$  \\
1.180 & 8 & 341(2)$\times10^{-3}$ & 2159(8)$\times10^{-3}$ & 1956(8)$\times10^{-3}$ & 0.950 & 20 & 1133(7)$\times10^{-4}$ & 239(4)$\times10^{-2}$ & 214(4)$\times10^{-2}$  \\
1.200 & 8 & 377(2)$\times10^{-3}$ & 2133(8)$\times10^{-3}$ & 1932(7)$\times10^{-3}$ & 0.960 & 20 & 1180(3)$\times10^{-4}$ & 262(2)$\times10^{-2}$ & 238(2)$\times10^{-2}$  \\
1.220 & 8 & 412(2)$\times10^{-3}$ & 2063(8)$\times10^{-3}$ & 1864(7)$\times10^{-3}$ & 0.970 & 20 & 1239(3)$\times10^{-4}$ & 295(2)$\times10^{-2}$ & 271(2)$\times10^{-2}$  \\
1.240 & 8 & 447(2)$\times10^{-3}$ & 1952(8)$\times10^{-3}$ & 1758(7)$\times10^{-3}$ & 0.980 & 20 & 1294(9)$\times10^{-4}$ & 325(7)$\times10^{-2}$ & 301(7)$\times10^{-2}$  \\
0.200 & 12 & 302(4)$\times10^{-5}$ & 470(6)$\times10^{-3}$ & 78(1)$\times10^{-3}$ & 1.000 & 20 & 1436(9)$\times10^{-4}$ & 412(8)$\times10^{-2}$ & 388(8)$\times10^{-2}$  \\
0.400 & 12 & 1275(9)$\times10^{-5}$ & 515(4)$\times10^{-3}$ & 173(1)$\times10^{-3}$ & 1.030 & 20 & 173(1)$\times10^{-3}$ & 58(1)$\times10^{-1}$ & 56(1)$\times10^{-1}$  \\
0.600 & 12 & 312(1)$\times10^{-4}$ & 615(3)$\times10^{-3}$ & 314(2)$\times10^{-3}$ & 1.050 & 20 & 195(1)$\times10^{-3}$ & 630(7)$\times10^{-2}$ & 608(7)$\times10^{-2}$  \\
0.800 & 12 & 645(3)$\times10^{-4}$ & 873(6)$\times10^{-3}$ & 606(5)$\times10^{-3}$ & 1.080 & 20 & 234(2)$\times10^{-3}$ & 631(7)$\times10^{-2}$ & 609(7)$\times10^{-2}$  \\
0.880 & 12 & 856(3)$\times10^{-4}$ & 1107(6)$\times10^{-3}$ & 852(5)$\times10^{-3}$ & 1.100 & 20 & 2631(8)$\times10^{-4}$ & 575(4)$\times10^{-2}$ & 554(3)$\times10^{-2}$  \\
0.900 & 12 & 921(3)$\times10^{-4}$ & 1189(6)$\times10^{-3}$ & 938(5)$\times10^{-3}$ & 1.120 & 20 & 2928(9)$\times10^{-4}$ & 507(4)$\times10^{-2}$ & 489(7)$\times10^{-2}$  \\
0.930 & 12 & 1034(3)$\times10^{-4}$ & 1354(7)$\times10^{-3}$ & 1107(6)$\times10^{-3}$ & 1.150 & 20 & 335(1)$\times10^{-3}$ & 414(5)$\times10^{-2}$ & 394(5)$\times10^{-2}$  \\
0.950 & 12 & 1114(10)$\times10^{-4}$ & 148(3)$\times10^{-2}$ & 124(2)$\times10^{-2}$ & 1.200 & 20 & 413(1)$\times10^{-3}$ & 291(4)$\times10^{-2}$ & 272(4)$\times10^{-2}$  \\
0.960 & 12 & 1169(4)$\times10^{-4}$ & 1580(9)$\times10^{-3}$ & 1338(9)$\times10^{-3}$ & 0.880 & 24 & 855(2)$\times10^{-4}$ & 1548(9)$\times10^{-3}$ & 1290(8)$\times10^{-3}$  \\
0.980 & 12 & 127(1)$\times10^{-3}$ & 179(3)$\times10^{-2}$ & 155(3)$\times10^{-2}$ & 0.900 & 24 & 920(3)$\times10^{-4}$ & 175(2)$\times10^{-2}$ & 149(1)$\times10^{-2}$  \\
1.000 & 12 & 139(2)$\times10^{-3}$ & 199(5)$\times10^{-2}$ & 175(4)$\times10^{-2}$ & 0.930 & 24 & 1039(2)$\times10^{-4}$ & 229(2)$\times10^{-2}$ & 203(2)$\times10^{-2}$  \\
1.030 & 12 & 164(2)$\times10^{-3}$ & 251(7)$\times10^{-2}$ & 228(6)$\times10^{-2}$ & 0.950 & 24 & 1133(3)$\times10^{-4}$ & 281(3)$\times10^{-2}$ & 257(3)$\times10^{-2}$  \\
1.050 & 12 & 185(2)$\times10^{-3}$ & 284(6)$\times10^{-2}$ & 262(5)$\times10^{-2}$ & 0.960 & 24 & 1182(3)$\times10^{-4}$ & 316(3)$\times10^{-2}$ & 292(3)$\times10^{-2}$  \\
1.080 & 12 & 222(3)$\times10^{-3}$ & 337(5)$\times10^{-2}$ & 315(5)$\times10^{-2}$ & 0.970 & 24 & 1240(3)$\times10^{-4}$ & 362(3)$\times10^{-2}$ & 338(3)$\times10^{-2}$  \\
1.100 & 12 & 249(3)$\times10^{-3}$ & 361(5)$\times10^{-2}$ & 339(5)$\times10^{-2}$ & 0.980 & 24 & 1302(3)$\times10^{-4}$ & 417(4)$\times10^{-2}$ & 393(4)$\times10^{-2}$  \\
1.120 & 12 & 2817(8)$\times10^{-4}$ & 366(1)$\times10^{-2}$ & 345(1)$\times10^{-2}$ & 1.000 & 24 & 1456(3)$\times10^{-4}$ & 555(3)$\times10^{-2}$ & 532(3)$\times10^{-2}$  \\
1.150 & 12 & 3300(9)$\times10^{-4}$ & 346(1)$\times10^{-2}$ & 325(1)$\times10^{-2}$ & 1.020 & 24 & 1637(5)$\times10^{-4}$ & 690(5)$\times10^{-2}$ & 667(5)$\times10^{-2}$  \\
1.180 & 12 & 3791(9)$\times10^{-4}$ & 305(1)$\times10^{-2}$ & 284(1)$\times10^{-2}$ & 1.030 & 24 & 1747(4)$\times10^{-4}$ & 745(4)$\times10^{-2}$ & 723(4)$\times10^{-2}$  \\
1.200 & 12 & 4104(8)$\times10^{-4}$ & 2753(9)$\times10^{-3}$ & 254(2)$\times10^{-2}$ & 1.050 & 24 & 1971(6)$\times10^{-4}$ & 782(4)$\times10^{-2}$ & 759(4)$\times10^{-2}$  \\
0.800 & 16 & 646(3)$\times10^{-4}$ & 967(6)$\times10^{-3}$ & 695(5)$\times10^{-3}$ & 1.070 & 24 & 2226(5)$\times10^{-4}$ & 736(4)$\times10^{-2}$ & 72(1)$\times10^{-1}$  \\
0.880 & 16 & 860(2)$\times10^{-4}$ & 1297(7)$\times10^{-3}$ & 1039(6)$\times10^{-3}$ & 1.080 & 24 & 235(2)$\times10^{-3}$ & 698(8)$\times10^{-2}$ & 678(8)$\times10^{-2}$  \\
0.900 & 16 & 922(2)$\times10^{-4}$ & 1419(8)$\times10^{-3}$ & 1166(7)$\times10^{-3}$ & 0.900 & 28 & 924(2)$\times10^{-4}$ & 190(1)$\times10^{-2}$ & 165(1)$\times10^{-2}$  \\
0.930 & 16 & 1037(3)$\times10^{-4}$ & 1689(10)$\times10^{-3}$ & 1442(9)$\times10^{-3}$ & 0.930 & 28 & 1039(2)$\times10^{-4}$ & 253(2)$\times10^{-2}$ & 228(2)$\times10^{-2}$  \\
0.950 & 16 & 1126(9)$\times10^{-4}$ & 195(4)$\times10^{-2}$ & 170(4)$\times10^{-2}$ & 0.950 & 28 & 1132(2)$\times10^{-4}$ & 325(2)$\times10^{-2}$ & 300(2)$\times10^{-2}$  \\
0.960 & 16 & 1171(4)$\times10^{-4}$ & 207(2)$\times10^{-2}$ & 183(2)$\times10^{-2}$ & 0.960 & 28 & 1186(2)$\times10^{-4}$ & 379(3)$\times10^{-2}$ & 354(3)$\times10^{-2}$  \\
0.970 & 16 & 1229(2)$\times10^{-4}$ & 2271(9)$\times10^{-3}$ & 2031(9)$\times10^{-3}$ & 0.970 & 28 & 1244(3)$\times10^{-4}$ & 444(4)$\times10^{-2}$ & 420(4)$\times10^{-2}$  \\
0.980 & 16 & 1278(10)$\times10^{-4}$ & 245(5)$\times10^{-2}$ & 221(5)$\times10^{-2}$ & 0.980 & 28 & 1308(3)$\times10^{-4}$ & 522(4)$\times10^{-2}$ & 498(4)$\times10^{-2}$  \\
1.000 & 16 & 143(1)$\times10^{-3}$ & 301(6)$\times10^{-2}$ & 278(6)$\times10^{-2}$ & 1.000 & 28 & 1463(3)$\times10^{-4}$ & 716(6)$\times10^{-2}$ & 691(5)$\times10^{-2}$  \\
1.030 & 16 & 167(2)$\times10^{-3}$ & 390(8)$\times10^{-2}$ & 367(8)$\times10^{-2}$  \\
\hline
\end{tabular}
\end{center}
\caption{\label{tab:alldata} Monte Carlo results for $\rho_m$, $\chi_1$ and $\chi_2$ as a function of $U$ and $L$. Being conservative, all errors are multiplied by a factor of two as discussed at the end of section \ref{sec5}.}
\end{table*} 

\begin{figure}[t]
\vspace{0.18in}
\includegraphics[width=0.45\textwidth]{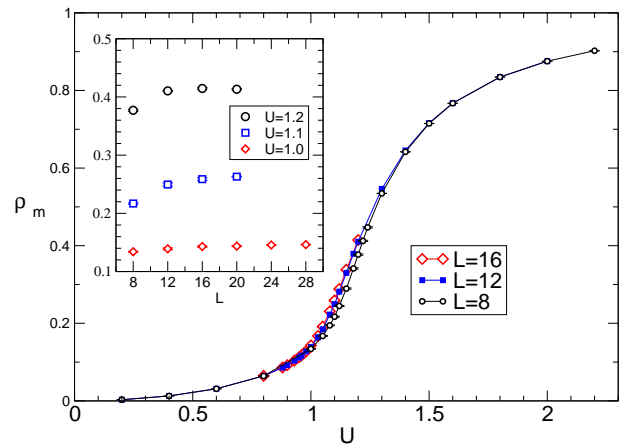}
\caption{\label{rhomono} The variation of the monomer density $\rho_m$ (a four-point condensate) as a function of $U$ at $L=8,12$ and $16$. The inset shows the change in $\rho_m$ as a function of $L$ at $U=1.0,1.1$ and $1.2$ where the variation is the maximum. By $L=16$ we find that $\rho_m$ has reached its thermodynamic limit at all values of $U$.}
\end{figure}

\begin{figure*}[t]
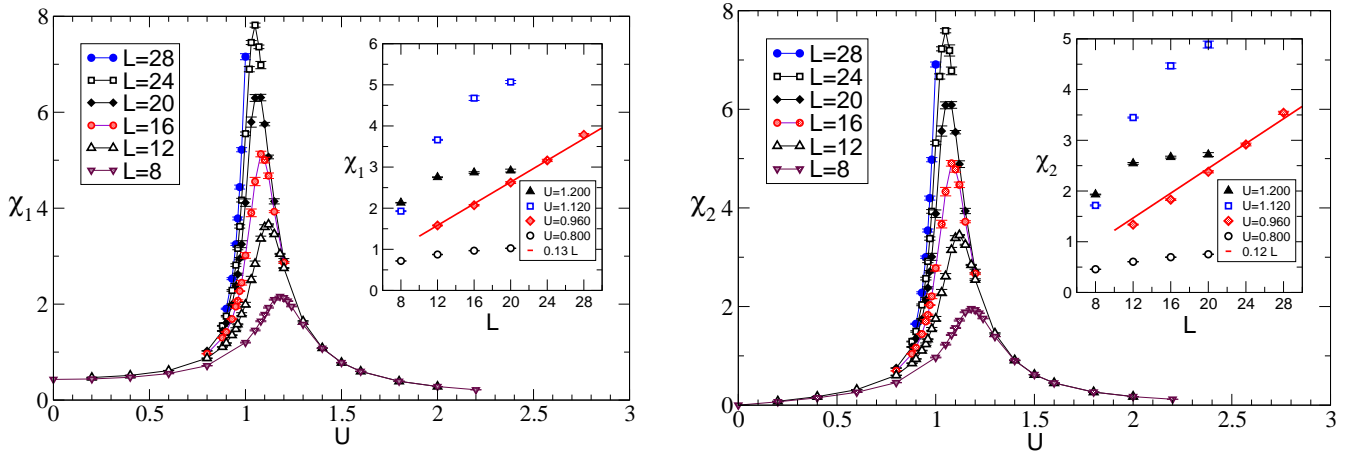

\vspace{0.18in}
\hbox{
\begin{minipage}[c]{0.48\textwidth}
\includegraphics[width=0.97\textwidth]{uu_cvU.eps}
\end{minipage}
\hspace{0.17in}
\begin{minipage}[c]{0.48\textwidth}
\includegraphics[width=\textwidth]{ud_cvU.eps}
\end{minipage}
}
\caption{\label{cvUL}  Plots of the susceptibilities $\chi_1$ (left) and $\chi_2$ (right) as a function of the coupling constant $U$ for lattice sizes ranging from $L=8$ to $L=28$. The inset shows the finite size scalings in the critical region. There is no sign of the $L^3$ divergence expected in the presence of a non-zero fermion bilinear condensate. A roughly linear divergence appears in the critical region consistent with a second order critical scaling.}
\end{figure*}

\section {Analysis and Results}
\label{sec6}

Based on weak and strong coupling analysis we have already argued in section \ref{sec3} that the model contains at least two phases: A PMW phase at weak couplings characterized by massless fermions and a PMS phase at strong couplings characterized by massive fermions without fermion bilinear condensates. While one mean field analysis suggested a direct first order transition between the two phases, another analysis found an intermediate phase with spontaneous symmetry breaking. In this section we analyze our Monte Carlo results and argue that our model in fact contains a single second order phase transition between the two phases. In table \ref{tab:alldata} we tabulate all our data. 

We first focus on the average monomer density $\rho_m$ defined in Eq.~(\ref{rhom}) as a function of $U$. This is plotted in Fig.~\ref{rhomono} for $L=8,12$ and $16$. We find the density to be a smooth function of $U$ for all values of $L$ and most importantly the thermodynamic limit is reached by $L=16$ for all values of $U$. There is no evidence for a first order transition. However, since there should at least be one transition as a function of $U$, the quick but smooth rise of the monomer density around $U \approx 1$ can be taken to be a signal for such a second order transition. The lack of any other feature in $\rho_m$ as a function of $U$ also provides evidence that there is only a single phase transition. 

Since $\rho_m$ is not a critical quantity, we need to look at other observables like the chiral susceptibilities $\chi_1$ and $\chi_2$ defined in Eq.~(\ref{susdefs}), in order to understand the properties of the phase transition. These susceptibilities couple to long wavelength modes of the theory and will diverge at a second order critical point. Another interesting feature of the definitions of $\chi_1$ and $\chi_2$ is that the disconnected component has not been subtracted. Hence in the presence of non-zero fermion bilinear condensates we expect both $\chi_1$ and $\chi_2$ to diverge as $L^3$. In Fig.~(\ref{cvUL}) we plot $\chi_1$ and $\chi_2$ as functions of $U$ for various values of $L$.  In the inset of Fig.~(\ref{cvUL}) we plot the finite size effects on the susceptibilities around $U \approx 1$ where such effects are maximum. We find that for a fixed $L$ both susceptibilities are smooth functions of $U$ with a clear peak around $U \approx 1$ as expected from $\rho_m$ data.  As $L$ increases, the location of the peak $U_{\rm peak}$ moves to the left and the value of the peak  $\chi_i^{\rm peak}$ increases.

Surprisingly there is no indication whatsoever for the $L^3$ divergence in the susceptibilities from Fig.~(\ref{cvUL}). As the inset shows, at both $U=0.8$ and $U=1.2$ the susceptibilities saturate for large $L$, while
at $U=0.96$, both the susceptibilities do seem to diverge but only linearly. As we discuss below, this divergence is consistent with the usual scaling at a second order critical point. Based on this evidence we conclude that both fermion bilinear condensates $\langle\phi_{x,1}\rangle$ and $\langle\phi_{x,2}\rangle$ vanish for all values of $U$. Due to the $SU(4)$ symmetry present in the model the same must be true for all the other condensates discussed in section \ref{sec2}. Finally, we note that both $\chi_1$ and $\chi_2$ are very similar for all values of $U$, except near $U=0$ where one can see from Fig.~(\ref{cvUL}) that $\chi_1\neq 0$ but $\chi_2 = 0$ as expected. 

We next quantify the divergence of $\chi_1$ and $\chi_2$ around $U\approx 1$ in order to verify that it is consistent with a second order transition. Defining $x = (U-U_c) L^{1/\nu}$, near a second order transition we expect both susceptibilities to satisfy the finite size scaling relations,
\begin{equation}
\chi_i(U,L) = L^{2-\eta} f_i(x),
\label{critscal}
\end{equation}
where $\eta$ and $\nu$ are the usual critical exponents and $f_i(x)$ are analytic functions for small values of $x$. In previous studies it was possible to use Eq.~(\ref{critscal}) by expanding $f(x)$ in a power series up to $x^4$, and fit the Monte Carlo data to it and thus extract the critical coupling and exponents \cite{PhysRevLett.108.140404,PhysRevD.88.021701}. Unfortunately, in our current study such an analysis seems to be quite unstable. It is possible that the function $f(x)$ cannot easily be approximated with a few terms in the range of the available data. Hence, we need to find a way to combine our data in the small $x$ region with some information from the large $x$ region using a more elaborate analysis. 

\begin{figure}[b]
\vskip0.2in
\includegraphics[width=0.45\textwidth]{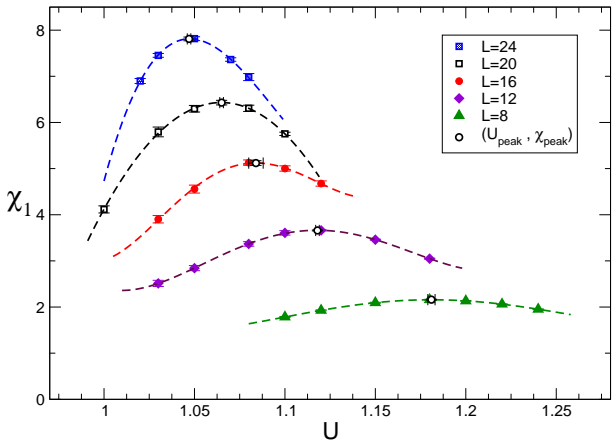}
\caption{\label{uupeak} Plots of $\chi_1$ as a function of $U$ for various values of $L$ near the peak. The dashed lines are fits to Eq.~(\ref{upqfit}) given in the table~\ref{tabpeakfit}.  The location of the peaks is shown with open circles.}
\end{figure}

\begin{table*}
\begin{tabular}{rrrrrrr||rrrrrrr}
\hline
L & $\chi_{1,\rm peak}$ & $U_{\rm peak}$ & $a$ & $b$ & $c$ & $\chi^2/DOF$ &
L & $\chi_{2,\rm peak}$ & $U_{\rm peak}$ & $a$ & $b$ & $c$ & $\chi^2/DOF$ \\
\hline
\hline
8 & 2.16(1) & 1.181(2) & -67(7)    & 20(60) & $2(1)\times 10^3$ & 0.03705  &
8 & 1.95(1) & 1.182(2) & -67(6) & 40(60) & $2(1) \times 10^3$ & 0.05666  \\
12 & 3.66(1) & 1.118(1) & -210(20) & 210(80) & $11(3)\times 10^3$ & 0.02933  &
12 & 3.45(1) & 1.118(1) & -210(20) & 220(80) & $10(3) \times 10^3$ & 0.04093  \\
16 & 5.12(5) & 1.084(4) & -460(60) & $2(2) \times 10^3$ & 4(3)$\times 10^4$ & 0.2583  &
16 & 4.90(4) & 1.084(4) & -450(60) & $2(2) \times 10^3$ & $4(3) \times 10^4$ & 0.2553  \\
20 & 6.43(5) & 1.065(1) & -550(30) & - & - & 0.06176  &
20 & 6.21(5) & 1.065(1) & -550(30) & - & - & 0.02792  \\
24 & 7.81(4) & 1.047(1) & -1030(80) & $7(4)\times 10^3$ & - & 0.1983  &
24 & 7.60(4) & 1.048(2) & -990(80) & $6(5)\times 10^3$ & - & 0.003722  \\ 
\hline
\end{tabular}
\caption{\label{tabpeakfit} Peak values of $\chi_1$ and $\chi_2$ and the value of $U$ where the peaks occur. These values are obtained by fitting Monte Carlo data to Eq.~(\ref{upqfit}). The fits for $\chi_1$ are shown in Fig.~\ref{uupeak} as an example.}
\end{table*}

Consider $\chi(U,L)$ as a function of $U$ for a fixed value of $L$.  From Fig.~{\ref{cvUL} we see that this function has a peak at some value $U=U_{\rm peak}$. On the other hand from Eq.~(\ref{critscal}) we notice that the peak occurs at the value $x=x_{\rm peak}$ where $d f(x)/d x = 0$. Although $x_{\rm peak}$ will not be small it will still satisfy the relation
\begin{equation}
U_{\rm peak} = U_c + \frac{x_{\rm peak}}{L^{1/\nu}}.
\label{upeak}
\end{equation}
Further, the value of $\chi$ at $U=U_{\rm peak}$ will be given by
\begin{equation}
\chi_i^{\rm peak} = L^{2-\eta} f_i(x_{\rm peak}).
\label{chipeak}
\end{equation}
Thus, if we know the values of $U_{\rm peak}$ and $\chi_{\rm peak}$ we can combine Eqs.~(\ref{upeak}) and (\ref{chipeak}) valid at large values of $x$ along with Eq.~(\ref{critscal}) valid at small values of $x$ and try to perform a combined fit. Such a combined fit seems to be more stable.

The large $x$ data is shown in Fig.~\ref{uupeak} and used to extract $U_{\rm peak}$ and $\chi_{\rm peak}$. This is accomplished by approximating the susceptibilities as a quartic polynomial of the form
\begin{equation}
 \chi = \chi_{\rm peak} + a (U-U_{\rm peak})^2 + b (U-U_{\rm peak})^3 + c (U-U_{\rm peak})^4
\label{upqfit}
\end{equation}
near the location of the peak. Table~\ref{tabpeakfit} gives our fitting results and the fits are shown as solid lines in Fig.~\ref{uupeak}. For the small $x$ data we consider four sets extracted from table~\ref{tab:alldata}, using two slightly different lattice sizes and two slightly different coupling regions. The first two sets consist of $0.93 \leq U \leq 1.0$ and the latter two sets focus on $0.95 \leq U \leq 1.0$. In each of these we choose one set containing all $L \geq 16$ data while the other contains only $L \geq 20$ data. These ranges are shown in the first column of table.~\ref{tabexp}.

Armed with the knowledge of $U_{\rm peak}$ and $\chi_{i,\rm peak}$ from table \ref{tabpeakfit} we have performed combined fits of Eqs.~(\ref{critscal},\ref{upeak},\ref{chipeak}) with each of the four sets of small $x$ data. Our results are tabulated in table~\ref{tabexp}. In the first two rows we combine the small $x$ data with only those values of large $x$ data that have the same range of $L$. However, in the third and the fourth rows we combine the small $x$ data with all the large $x$ data except for $L=8$. This is because the $U_{\rm peak}$ data fits remarkably well to Eq.~(\ref{upeak}) as an individual fit for all values of $L\geq 12$. Hence we wanted to explore if emphasizing that feature in the combined fit yielded different results. Indeed, as seen from table~\ref{tabexp}, the critical exponents do change significantly if we emphasize the scaling from large $x$ data. The best combined fit, in terms of the lowest $\chi^2/DOF$, is the one where we allow only lattice sizes $L\geq 20$ (second row of the table). However, if we include the large $x$ data at $L=12,16$ and drop the data at $U=0.93$ the $\chi^2/DOF$ goes up slightly but the fit continues to be reasonable (fourth row of the table). Including the $L=16$ data makes the fit worse but things don't completely break down. Remarkably, the critical point is stable among all the fits and we estimate it to be $U_c = 0.958(2)$.  In contrast there is a large systematic error in the critical exponents and they seem very sensitive to the range of couplings and whether we emphasize the large $x$ data or not. For these reasons we can only estimate them in a range at the moment: $\eta = 0.88\ -\ 0.94$ and $\nu = 0.9\ -\ 1.25$. Further calculations on larger lattices along with measurements of other observables will be necessary to determine them accurately. This is currently being done and we hope to accomplish it in the near future.

\begin{table*}
\begin{center}
\begin{tabular}{|c||c|c|c|c|c|c|c|}
\hline
Fit Range of $U$ and $L$
&  $ \eta $ & $ \nu $ & $ U_c $ & $x_{\rm peak}$ & $f_1(x_{\rm peak})$ &  $f_1(x_{\rm peak})$ & $\chi^2 $ \\
\hline
$U: 0.93-1.0$,  $ L \geq 16$ & 0.940(5)   &  0.93(3)& 0.957(1)  & 2.6(1) & 0.28(1)& 0.27(1) & 2.4\\
$U: 0.93-1.0$, $ L \geq 20$  &0.940(9)    & 0.95(5)& 0.957(1) &  2.5(1) & 0.27(3) & 0.26(3) & 1.1\\
$U:0.95-1.0$, $ L\geq 16$(*)  & 0.884(1)   &  1.21(3)  &0.959(1)  & 1.24(5) & 0.228(3) & 0.217(3) &2.4\\
$U:0.95-1.0$, $L \geq 20$(*)  & 0.884(1)    & 1.24(2)  &0.958(1)    & 1.20(5) & 0.228(3) & 0.217(3) &1.9\\
\hline
\end{tabular}
\end{center}
\caption{\label{tabexp} Results for the critical exponents $\eta $, $\nu$ and the critical coupling $U_c$ from combined fits of four data sets as explained in the text. The (*) in the last two rows indicate that data in table \ref{tabpeakfit} at $L = 12,16$ were included in the fit, unlike the first two fits where data in table \ref{tabpeakfit} from smaller lattice sizes were dropped consistently.}.
\end{table*}

\begin{figure*}
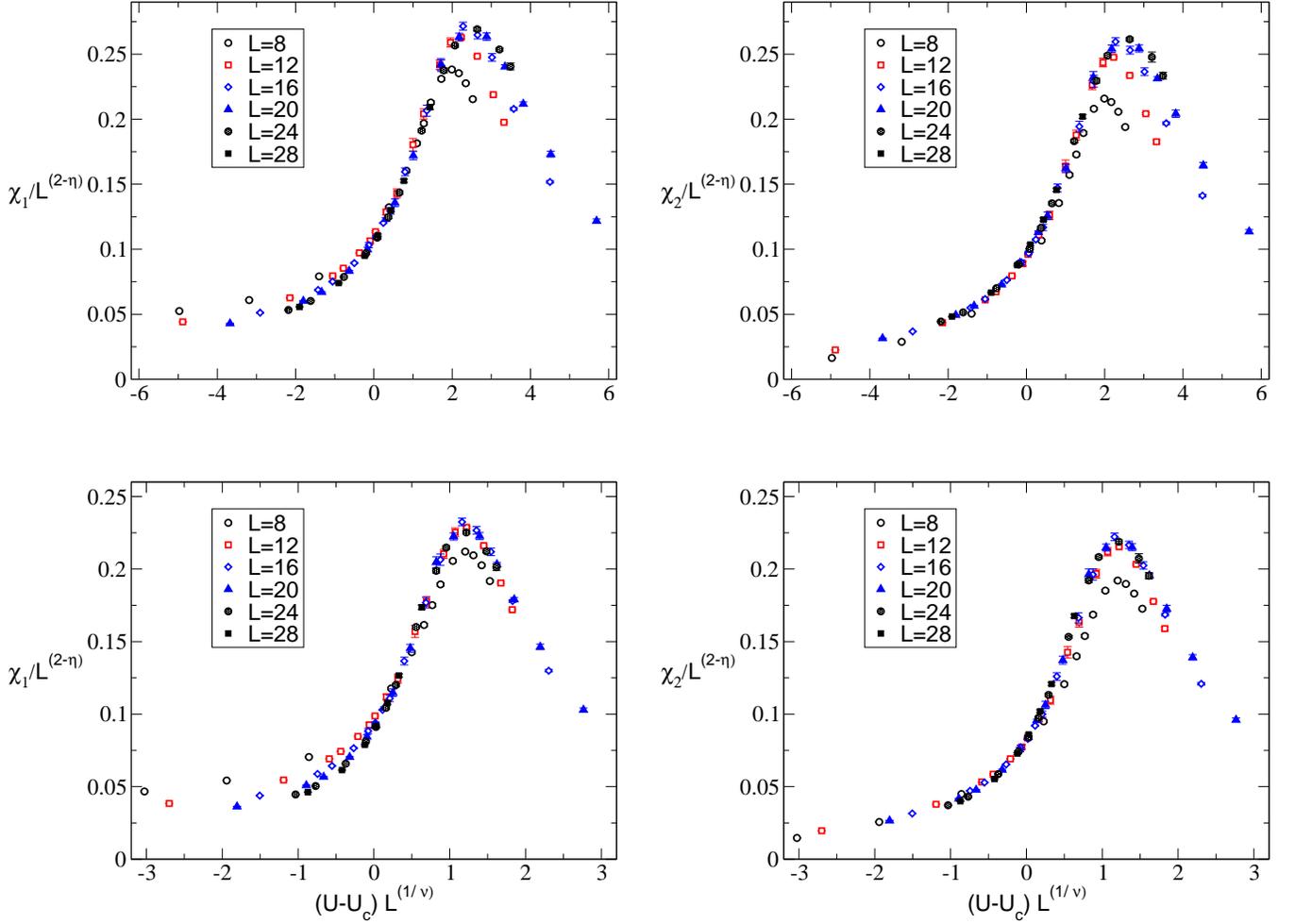

\vspace{0.5in}
\hbox{
\begin{minipage}[c]{0.48\textwidth}
\includegraphics[width=\textwidth]{univf1_uu.eps}
\end{minipage}
\hspace{0.17in}
\begin{minipage}[c]{0.48\textwidth}
\includegraphics[width=\textwidth]{univf1_ud.eps}
\end{minipage}
}
\vspace{0.2in}
\hbox{
\begin{minipage}[c]{0.48\textwidth}
\includegraphics[width=\textwidth]{univf2_uu.eps}
\end{minipage}
\hspace{0.17in}
\begin{minipage}[c]{0.48\textwidth}
\includegraphics[width=\textwidth]{univf2_ud.eps}
\end{minipage}
}
\caption{\label{critfn} Universal scaling plots of $\chi/L^{2-\eta}$ as a function of $(U-U_c) L^{1/\nu}$ using all Monte Carlo data in the critical region. The top two figures use $\eta = 0.94$, $\nu = 0.95$, and $U_c=0.957$ while the bottom two figures use $\eta=0.884$, $\nu=1.24$ and $U_c = 0.958$.}
\end{figure*}

If our estimate of the critical quantities are meaningful, all of our data in the critical region including those that were not used in the analysis must follow the critical scaling form given in Eq.~(\ref{critscal}). In Fig.~\ref{critfn} we plot $\chi/L^{2-\eta}$ as a function of $(U-U_c) L^{1/\nu}$ for both $\chi_1$ and $\chi_2$ using the values from the two best fits (second and fourth rows of table \ref{tabexp}). Using the second row values (top two figures) we find good scaling in the small $x$ region but the data becomes scattered in the large $x$ region unless $L \geq 20$. On the other hand with the fourth row values (bottom two figures), a good scaling is observed in the large $x$ region, but the data becomes more scattered in the small $x$ region especially for $\chi_1$. Interestingly, we find that $\chi_2$ with the fourth row values (bottom right figure) shows the best scaling (to the eye) if we ignore the $L=8$ data. Based on this we suspect that taking the lowest $\chi^2/DOF$ to extract the critical exponents may be a bit premature. Although, we are unable to give accurate estimates for the critical exponents in this work, we do believe the universal plots shown in Fig.~(\ref{critfn}) provide strong evidence for a second order transition separating the PMW and the PMS phase.

\section{Local Order Parameters}
\label{sec7}

An interesting aspect of the phase transition we have uncovered here, is the absence of an obvious local lattice order parameter that distinguishes the two phases. By definition a local lattice order parameter $O_x$ is made with Grassmann fields in the vicinity of the lattice site $x$. It is zero in one phase for a symmetry reason, but becomes non-zero in the other phase because the symmetry is spontaneously broken. A simple intuitive argument shows that fermion bilinear order parameters cannot exist. First we note that in a finite volume by definition we must have $\langle O_x \rangle = 0$ for symmetry reasons. In order to study whether the symmetry can break spontaneously, one has to compute the behavior of the two point correlation function of order parameters at large separations,
\begin{equation}
\lim_{|x-y| \rightarrow \infty} \langle O_x\ O_y \rangle.
\end{equation}
If the symmetry is spontaneously broken the above expression becomes non-zero. At weak couplings, since $U$ couples to an irrelevant operator, the physics is governed by the $U=0$ fixed point where we know that fermion bilinear order parameters do not exist. At the other extreme, when $U$ is very large, the whole lattice is filled with monomers and no lattice symmetries of the interacting theory are broken in this trivial state. Further, if we compute the above two point correlation function, as discussed in section \ref{sec4} we expect $x$ and $y$ to be in different free fermion bags and the calculation reduces to computing $\langle O_x\rangle$ and $\langle O_y \rangle$ in two distant fermion bags one containing $x$ and the other containing $y$. Monomers fill the remaining lattice sites. Each of these calculation is very similar to the calculation of $\langle O_x \rangle$ in a finite volume, except that the fermion bag has an arbitrary shape with Dirichlet boundary conditions. If the boundaries do not break a symmetry, then we must have $\langle O_x \rangle = 0$. If the boundaries do break the symmetry we can restore the symmetry by summing over fermion bag configurations obtained by symmetry transformations. We can do this because there is only a single fermion bag and the remaining lattice sites are filled with monomers. The other fermion bag containing the site $y$ is far away. This again implies that $\langle O_x \rangle = 0$.
In the above argument we have assumed that the integration measure remains symmetric under the symmetry transformations. Chiral symmetries can be broken by boundary effects, because the measure becomes non-invariant. However, in our model the measure remains invariant under the $SU(4)$ symmetry transformations inside a fermion bag. Hence in our model $\langle O_x \rangle = 0$ for all fermion bilinear lattice order parameter at least for sufficiently large values of $U$. When these facts are combined with the assumption that there is only a single phase transition, we find that there cannot be a fermion bilinear order parameter that distinguishes between the phases.

Of course the above arguments do not rule out a four-fermion order parameter, the simplest being $\rho_m$. But it cannot be an order parameter in the strict sense of the word since it is non-zero for all values of $U$ except $U=0$. However, it does play the role of an order parameter in the sense that it changes quite rapidly as one passes through the phase transition. Hence we refer to it as a pseudo order parameter. Since the small $U$ theory and the $U=\infty$ theory seem to have exactly the same lattice symmetries, we are tempted to conclude that no lattice symmetries are broken as a function of $U$. Yet there are massless particles at weak couplings, which are absent at strong couplings. The situation seems to be similar to certain metal insulator transitions where there are no clear order parameters that govern the phase transition \cite{mitrans}. 
\section{Conclusions}
\label{sec8}

In this work we have provided strong evidence that a simple four-fermion model containing two flavors of staggered fermions on a cubic lattice contains a phase where a non-zero fermion mass arises although all fermion bilinear condensates vanish. While such an exotic scenario of mass generation was known before, previous work had suggested that the exotic phase was only a lattice artifact since fermion masses could not be made small compared to the lattice spacing. In contrast our work shows that one may indeed be able to make fermions light by tuning close to the second order critical point that exists in the model. We locate the critical point with an error of about a percent. Although we were able to perform calculations up to lattice sizes of $L=28$, scaling seems to set in only for $L \geq 20$, unlike other staggered four-fermion models that were solved recently, where the data begin to show scaling behavior even for $L \geq 12$ \cite{PhysRevLett.108.140404,PhysRevD.88.021701}. For this reason we were only able to bound the critical exponents within a range. Our rough estimates are $0.95 \leq \nu \leq 1.2$ and $0.88 \leq \eta \leq 0.94$. Larger lattice calculations along with new observables are necessary to provide a more complete picture of the critical behavior. This work is currently under progress. We have also argued that in our model there is no symmetry that distinguishes the massless phase from the massive phase. This suggests that fermion mass generation in our model is related only to dynamics and not to symmetries. The quantity that comes close to a definition of the order parameter is the four-fermion condensate or the monomer density $\rho_m$. Although it is non-zero in both the phases it changes rapidly over a small region of the couplings.

Our work can be extended in different directions. For example, it is possible to explore if a similar second order critical point exists in four space-time dimensions. Mean field theory, which becomes accurate in large number of dimensions, suggests that the transition would become first order at sufficiently large number of dimensions. Is four large enough? We plan to return to this question in a future publication. Another possible direction is to view our model within the context of Yukawa models with a variety of symmetries. From this perspective our critical point has many relevant and marginal directions that break a variety of symmetries. It would be interesting to compute the critical exponents associated with all these directions. It is also interesting to explore what would happen if the $SU(4)$ symmetry of our model is gauged.

Finally, the quantum field theoretic description of the second order critical point that we have found remains unknown. As we mentioned in the introduction, an exotic transition very similar to ours was recently discovered in an extended Hubbard model on a bilayer honeycomb lattice \cite{Slagle:2014vma}. It was argued that the critical point there could be viewed as a multi-critical point where three different topological transitions meet. Interestingly, both the models contain the same number of massless fermions in the weak coupling phase. It is also easy to argue that a simpler model on the honeycomb lattice with an $SU(4)$ symmetry, than the one considered by the authors, shows a similar exotic phase transition. Hence, we believe the two transitions are closely related and perhaps even belong to the same universality class. If true, this should mean that our staggered fermion model can be deformed to introduce topological phase transitions as in the honeycomb lattice model. Such an extension could shed further insight into staggered fermions and its connections to honeycomb lattice models, while at the same time helping us uncover the field theory that governs the critical point.

\section*{Acknowledgments}
We would like to thank M.~Golterman for pointing us to the lattice literature on the subject. We also thank H.~Geis, Ph.~deForcrand, S.~Hands, T.~,Mehen, R.~Springer, R.~Plesser, C.~Xu and U.-J.~Wiese for helpful discussions at various stages of this work. The material presented here is based upon work supported by the U.S. Department of Energy, Office of Science, Nuclear Physics program under Award Number DE-FG02-05ER41368.

\bibliography{ref,pms}

\appendix

\section{Testing the Monte Carlo Algorithms}
\label{appx}

In order to test our Monte Carlo algorithms we have performed a series of checks which we describe here. As discussed in section \ref{sec5} we developed three algorithms to perform these checks: A block algorithm: (ALG1), a worm algorithm (ALG2) and a heat-bath sweep algorithm (ALG3). Among these three, the worm algorithm is the most efficient and has been used for our production runs. However, we can run the worm algorithm in two ways: perform many sweeps on a single core (ALG2S), or perform a few sweeps on hundreds of parallel cores each starting from an equilibrated configuration (ALG2P). Clearly, the latter is very efficient and we show here that it is a reliable approach. Among the three algorithms, ALG3 is the most time consuming but has the lowest fluctuations. Also it is the only algorithm that works on a $2^3$ lattices for technical reasons. Since we can compute everything analytically on this small lattice we can test ALG3 against exact results and use it as a benchmark algorithm to test others.

\begin{table}[b]
\begin{tabular}{|c||c||c|c||c|c|}
\hline
U & L & \multicolumn{2}{|c||}{$\rho_m$} & \multicolumn{2}{|c||}{$\chi_1$} \\
\hline
& & Exact & ALG1 & Exact & ALG1 \\
\hline
0.8 & 6 & $0.015236...$ & $0.01523(2)$ & 0.44183... & 0.4421(05) \\
1.0 & 6 & $0.016367...$ & $0.01636(4)$ & 0.44489... & 0.4450(10) \\
0.8 & 8 & $0.007193...$ & $0.00720(1)$ & 0.45619... & 0.4559(02)\\  
\hline
\end{tabular}
\caption{\label{tab:pert} Comparison between a perturbative calculation containing up to four-monomers (i.e., up to $U^4$) and results from ALG1 which was also restricted to the same monomer sectors.}
\end{table}
 
\begin{table*}[b]
\centering
\begin{tabular}{|c||c|c||c|c||c|c|}
\hline
U & \multicolumn{2}{|c||}{$\rho_m$} & \multicolumn{2}{|c||}{$\chi_1$} & \multicolumn{2}{|c|}{$\chi_2$} \\
\hline
& Exact & ALG3 & Exact & ALG3 & Exact & ALG3 \\
\hline
0.1 & 0.000370... & 0.00037(02) & 0.166728... & 0.16673(01) & 0.003703... & 0.00369(02) \\
0.5 & 0.009517... & 0.00952(01) & 0.168166... & 0.16817(01) & 0.018510... & 0.01851(02) \\
0.8 & 0.025400... & 0.02540(04) & 0.170310... & 0.17032(02) & 0.029540... & 0.02957(04) \\
1.0 & 0.041188... & 0.04118(02) & 0.172054... & 0.17206(01) & 0.036757... & 0.03675(02) \\
1.2 & 0.061937... & 0.06192(03) & 0.173834... & 0.17383(01) & 0.043726... & 0.04372(02) \\
1.5 & 0.104086... & 0.10413(04) & 0.175961... & 0.17598(01) & 0.053285... & 0.05328(02) \\
2.0 & 0.208466... & 0.20836(05) & 0.174920... & 0.17491(01) & 0.064497... & 0.06448(01) \\
3.0 & 0.500000... & 0.49996(07) & 0.142857... & 0.14287(01) & 0.059523... & 0.05954(01) \\
5.0 & 0.838548... & 0.83851(05) & 0.063477... & 0.06348(04) & 0.021941... & 0.02195(05) \\
\hline
\end{tabular}
\caption{\label{tab:222} Comparison between exact results and those from Monte Carlo calculations using ALG3, on a $2^3$ lattice for the three observables $\rho_m$, $\chi_1$ and $\chi_2$.}
\end{table*}

In order to compute the exact results on a $2^3$ lattice we write the partition function as
\begin{equation}
Z = \sum_{[n]} U^{N_m}\ g([n]) \big(\mathrm{Det}(W([n]))\big)^2,
\end{equation}
where the sum is over a class of monomer configurations $[n]$, not counting configurations with the same number number of monomers that are obtainable by rotations and (or) reflections. The number of configurations within a given class (degeneracy) is denoted as $g([n])$, each class contains $N_m$ monomers and $\mathrm{Det}(W([n]))$ is the free fermion bag weight for a single staggered fermion restricted to the bag. Table~\ref{tab:zconf} gives the various possible equivalence classes along with their degeneracy factors $g([n])$, the fermion bag weight from $\mathrm{Det}(W([n]))$, and the corresponding $N_m$ values. Using these we find that the partition function is given by
\begin{equation}
Z = 6561 + 972 U^2 + 126 U^4 + 12 U^6 + U^8
\end{equation}
The average monomer density can then be easily computed and is given by
\begin{equation}
\rho_m  = \frac{1}{8 Z} \ ( 1944 U^2 + 504 U^4 + 72 U^6 + 8 U^8).
\end{equation}
Note that it is zero for small $U$ and approaches one for large $U$. In order to compute the two susceptibilities defined in Eqs.~(\ref{susdefs}), we consider two monomer configurations $n_1$ and $n_2$ that are naturally defined for each flavor through the knowledge of the location of the two half monomers. We then define $\mathrm{Det}(W_1)$ and $\mathrm{Det}(W_2)$ as the fermion bag weights for the two flavors respectively. With these definitions we see that
\begin{eqnarray}
\chi_1 &=& \frac{1}{2Z}\ \sum_{[n_1,n_2]} U^{N_m}\ g_1 \ \mathrm{Det}(W_1W_2)
\\
\chi_2 &=& \frac{1}{2Z}\ \sum_{[n_1,n_2]} U^{N_m}\ g_2\ \mathrm{Det}(W_1W_2)
\end{eqnarray}
 where $[n_1,n_2]$ refer to a class of configurations of monomers with two half-monomer insertions which are shown in table \ref{tab:sus1} and \ref{tab:sus2}, along with the degeneracy factors $g_1$ and $g_2$ and the fermion bag weights. Substituting the values in these tables we find
\begin{eqnarray}
\chi_1 &=& \frac{1}{2Z}\ (2187 + 405 U^2 + 45 U^4 + 3 U^6) 
\\
\chi_2 &=& \frac{1}{2Z}\ (486 U + 72 U^3 + 6 U^5) 
\end{eqnarray}
Table \ref{tab:222} gives a comparison of the three observables computed exactly using the above relations and through ALG3. Our algorithm accurately reproduces the results for various values of $U$.

Another class of checks that we have performed involves calculations of observables exactly on slightly larger lattices, but in perturbation theory up to order $U^4$. In this case we were able to study lattices up to $8^3$. It is also easy to restrict the monomer number to the same order in the algorithms by simply adding a few lines to the entire code. We used this approach to test ALG1. Table \ref{tab:pert} gives a comparison between ALG1 and exact perturbative results.

Finally we compared all three algorithms at various couplings at an accuracy of one percent or less. Table \ref{tab:comp} gives these comparisons.
\begin{table*}[h]
\begin{center}
\begin{tabular}{|l|l|l|l|l|l|l|l|l|l|l|l|l|}
\hline
L & U & \multicolumn{4}{|c|}{$\rho_m$} & \multicolumn{4}{|c|}{$\chi_1$} & \multicolumn{3}{|c|}{$\chi_2$} \\
\hline
& & ALG1 & ALG2S & ALG2P  & ALG3 & ALG1 & ALG2S & ALG2P & ALG3 & ALG2S & ALG2P & ALG3 \\
\hline
4 & 0.95 & 0.0915(4) & 0.0922(2) &    N/A   & 0.0922(1)  &  0.4533(7) &  0.453(1) & N/A  & 0.4543(3) &  0.2386(6) &  N/A &  0.2395(5) \\
4 & 1.05 & 0.1237(6) & 0.1236(2) &     N/A  & 0.1236(1)  &  0.4922(9) &  0.492(1) & N/A  & 0.4920(3) &  0.2857(7) &  N/A &  0.2853(4) \\
4 & 1.20 & 0.1936(9) & 0.1939(3) &     N/A   & 0.1946(2)  & 0.567(1)    & 0.564(1)  & N/A  & 0.5662(4) &  0.3707(7) &  N/A &  0.3721(4) \\
8 & 0.95 & 0.1097(1) & 0.1096(1) & 0.1098(1) & 0.1098(1)   & 1.017(1)  &  1.017(2) &  1.017(1) & 1.017(2) &   0.778(1) & 0.7781(4) & 0.777(3) \\
8 & 1.05 & 0.1685(4) & 0.1680(1) & 0.1684(3) & 0.1678(3)   & 1.467(5)  & 1.458(3)  &  1.461(3) & 1.458(4) &  1.234(3) & 1.236(3)  & 1.232(4) \\
8 & 1.20 & 0.3772(8) & 0.3751(7) & 0.375(1)  & 0.3769(7)  & 2.134(6) &  2.14(1)   &  2.128(5) & 2.137(3)  &   1.936(9) & 1.928(5) & 1.936(3) \\
12 & 0.95 & 0.111(1) & 0.1112(4) & 0.1114(5) & 0.1119(1) & 1.46(4) & 1.49(1) &  1.48(1) & 1.497(4) & 1.25(1) & 1.24(1) & 1.254(4) \\
16 & 0.95 & 0.1131(9) & 0.1129(3) & 0.1126(4) & N/A & 2.06(7) &  1.95(1) &  1.95(2)   & N/A &  1.70(1) & 1.70(2) & N/A \\
16 & 1.00 & 0.142(1) & 0.1428(4) & 0.1429(6)   & N/A & 2.96(8) &  3.00(2) &  3.01(3)   & N/A &  2.76(2) & 2.78(3) & N/A \\
20 & 0.95 & 0.1128(7) & 0.1133(4) & 0.1133(3)  & N/A & 2.36(6) &  2.37(3) &  2.39(2)   & N/A &   2.13(3) & 2.14(2) & N/A \\
20 & 1.00 & 0.143(1) & 0.1434(5) & 0.1436(4)   &  N/A & 4.1(1) &  4.08(4) &  4.12(4)   & N/A &  3.85(4) & 3.88(4) & N/A \\
\hline
\end{tabular}
\end{center}
\caption{\label{tab:comp}Comparison between results from the three different algorithms: the block algorithm (ALG1), the worm algorithm (ALG2) and a heat-bath sweep algorithm (ALG3). For the worm algorithm we also compare between a single core run with many sweeps (ALG2S) and a parallel core run on hundreds of cores, but each core only running for a few sweeps (ALG2P). N/A indicates the comparison is not available.}
\end{table*}

\begin{table*}[h]
\begin{tabular}{|c|c|c|c||c|c|c|c||c|c|c|c|}
\hline\hline
$[n]$ & $g([n])$ & $\mathrm{Det}(W([n]))$ & $N_m$ & 
$[n]$ & $g([n])$ & $\mathrm{Det}(W([n]))$ & $N_m$ &
$[n]$ & $g([n])$ & $\mathrm{Det}(W([n]))$ & $N_m$ \\
\hline 
&&&&&&&&&&& \\
\begin{minipage}[c]{0.12\textwidth}
\includegraphics[width=0.8\textwidth]{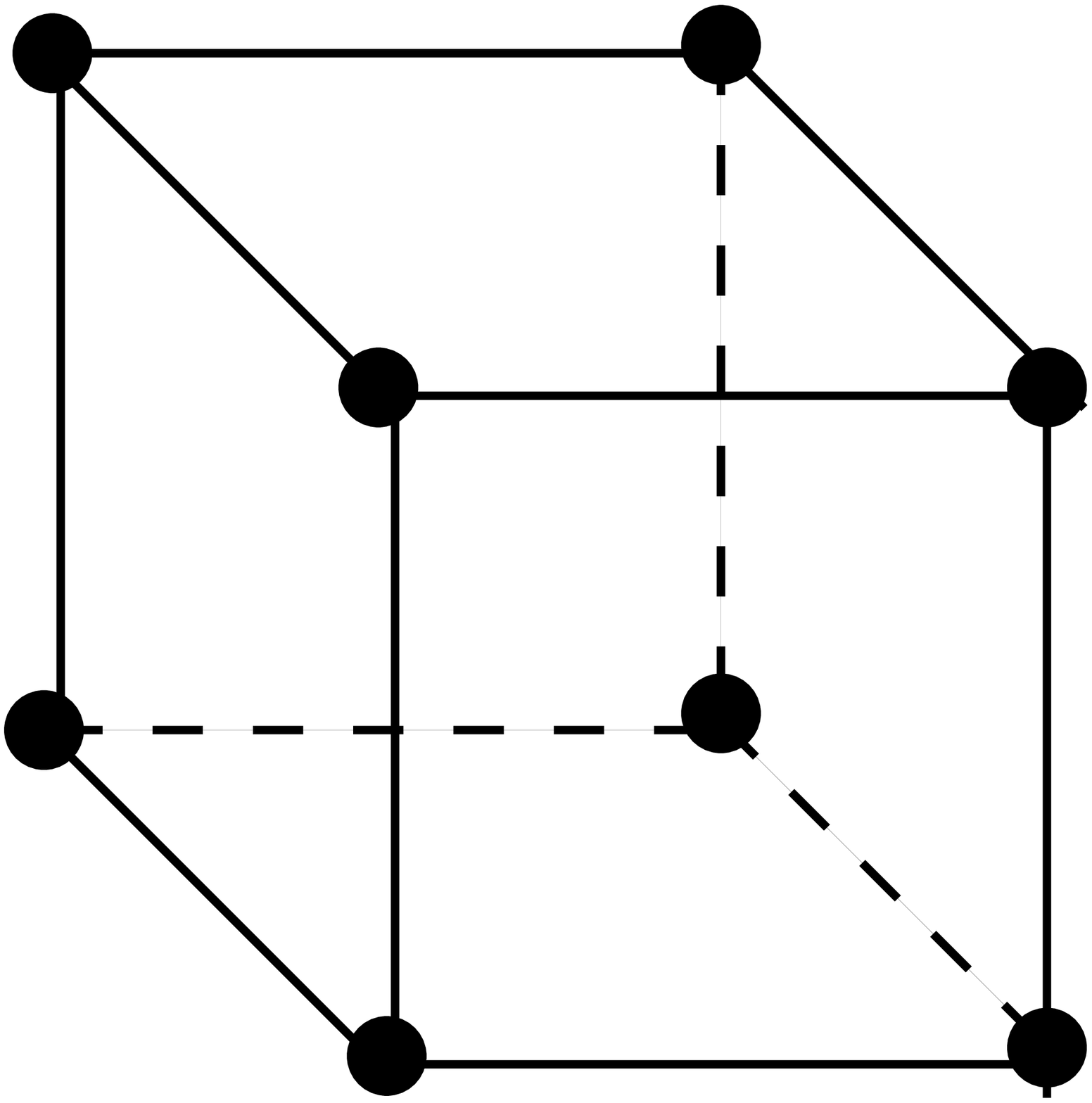}
\end{minipage}
& 1 & 81 & $0$ & 
\begin{minipage}[c]{0.12\textwidth}
\includegraphics[width=0.8\textwidth]{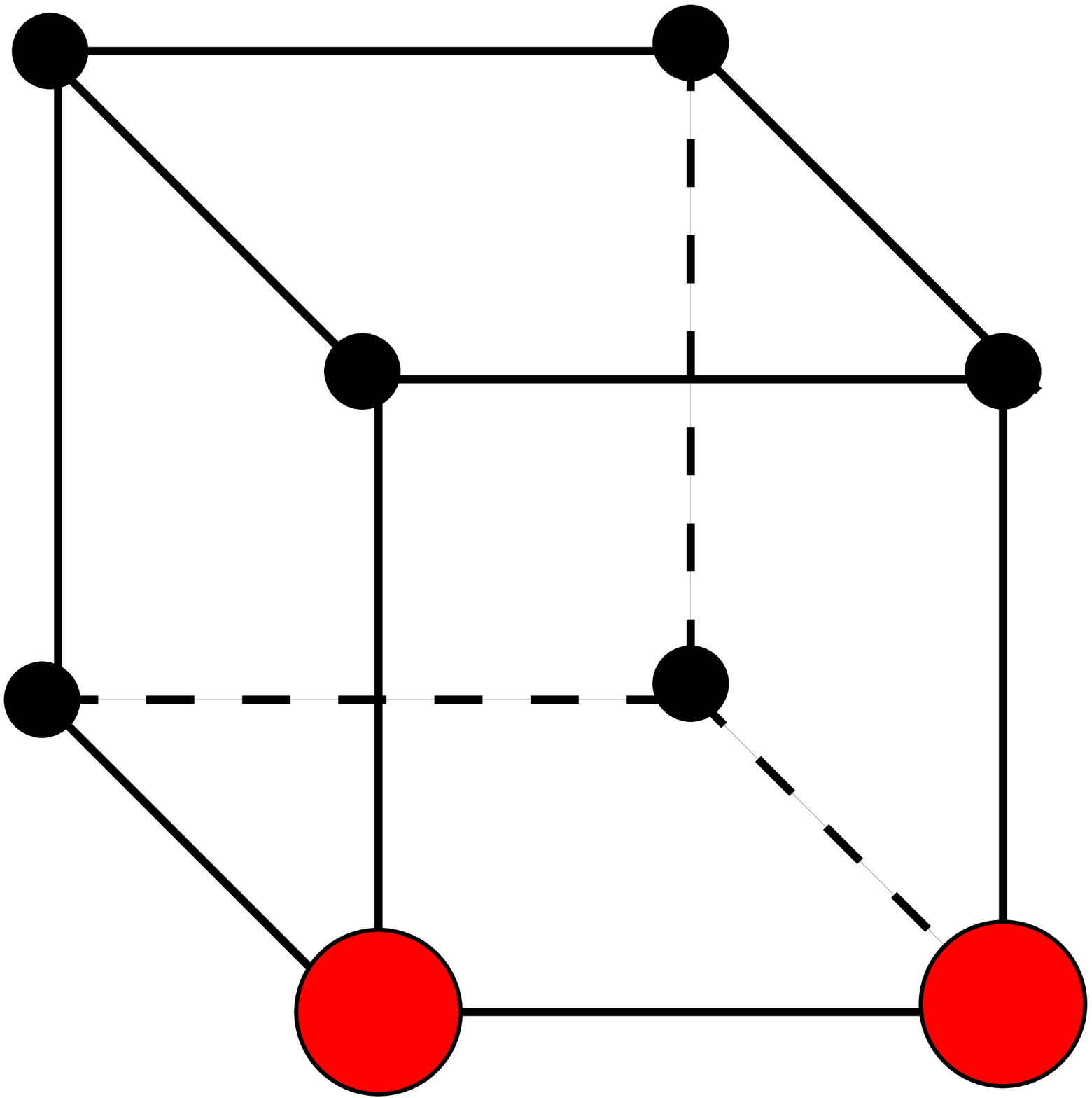}
\end{minipage}
& 12 & 9 & $2$ &
\begin{minipage}[c]{0.12\textwidth}
\includegraphics[width=0.8\textwidth]{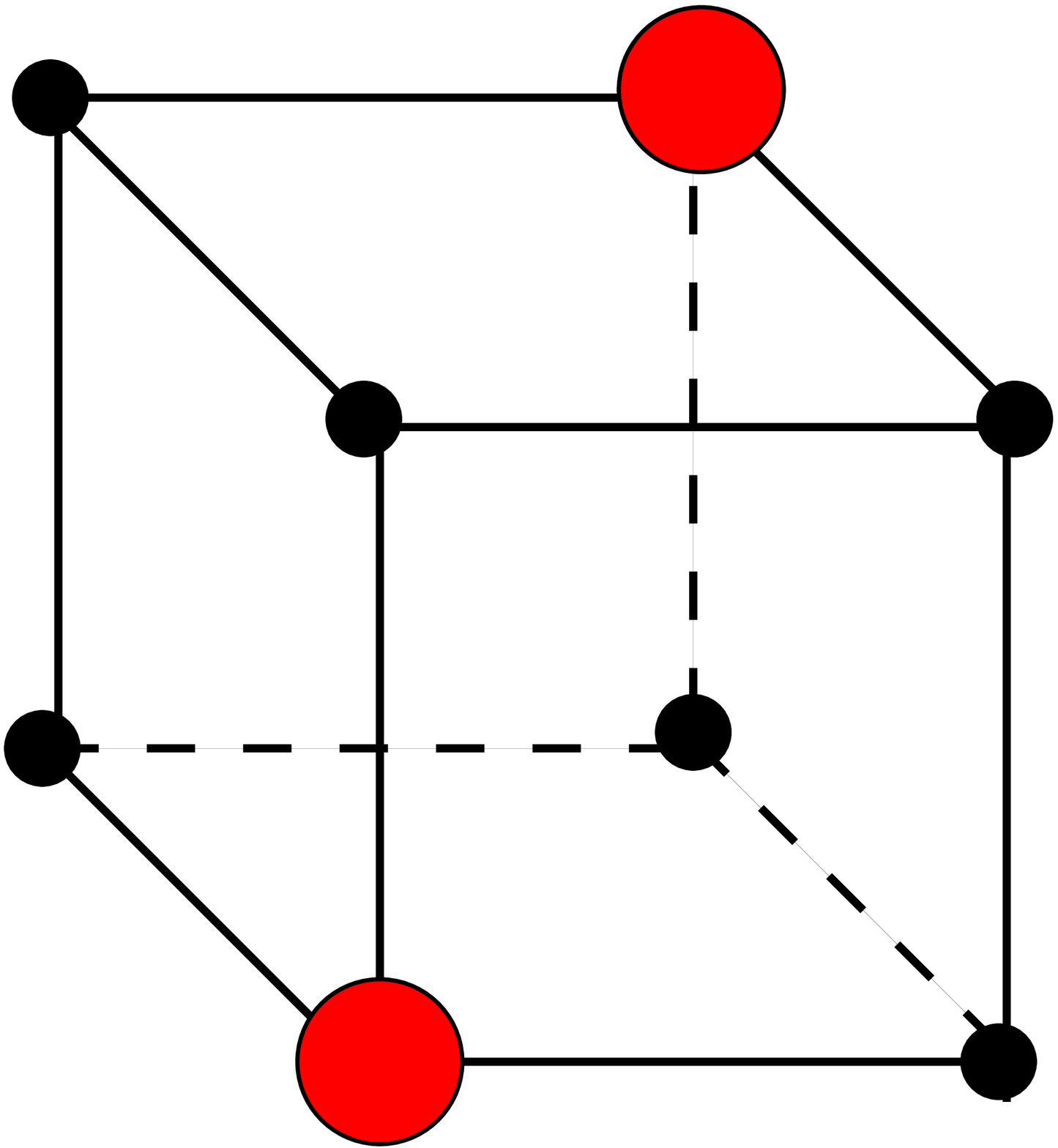}
\end{minipage}
& 4 & 0 & $2$ \\
&&&&&&&&&&& \\
\begin{minipage}[c]{0.12\textwidth}
\includegraphics[width=0.8\textwidth]{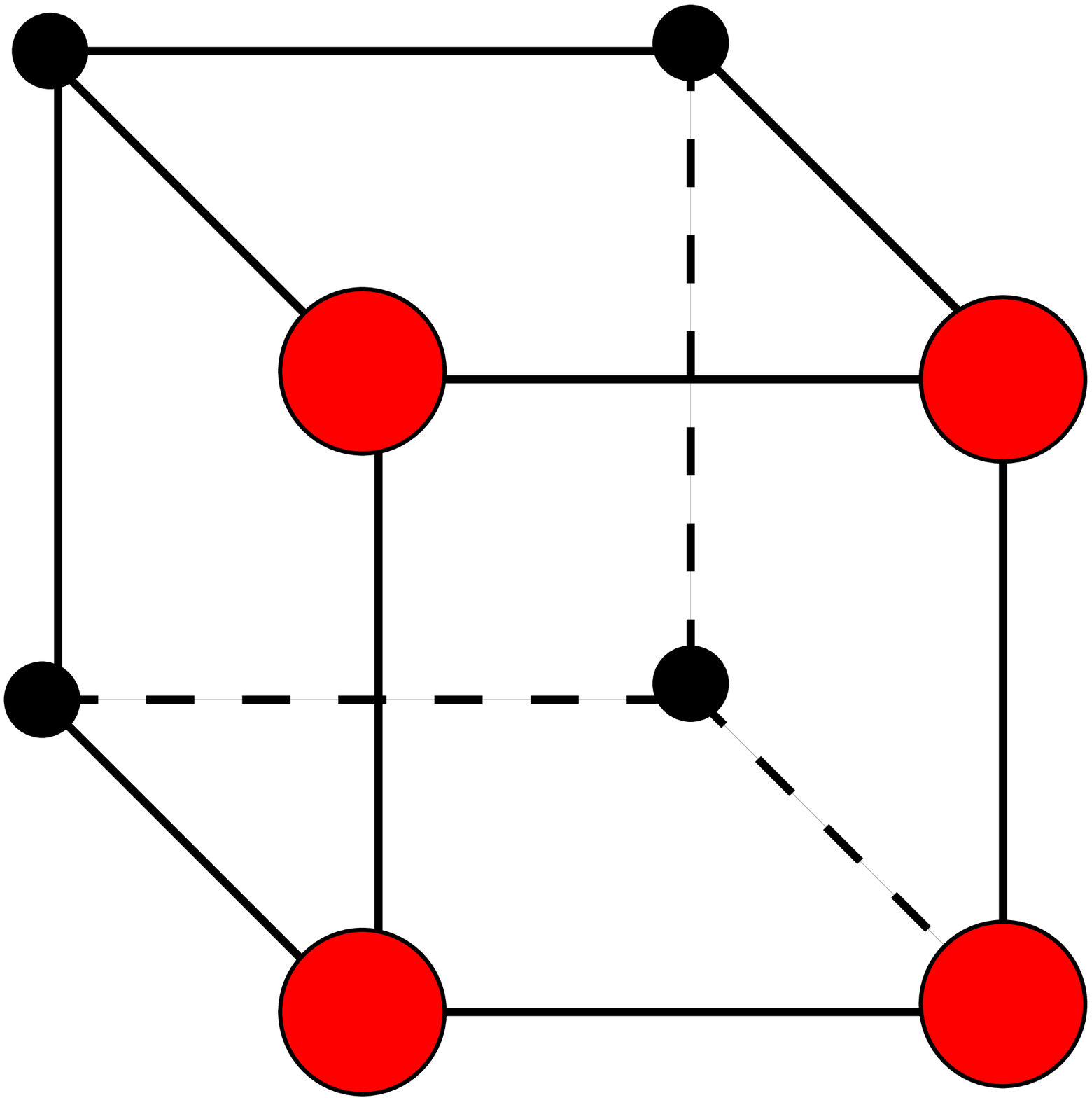}
\end{minipage}
& 6 & 4 & $4$ &
\begin{minipage}[c]{0.12\textwidth}
\includegraphics[width=0.8\textwidth]{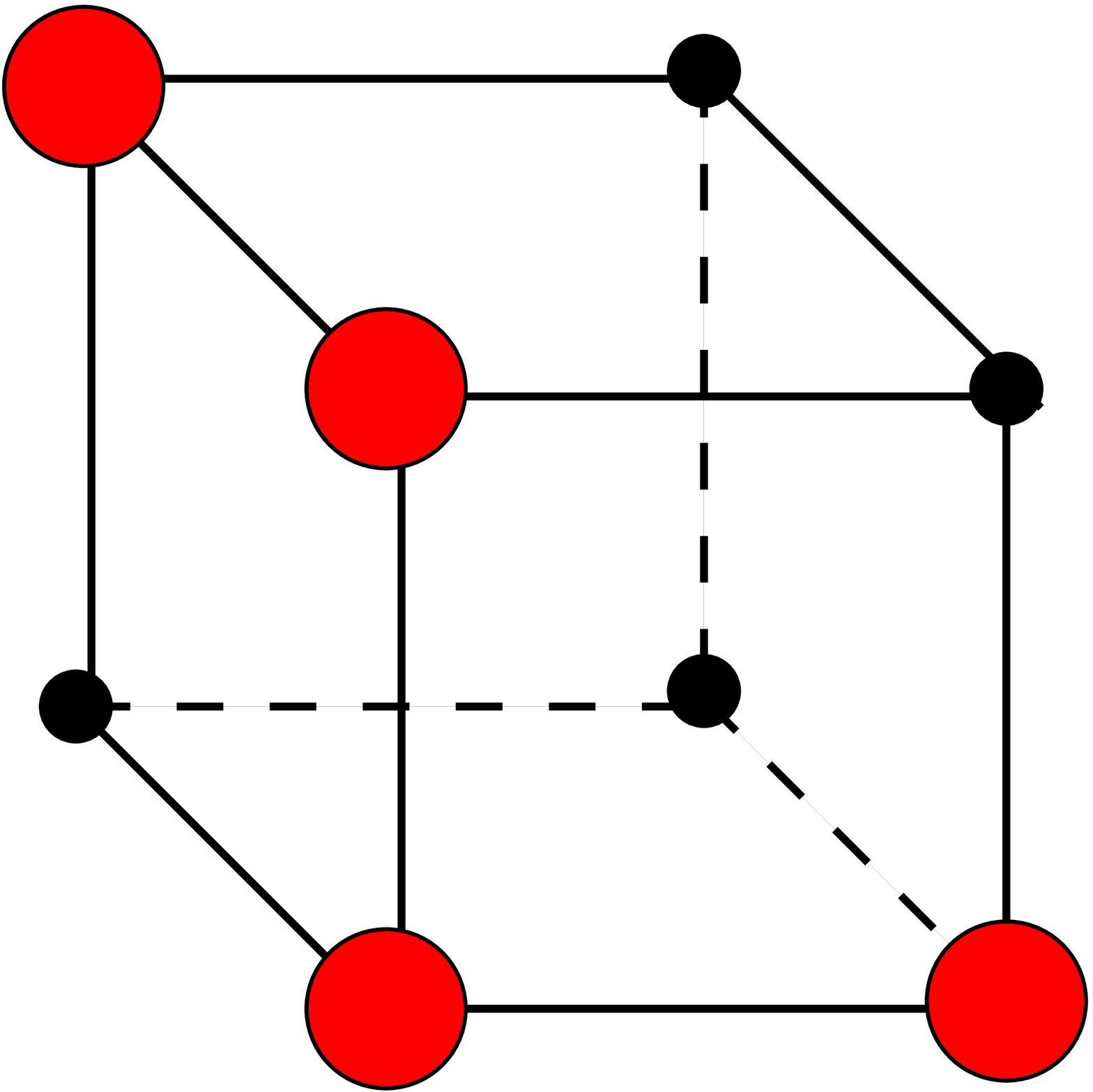}
\end{minipage}
& 24 & 1 & $4$ &
\begin{minipage}[c]{0.12\textwidth}
\includegraphics[width=0.8\textwidth]{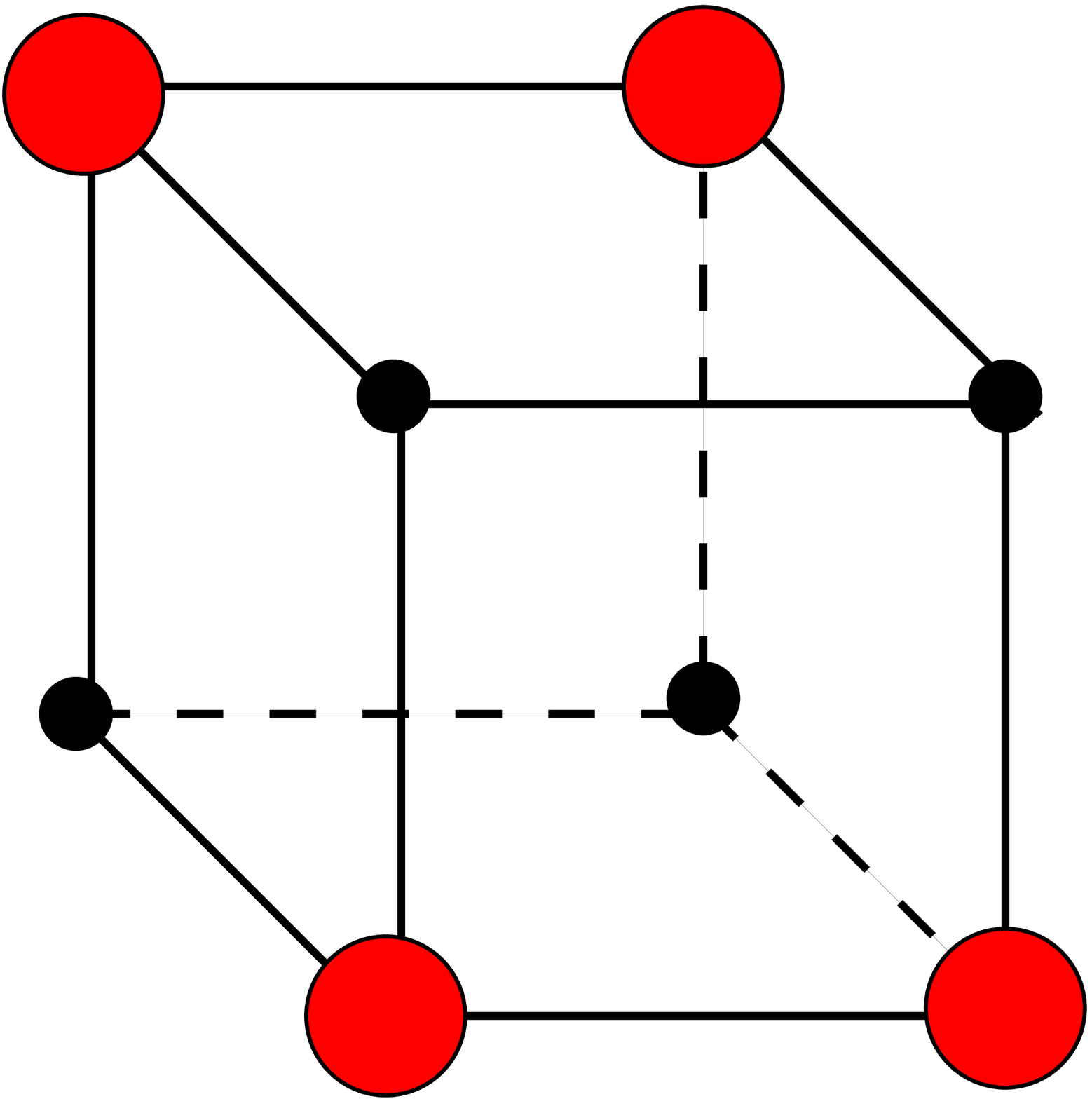}
\end{minipage}
& 6 & 1 & $4$ \\
&&&&&&&&&&& \\
\begin{minipage}[c]{0.12\textwidth}
\includegraphics[width=0.8\textwidth]{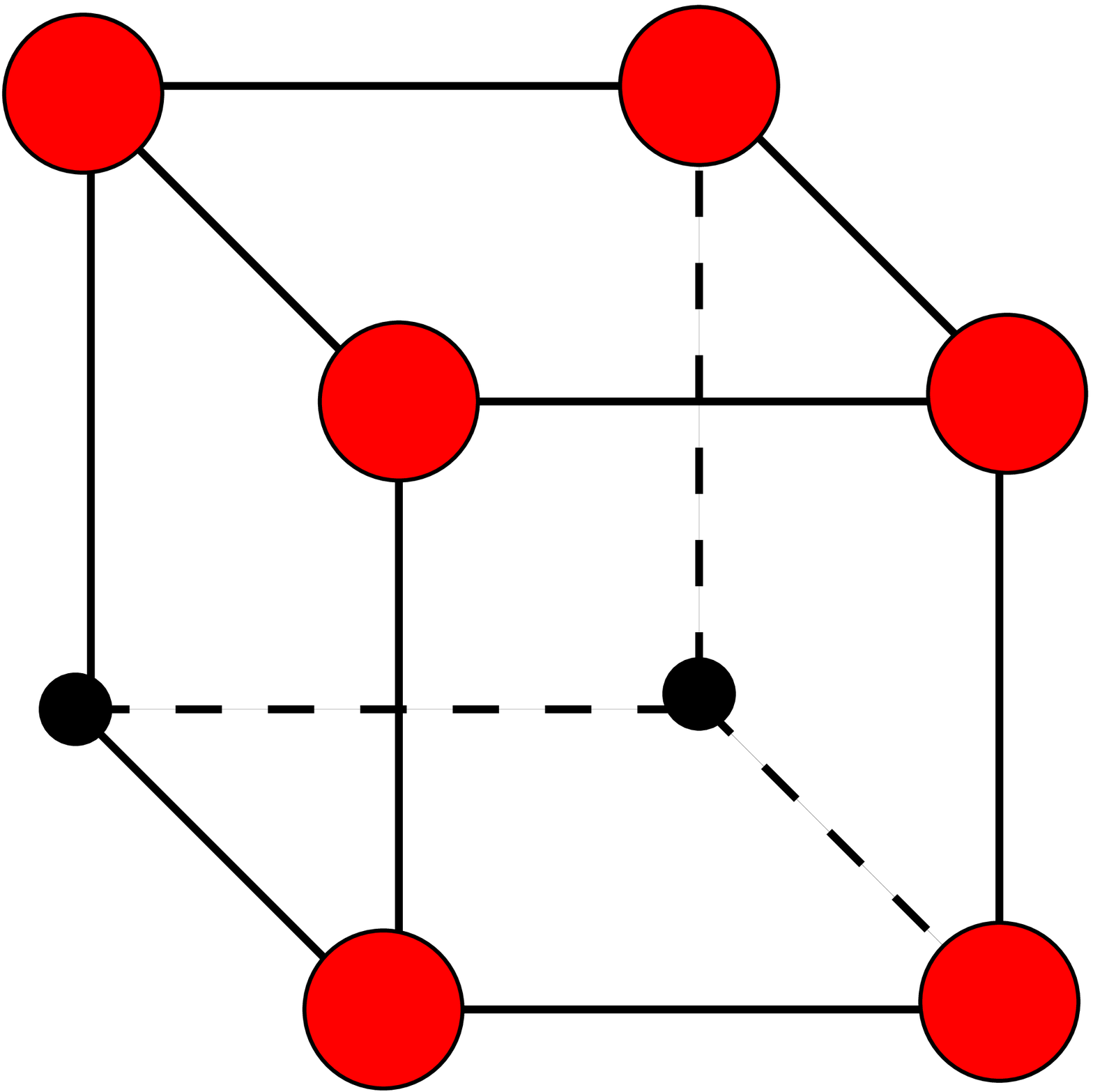}
\end{minipage}
& 12 & 1 & $6$ &
\begin{minipage}[c]{0.12\textwidth}
\includegraphics[width=0.8\textwidth]{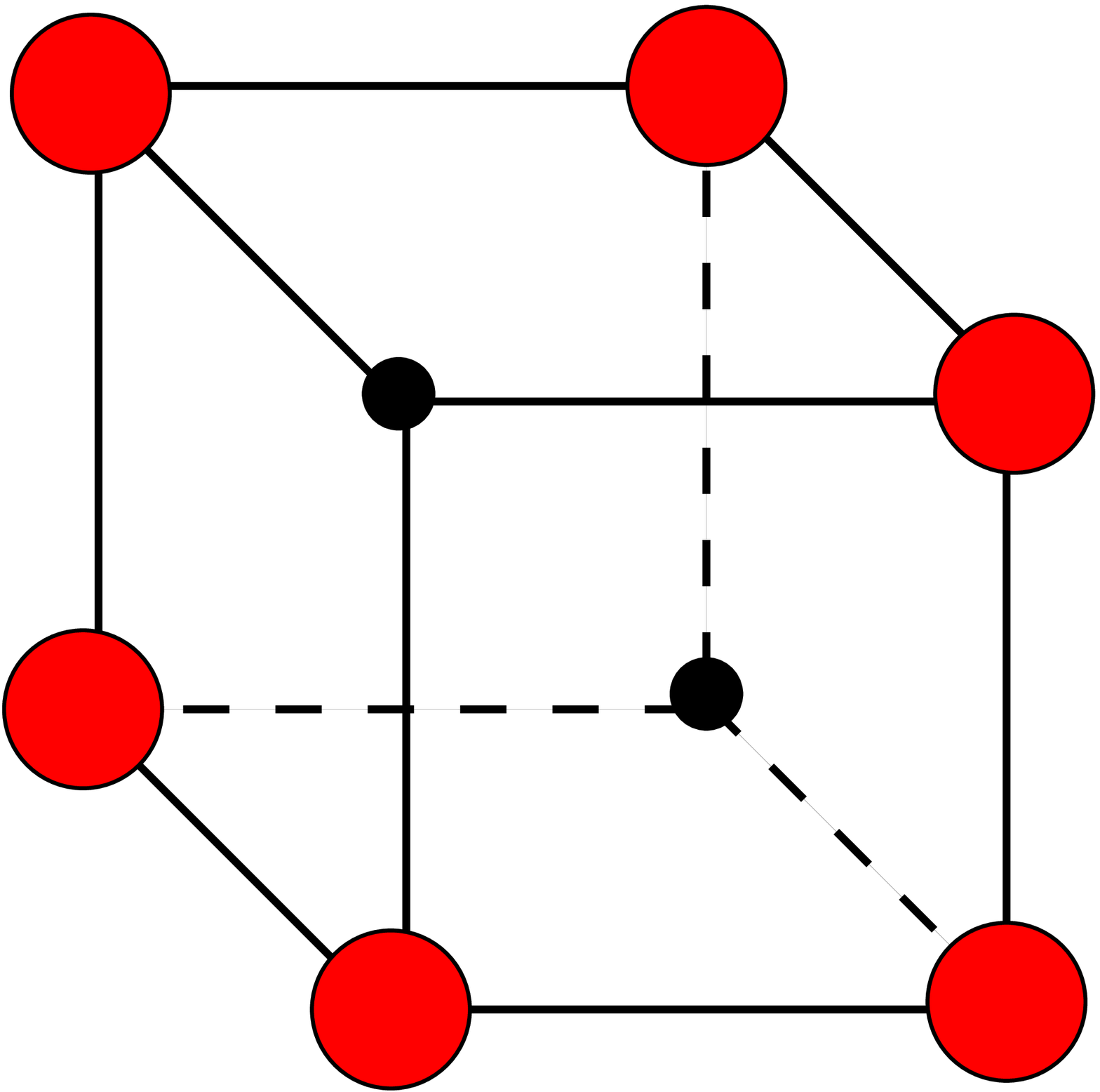}
\end{minipage}
& 4 & 0 & $6$ &
\begin{minipage}[c]{0.12\textwidth}
\includegraphics[width=0.8\textwidth]{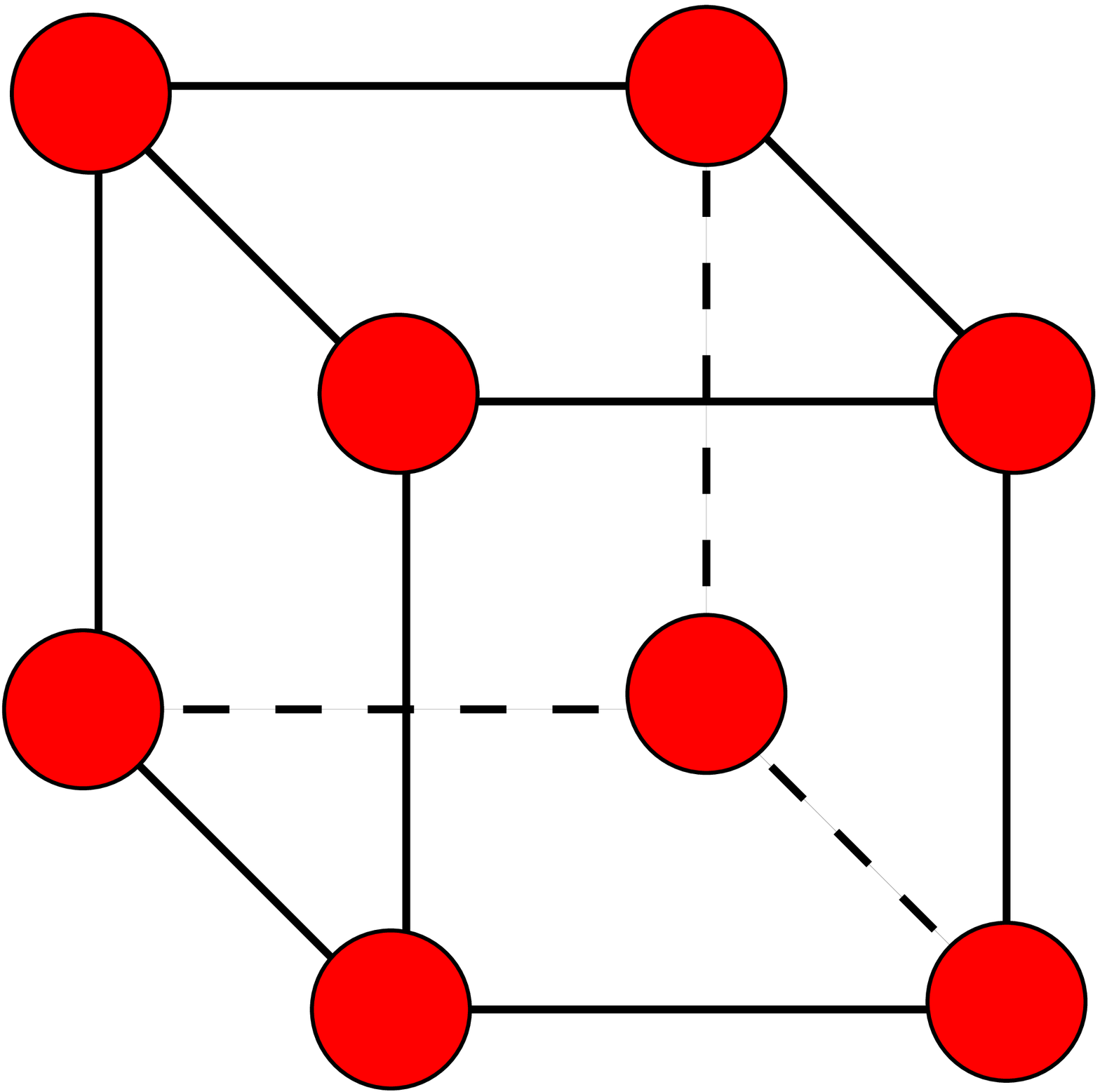}
\end{minipage}
& 1 & 1 & $8$ \\
&&&&&&&&&&& \\ 
\hline
\hline
\end{tabular}
\caption{\label{tab:zconf} Configuration classes $[n]$ that contribute to the partition function on a $2^3$ lattice.  The degeneracies $g([n])$, the fermion bag determinant for each flavor $\mathrm{Det}(W([n]))$ are also given.}
\end{table*}

\begin{table*}[h]
\begin{tabular}{|c|c|c|c||c|c|c|c||c|c|c|c|}
\hline
$[n_1,n_2]$ & $g_1$ & $\mathrm{Det}(W_1W_2)$ & $N_m$ &
$[n_1,n_2]$ & $g_1$ & $\mathrm{Det}(W_1W_2)$ & $N_m$ &
$[n_1,n_2]$ & $g_1$ & $\mathrm{Det}(W_1W_2)$ & $N_m$ \\
\hline 
&&&&&&&&&&&\\
\begin{minipage}[c]{0.12\textwidth}
\includegraphics[width=0.8\textwidth]{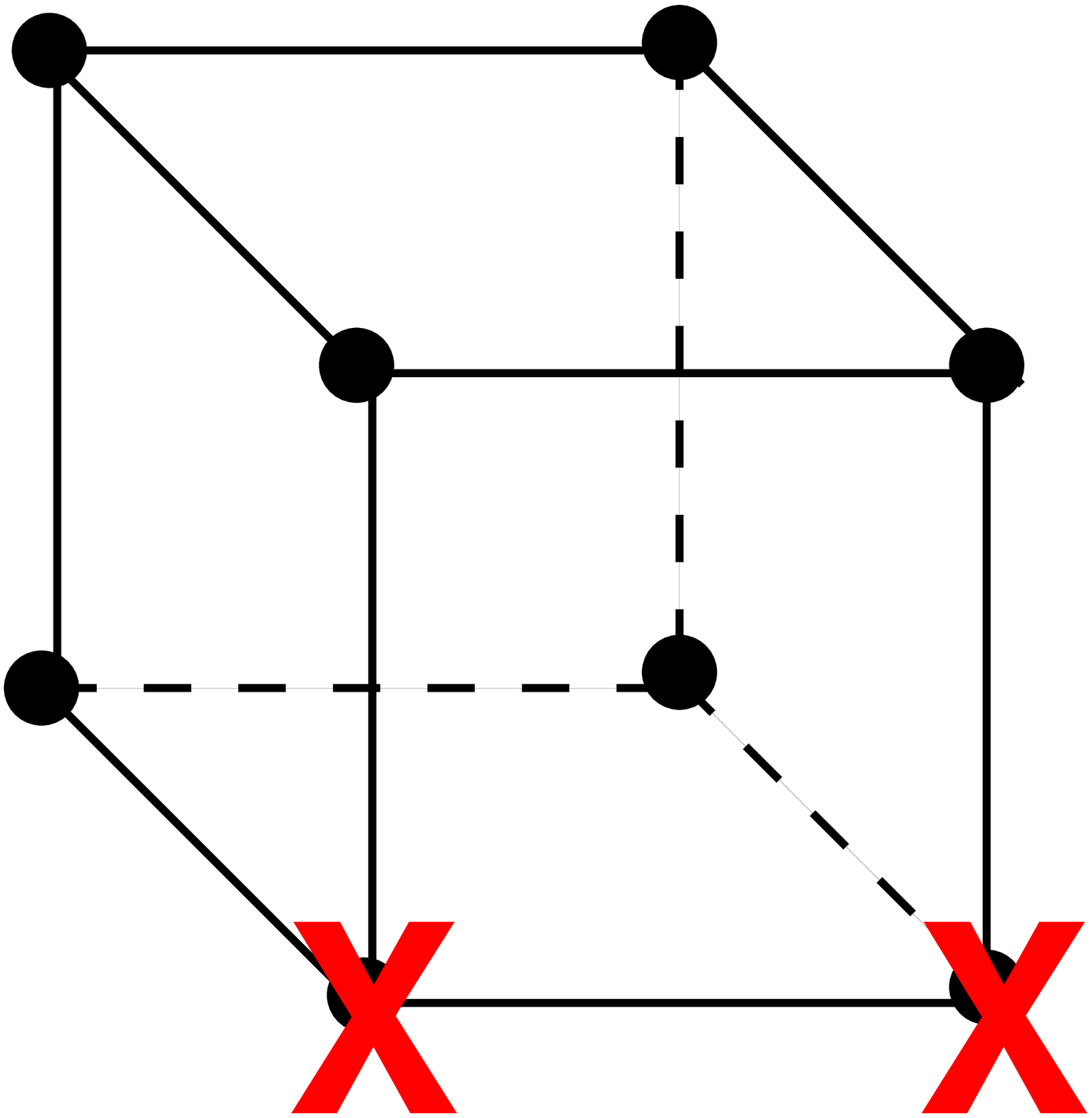}
\end{minipage}
& 3  & $729$ & 0 &
\begin{minipage}[c]{0.12\textwidth}
\includegraphics[width=0.8\textwidth]{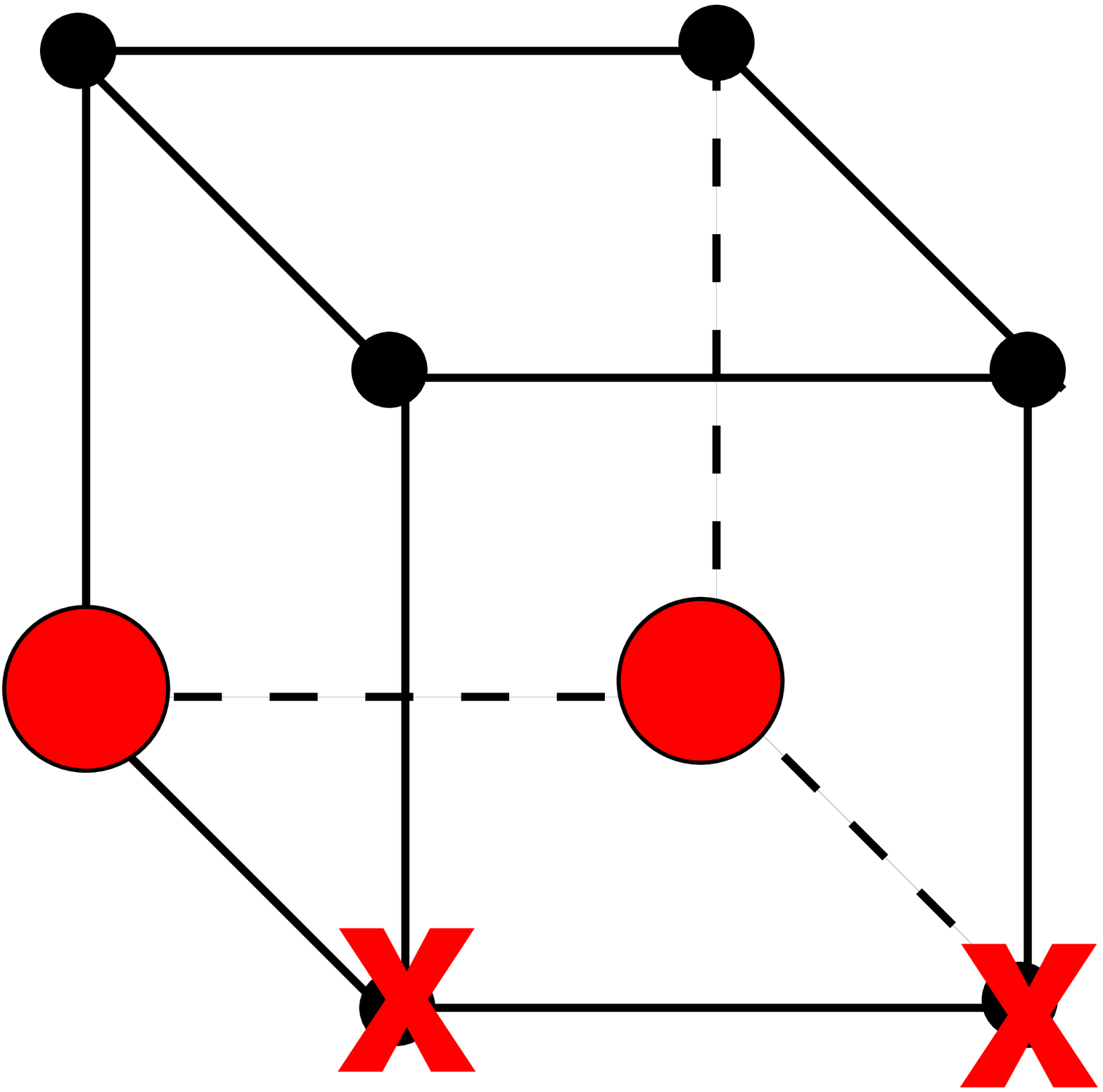}
\end{minipage}
& 6 & $36$ & 2 &
\begin{minipage}[c]{0.12\textwidth}
\includegraphics[width=0.8\textwidth]{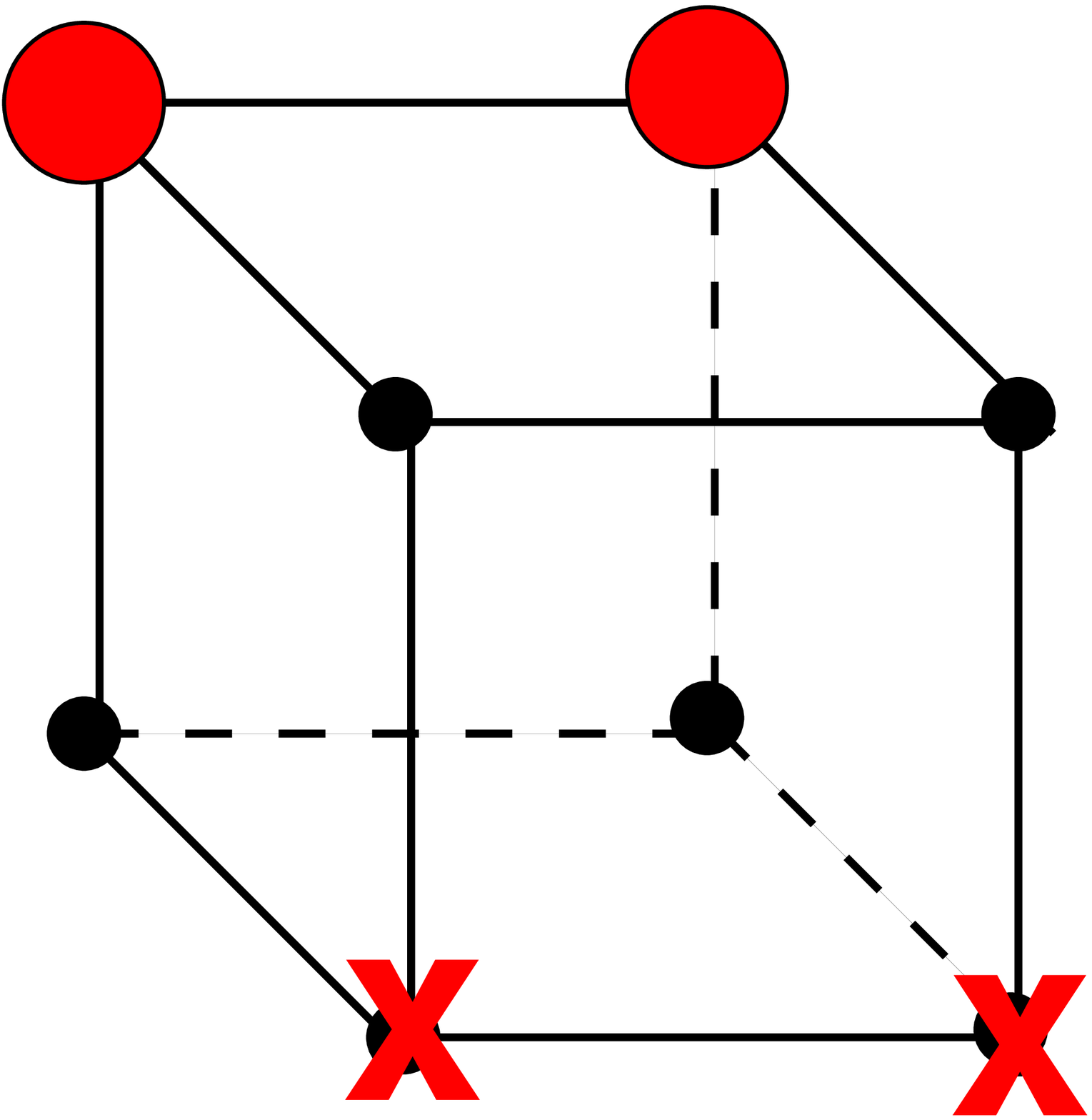}
\end{minipage}
& 3 & $9$ & 2 \\
&&&&&&&&&&&\\
\begin{minipage}[c]{0.12\textwidth}
\includegraphics[width=0.8\textwidth]{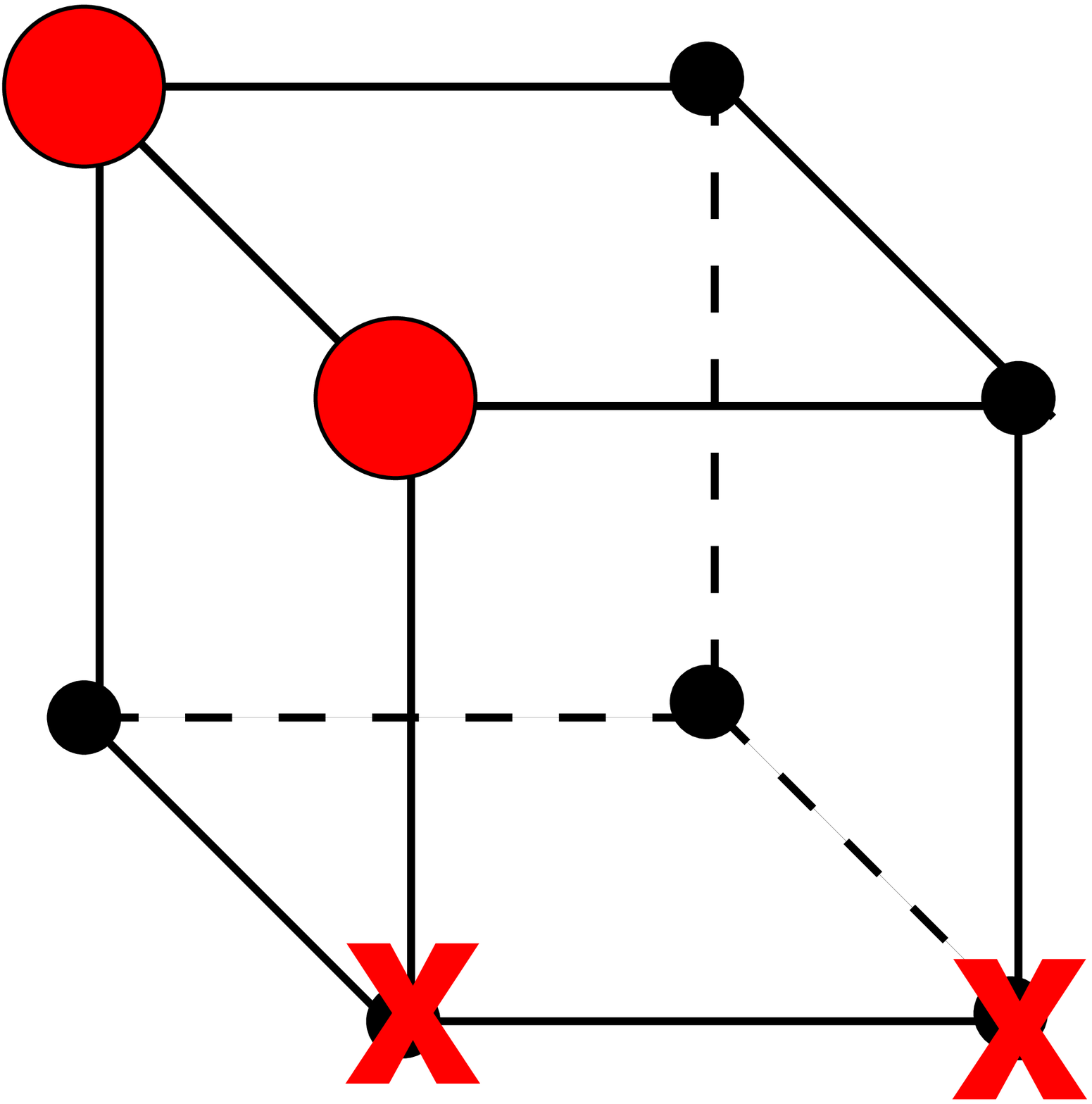}
\end{minipage}
& 12 & $9$ & 2 &
\begin{minipage}[c]{0.12\textwidth}
\includegraphics[width=0.8\textwidth]{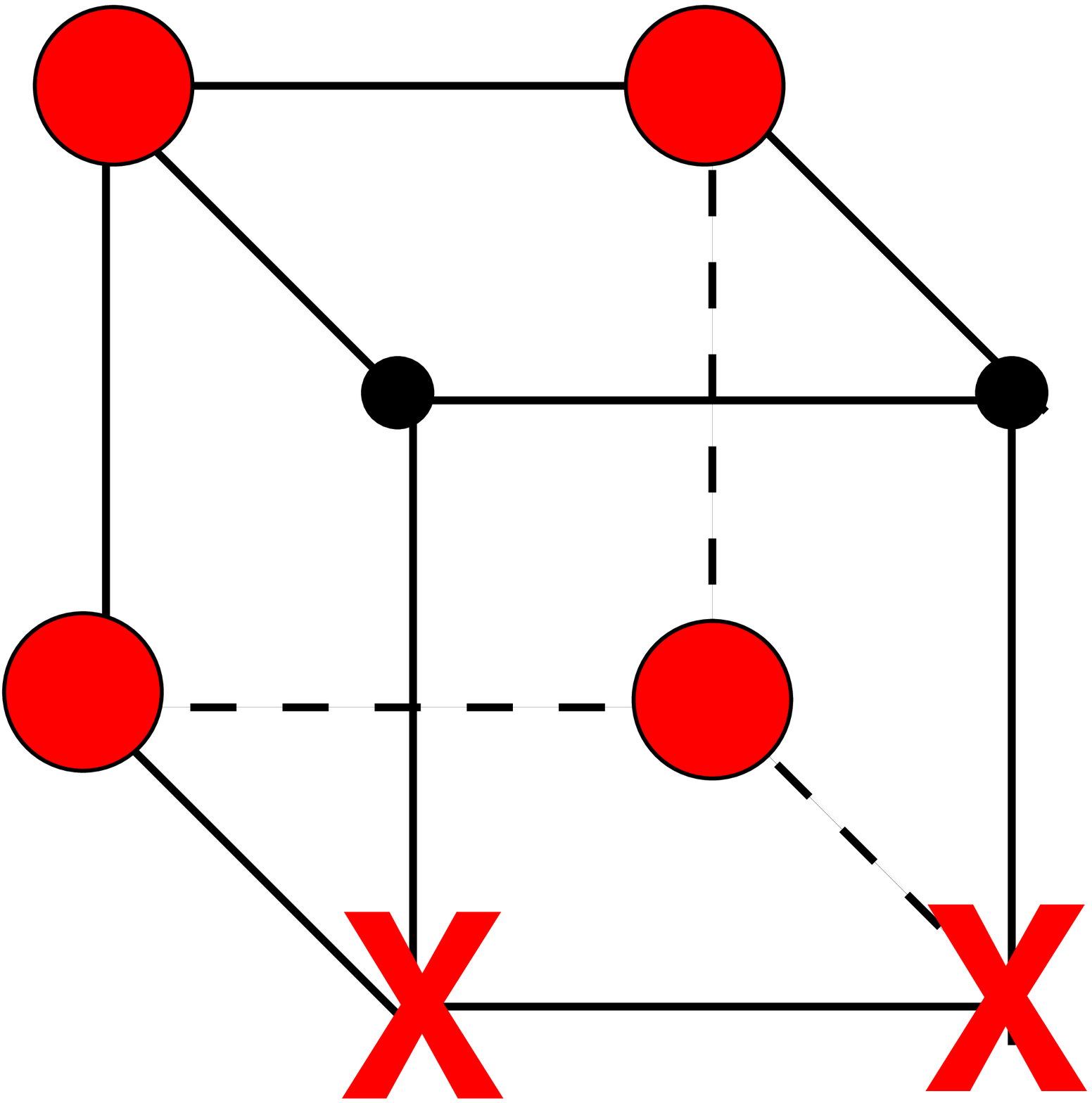}
\end{minipage}
& 6 & $4$ & 4 &
\begin{minipage}[c]{0.12\textwidth}
\includegraphics[width=0.8\textwidth]{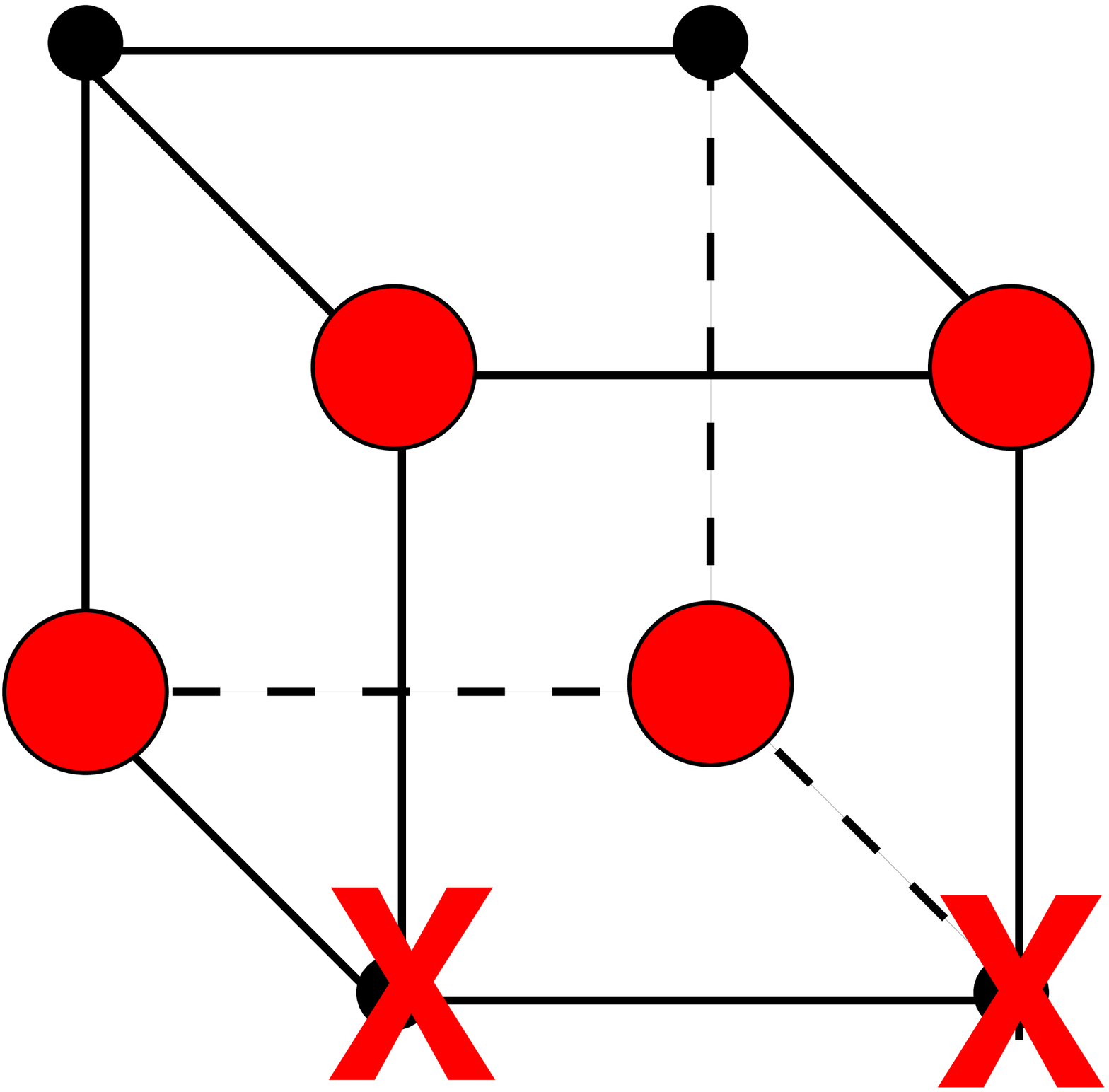}
\end{minipage}
& 3 & $1$ & 4 \\
&&&&&&&&&&&\\
\begin{minipage}[c]{0.12\textwidth}
\includegraphics[width=0.8\textwidth]{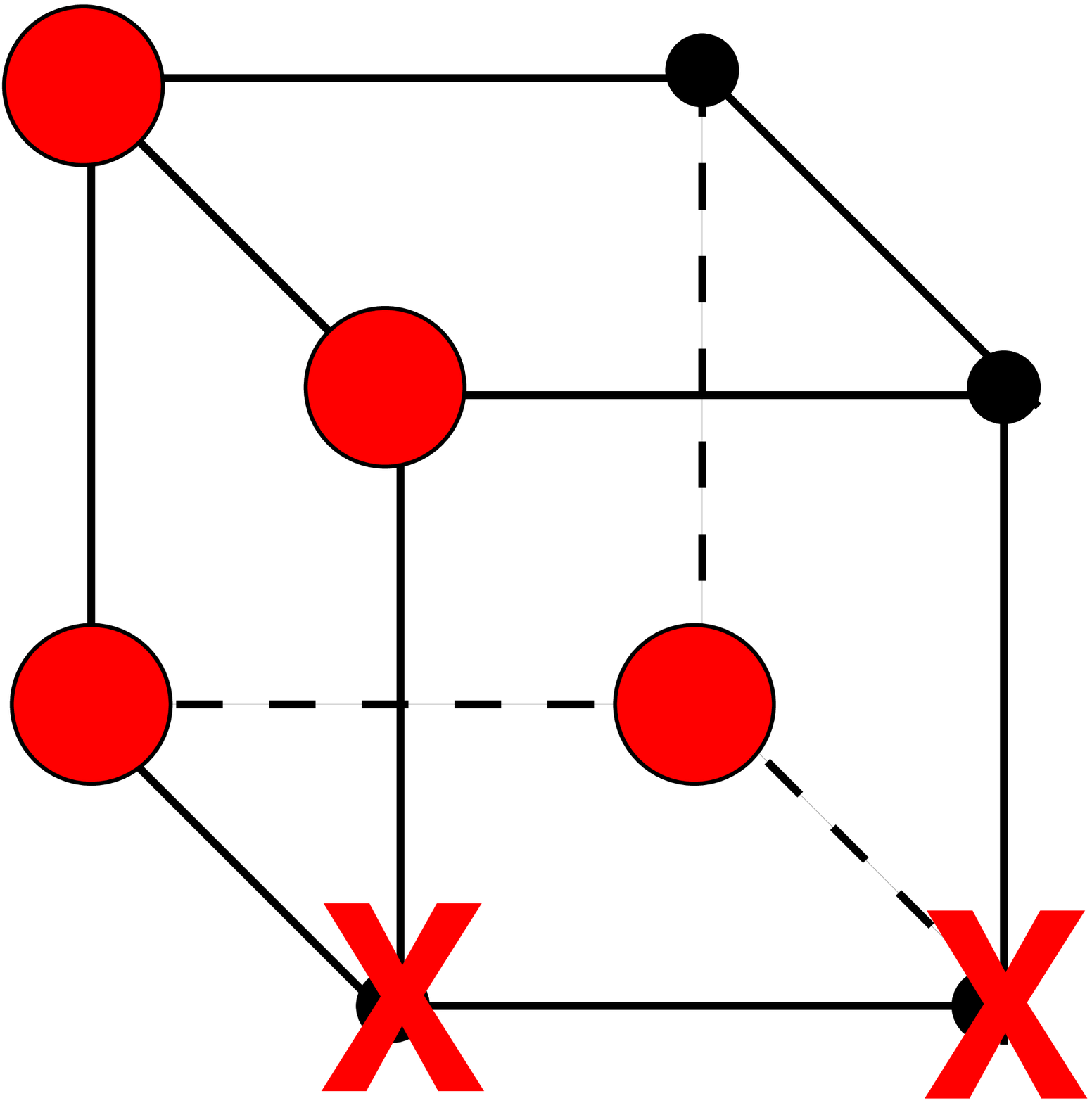}
\end{minipage}
& 12 & $1$ & 4 &
\begin{minipage}[c]{0.12\textwidth}
\includegraphics[width=0.8\textwidth]{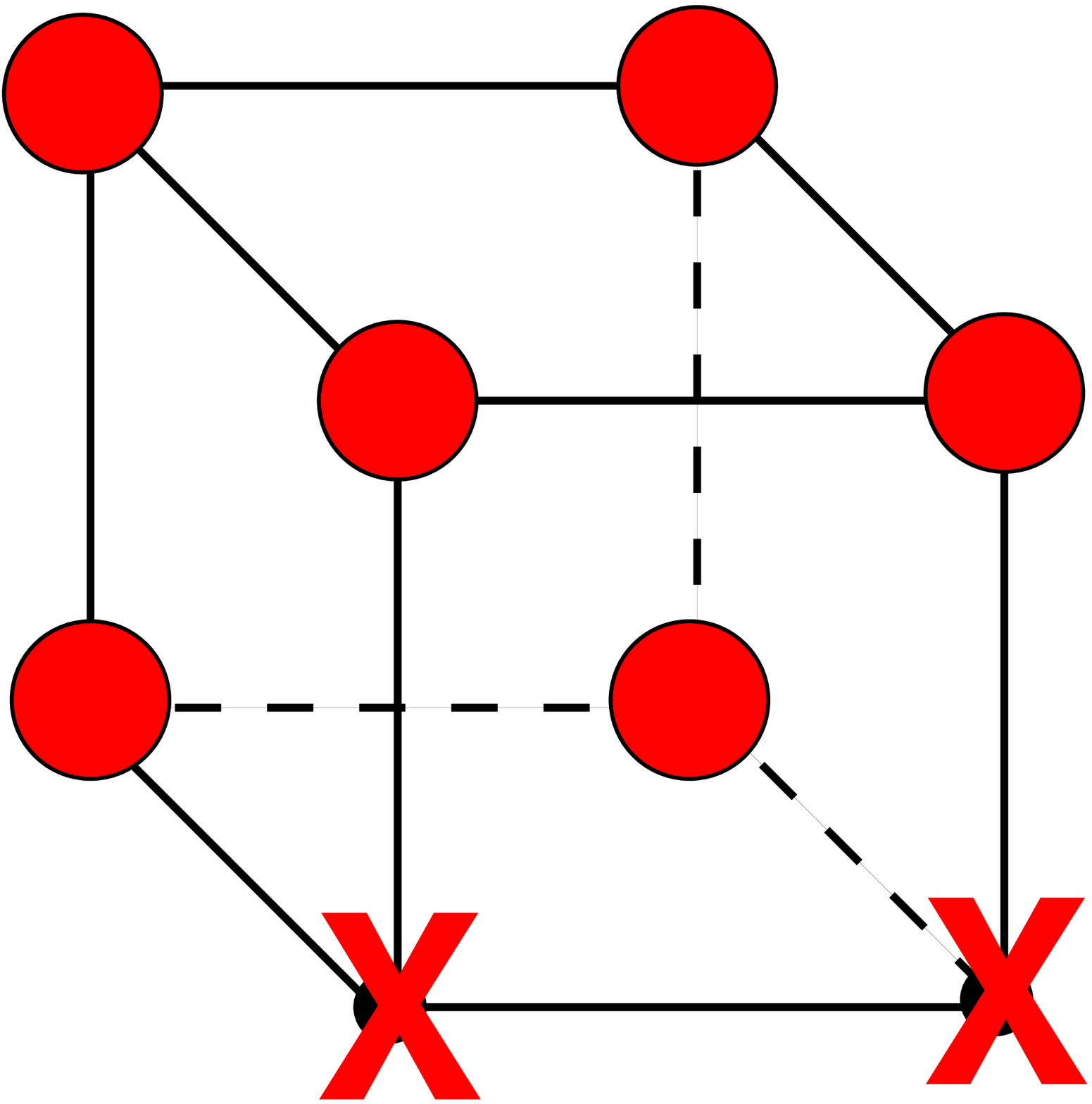}
\end{minipage}
& 3 & $1$ & 6 &
\begin{minipage}[c]{0.12\textwidth}
\includegraphics[width=0.8\textwidth]{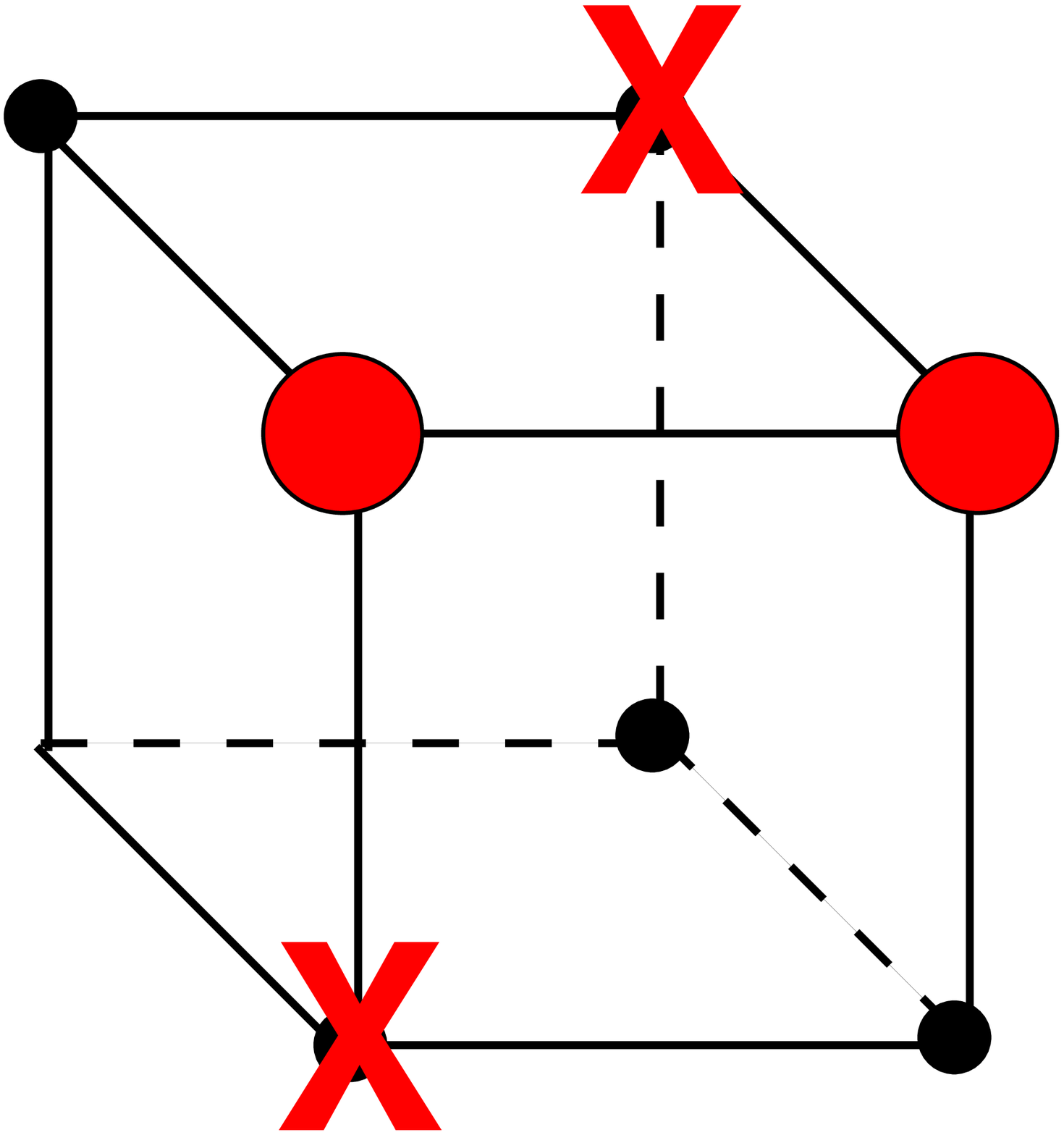}
\end{minipage}
&6 & $9$ & 2\\
&&&&&&&&&&&\\
\begin{minipage}[c]{0.12\textwidth}
\includegraphics[width=0.8\textwidth]{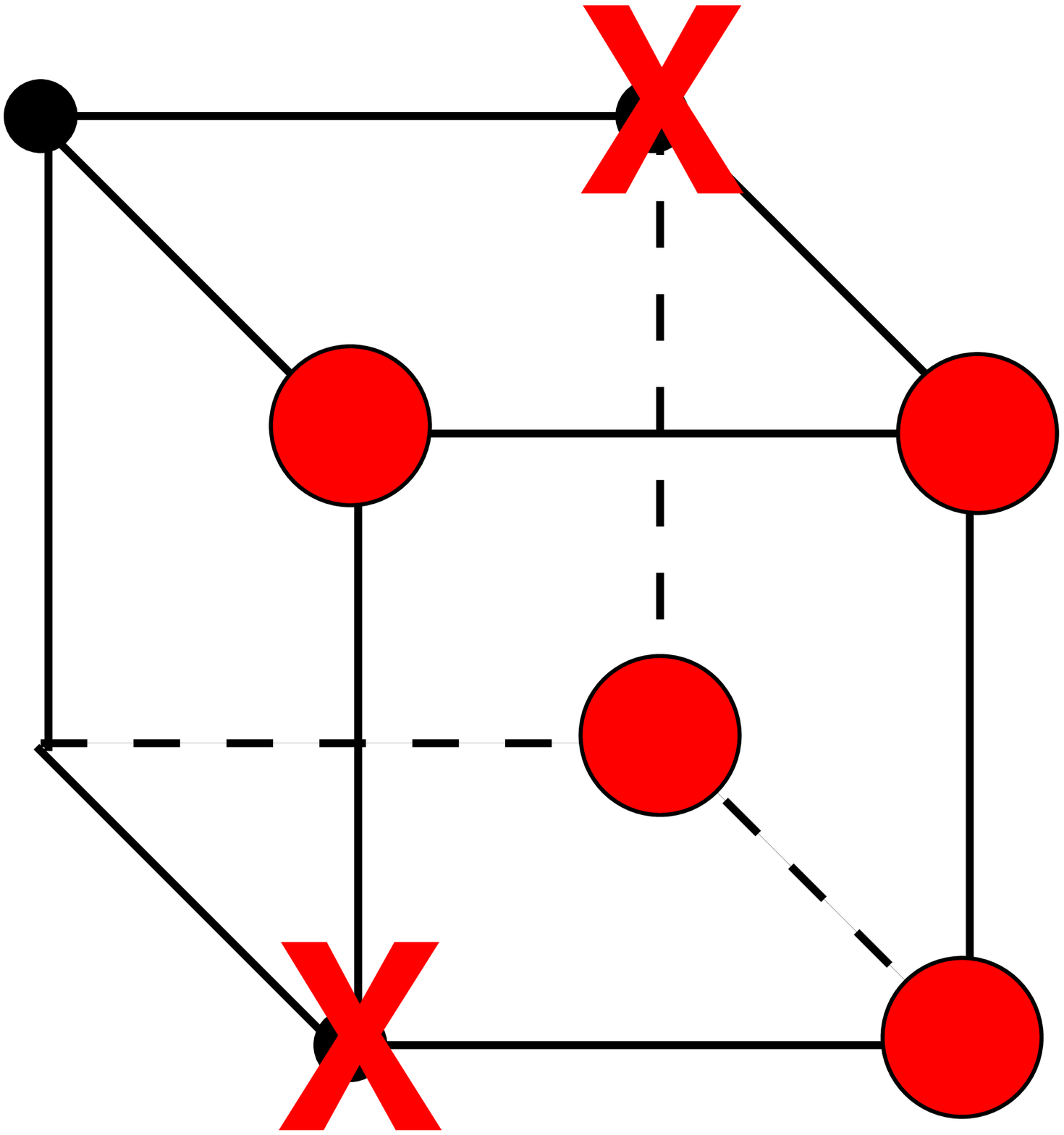}
\end{minipage}
&6 & 1 & 4 &
\begin{minipage}[c]{0.12\textwidth}
\includegraphics[width=0.8\textwidth]{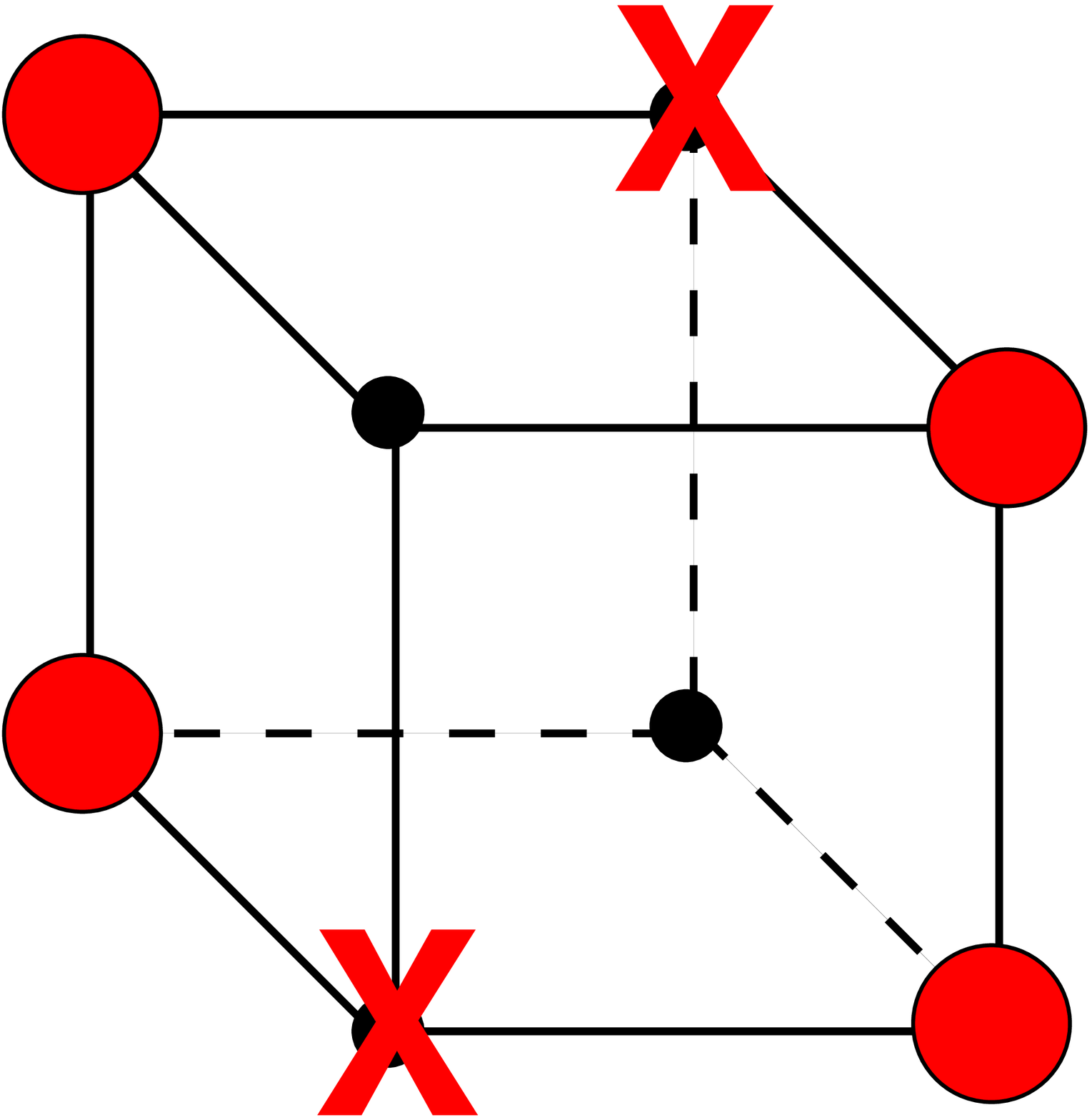}
\end{minipage}
&1 &0 &4 &
\begin{minipage}[c]{0.12\textwidth}
\includegraphics[width=0.8\textwidth]{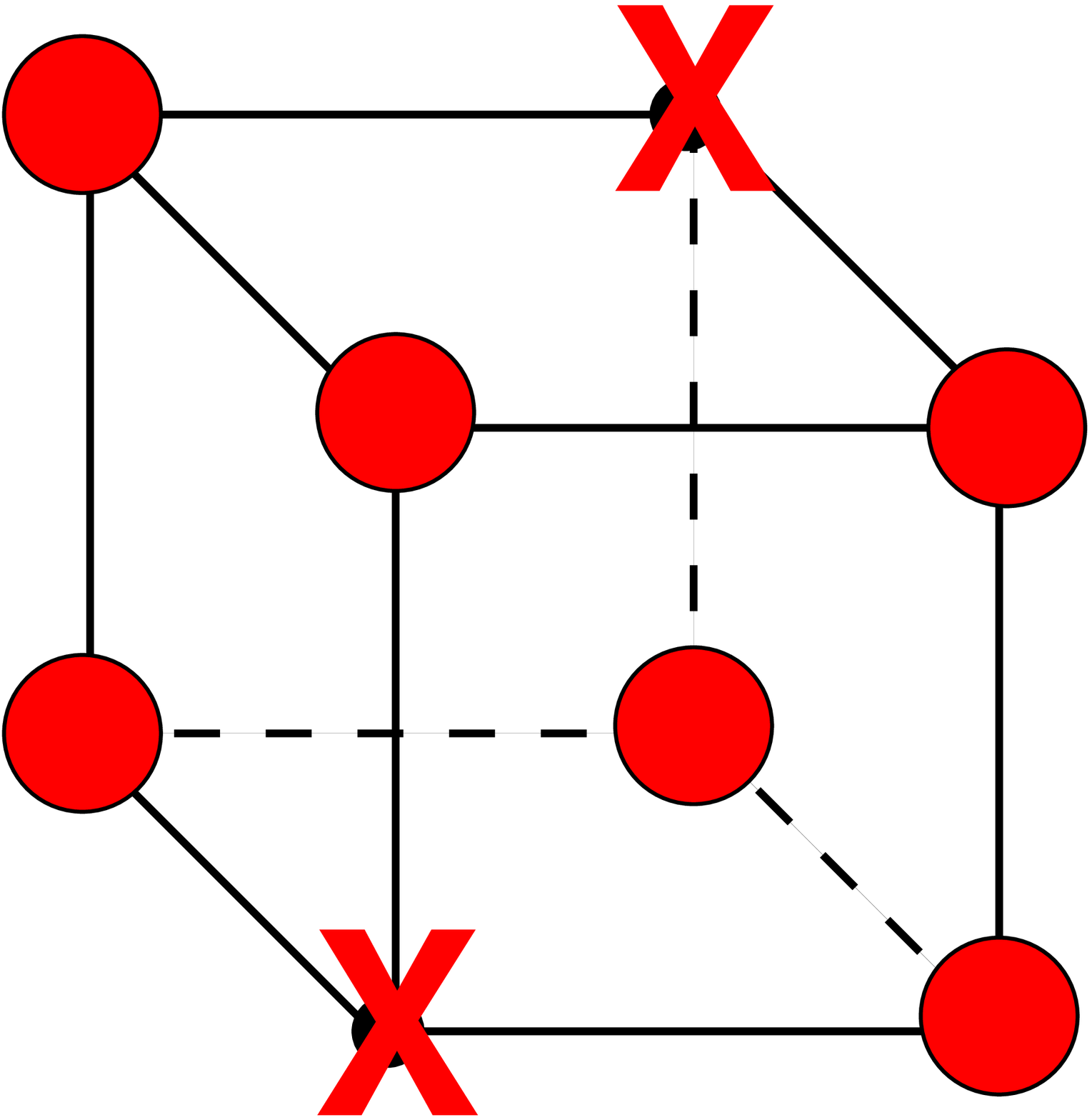}
\end{minipage}
&1 & 0& 6 \\
&&&&&&&&&&&\\
\hline
\end{tabular}
\caption{\label{tab:sus1} Configuration classes $[n_1,n_2]$ that contribute to $\chi_1$ on a $2^3$ lattice.  The degeneracy $g_1$, the fermion bag weight $\mathrm{Det}(W_1W_2)$ and the total monomer number $N_m$ are also given.}
\end{table*}

\begin{table*}[ht]
\begin{tabular}{|c|c|c|c||c|c|c|c||c|c|c|c|}
\hline
$[n_1,n_2]$ & $g_1$ & $\mathrm{Det}(W_1W_2)$ & $N_m$ &
$[n_1,n_2]$ & $g_1$ & $\mathrm{Det}(W_1W_2)$ & $N_m$ &
$[n_1,n_2]$ & $g_1$ & $\mathrm{Det}(W_1W_2)$ & $N_m$ \\
\hline 
&&&&&&&&&&&\\
\begin{minipage}[c]{0.12\textwidth}
\includegraphics[width=0.8\textwidth]{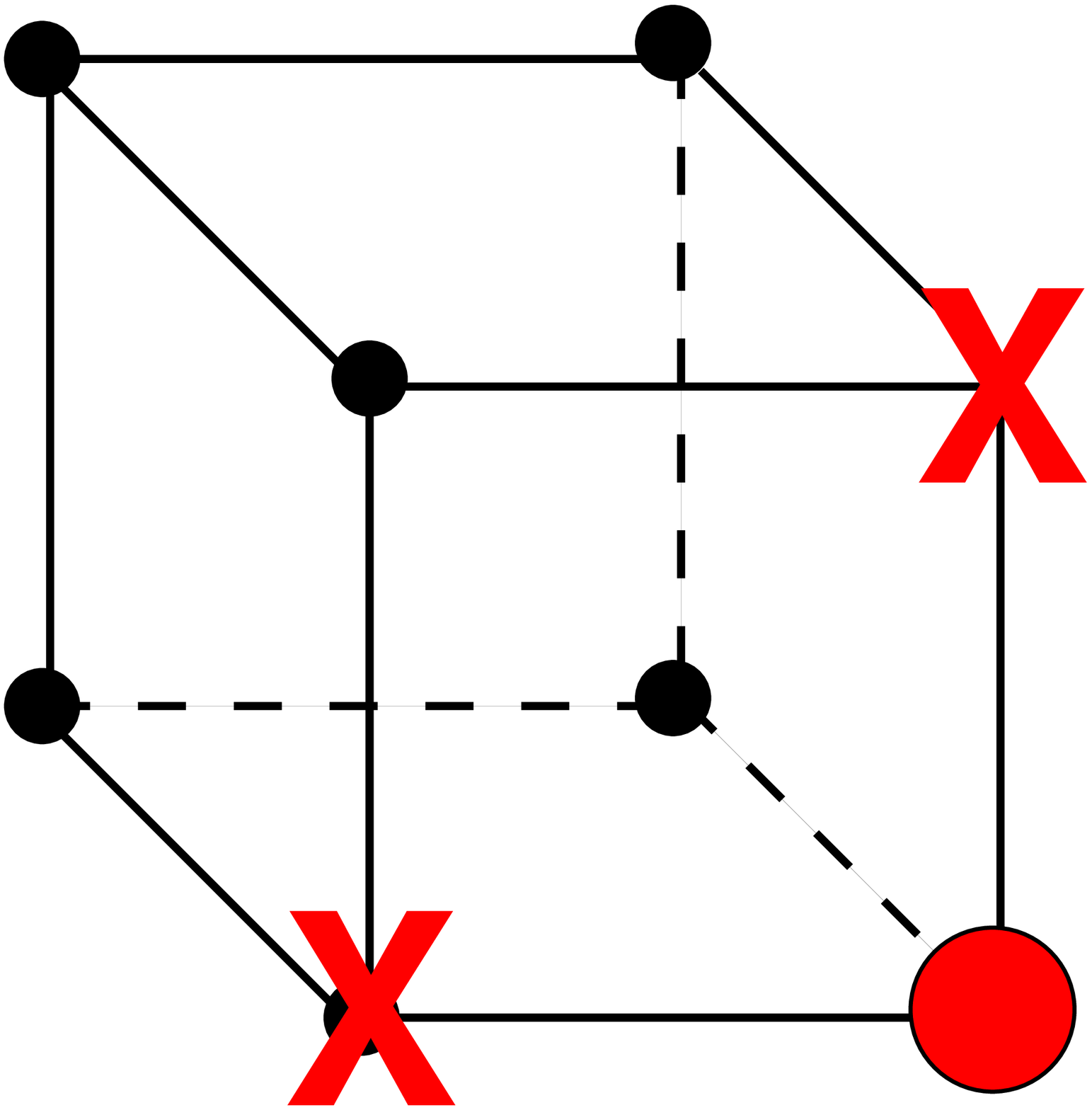}
\end{minipage}
& 6  & $81$ & 1 &
\begin{minipage}[c]{0.12\textwidth}
\includegraphics[width=0.8\textwidth]{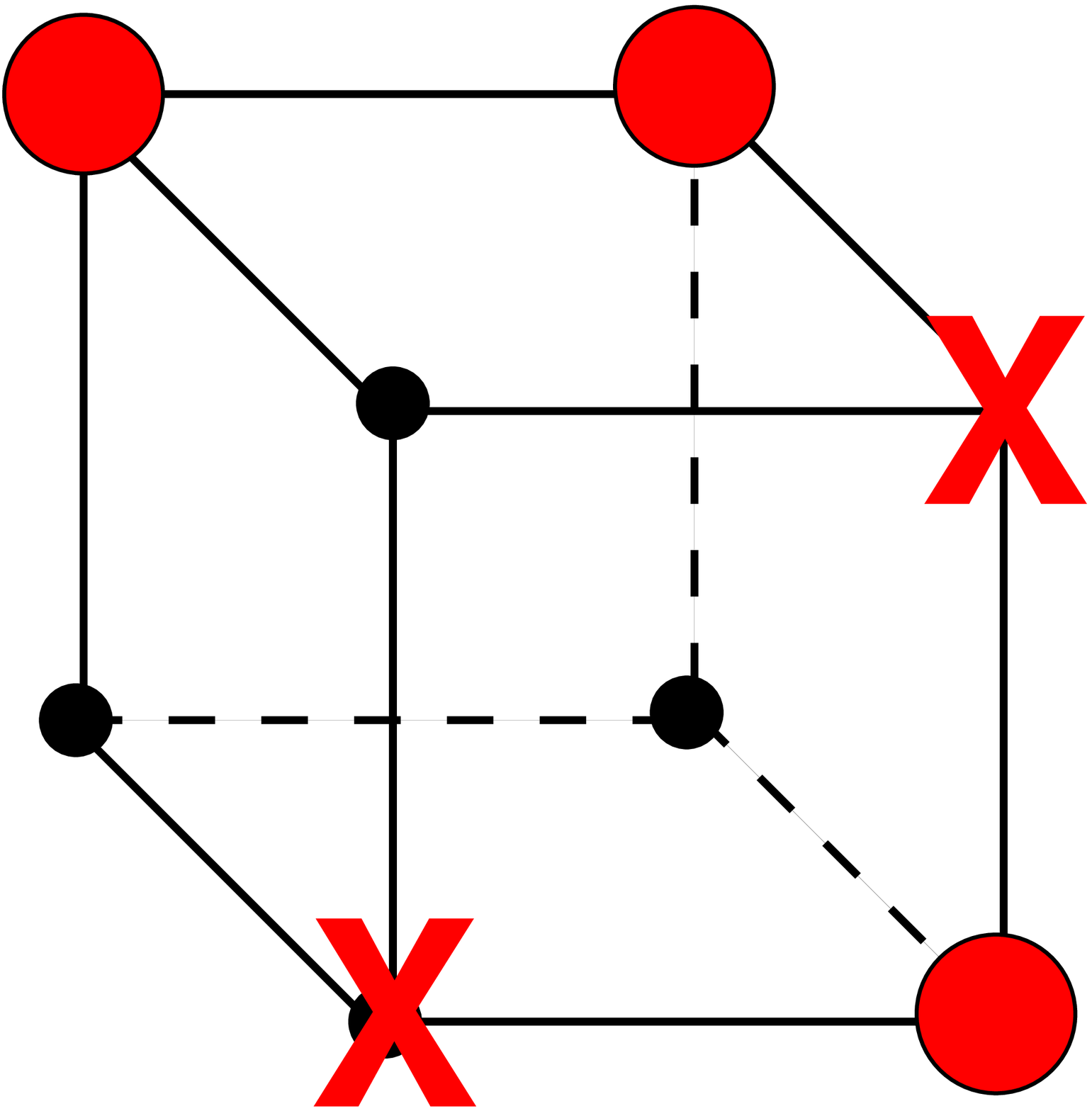}
\end{minipage}
& 6 & $1$ & 3 &
\begin{minipage}[c]{0.12\textwidth}
\includegraphics[width=0.8\textwidth]{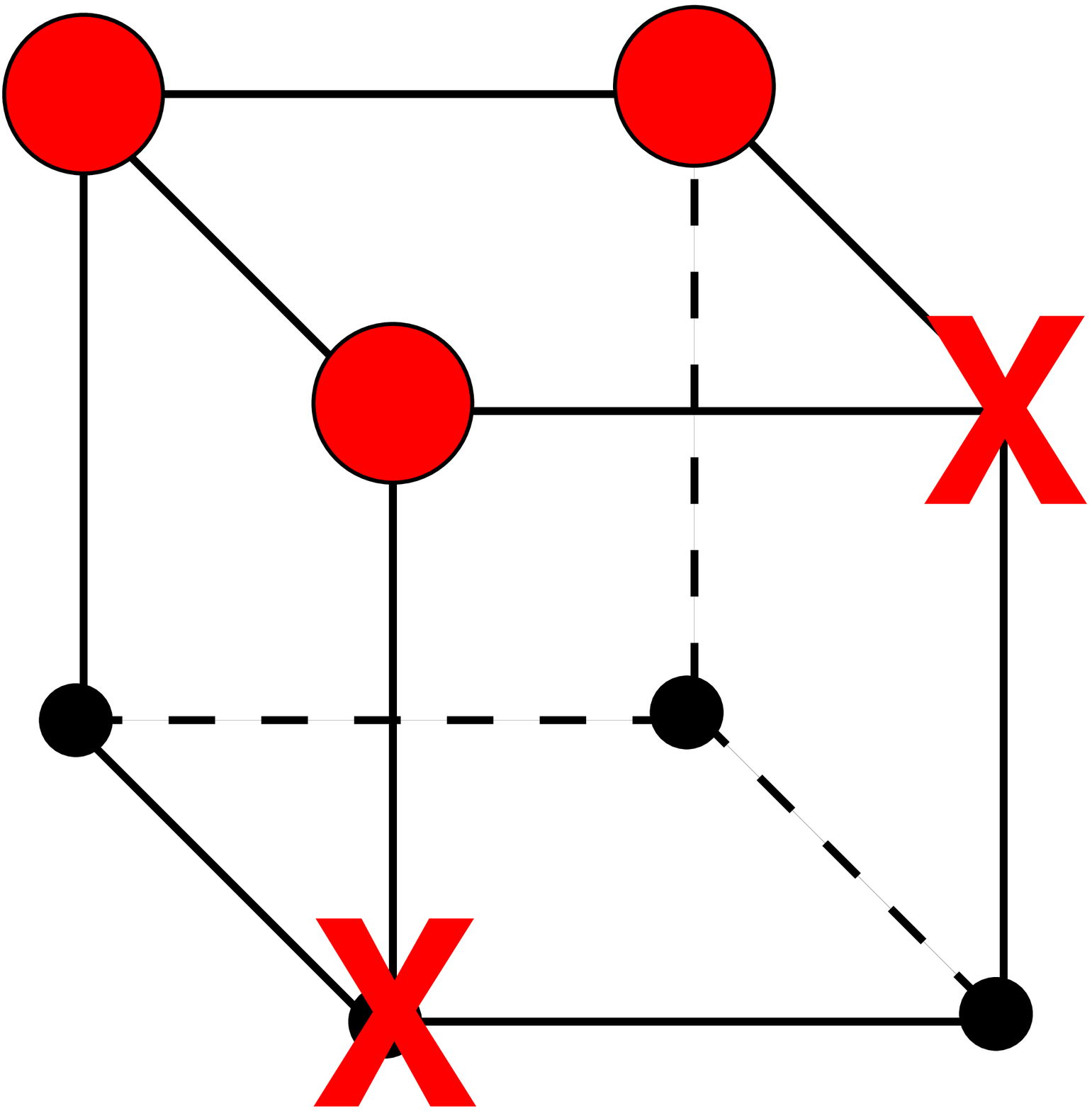}
\end{minipage}
& 12 & $4$ & 3 \\
&&&&&&&&&&&\\
\begin{minipage}[c]{0.12\textwidth}
\includegraphics[width=0.8\textwidth]{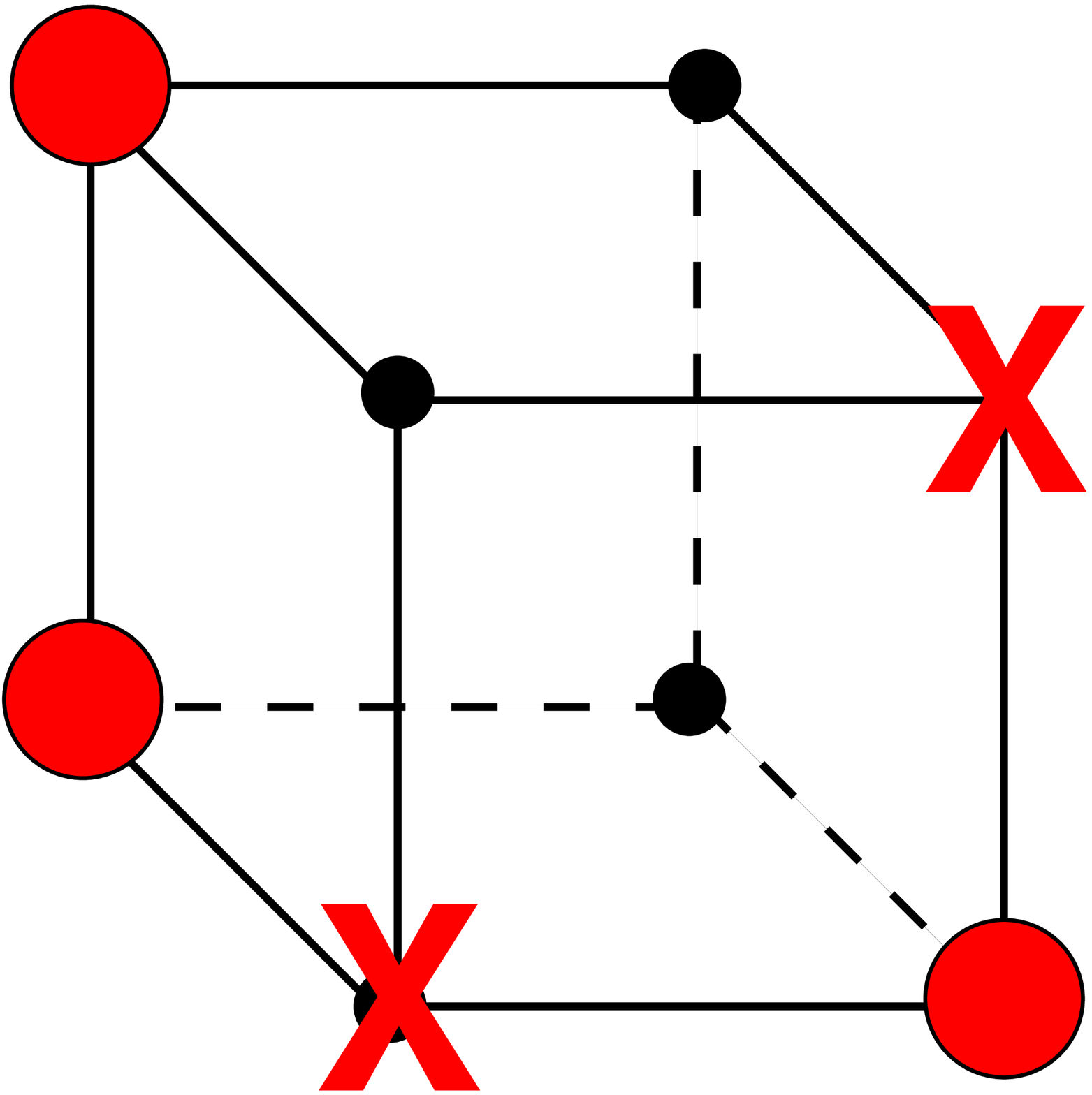}
\end{minipage}
& 6 & $1$ & 3 &
\begin{minipage}[c]{0.12\textwidth}
\includegraphics[width=0.8\textwidth]{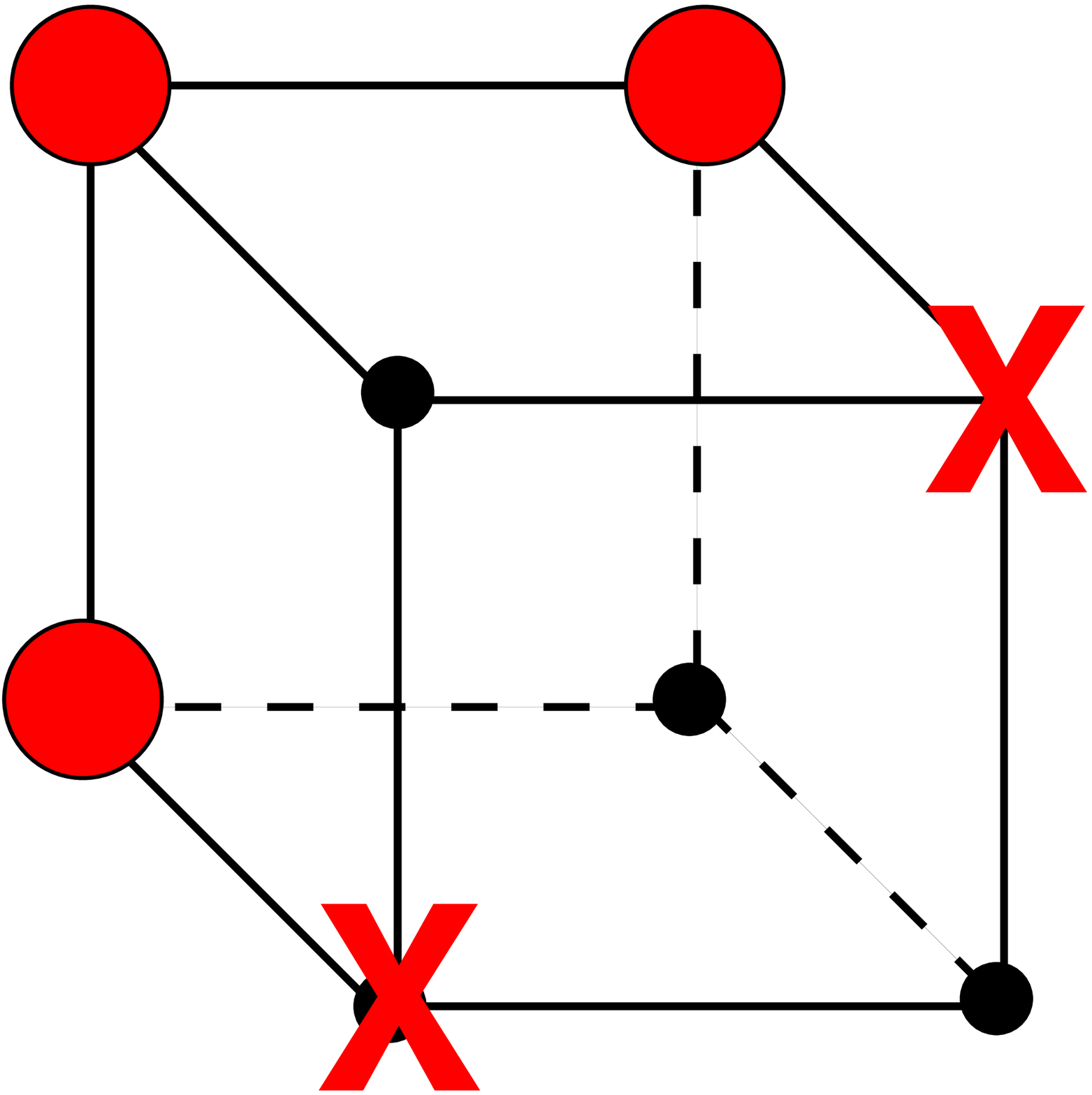}
\end{minipage}
& 6 & $1$ & 3 &
\begin{minipage}[c]{0.12\textwidth}
\includegraphics[width=0.8\textwidth]{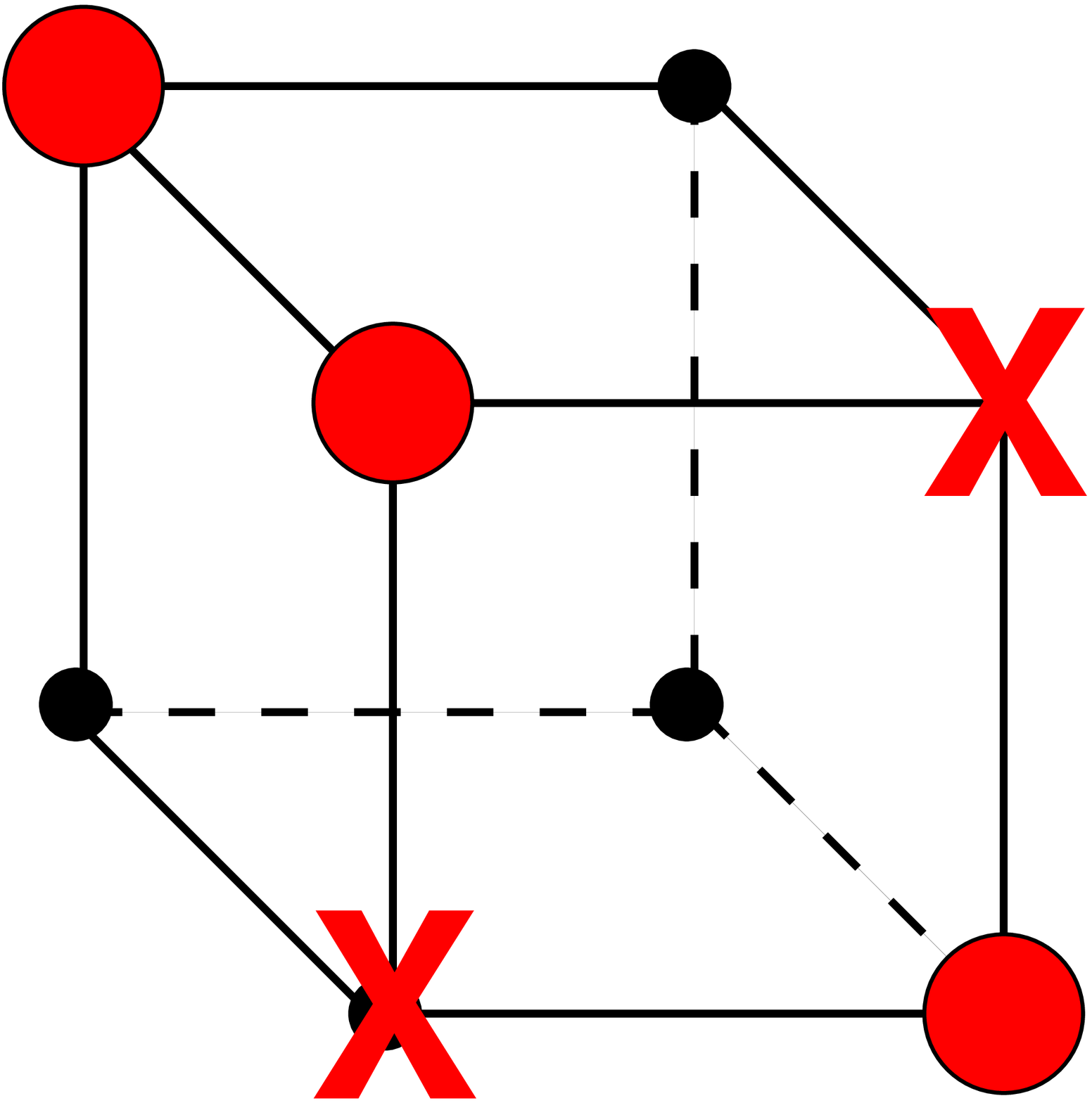}
\end{minipage}
& 6 & $1$ & 3 \\
&&&&&&&&&&&\\
\begin{minipage}[c]{0.12\textwidth}
\includegraphics[width=0.8\textwidth]{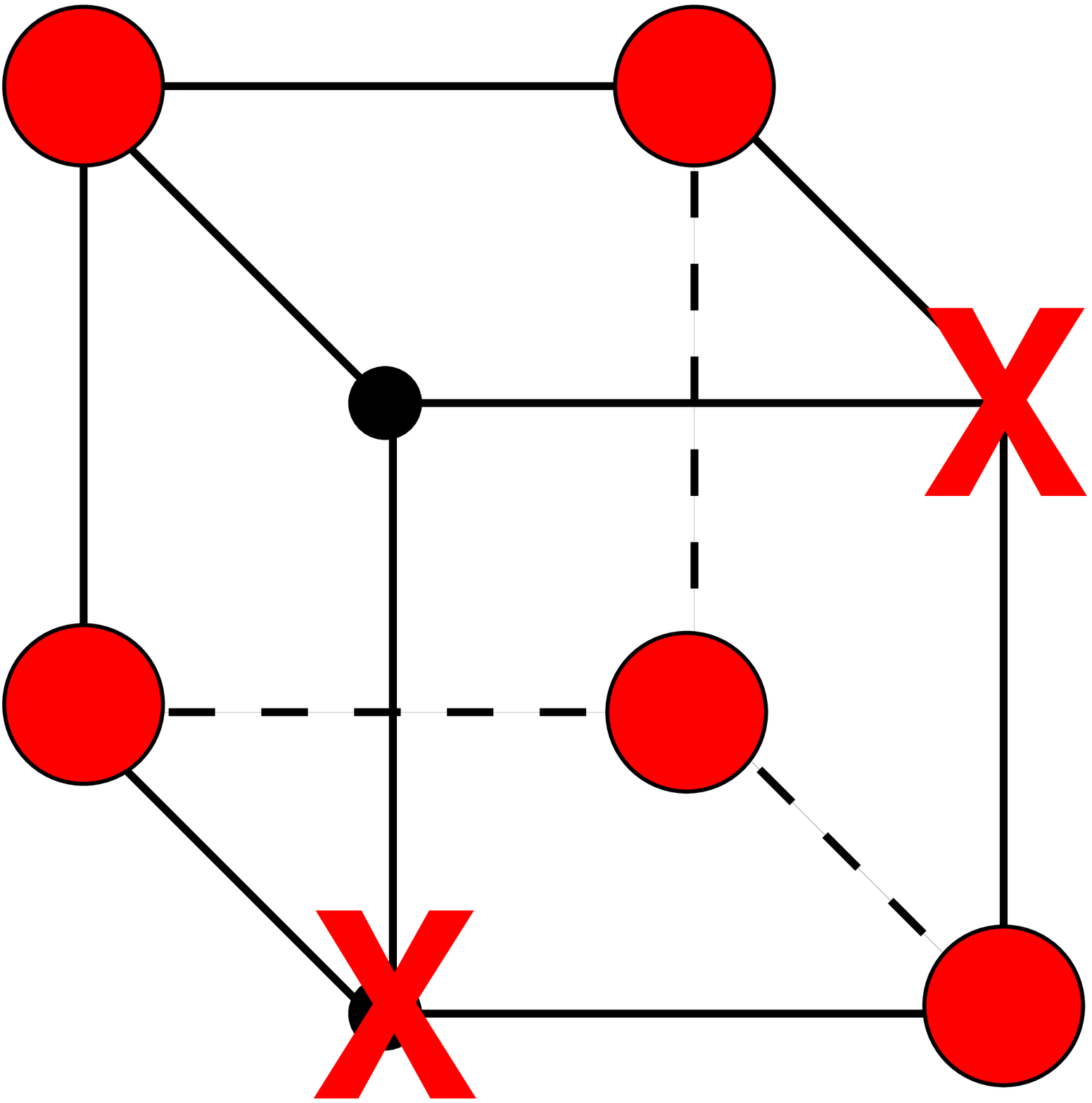}
\end{minipage}
& 6 & $1$ & 5 &
\begin{minipage}[c]{0.12\textwidth}
\includegraphics[width=0.8\textwidth]{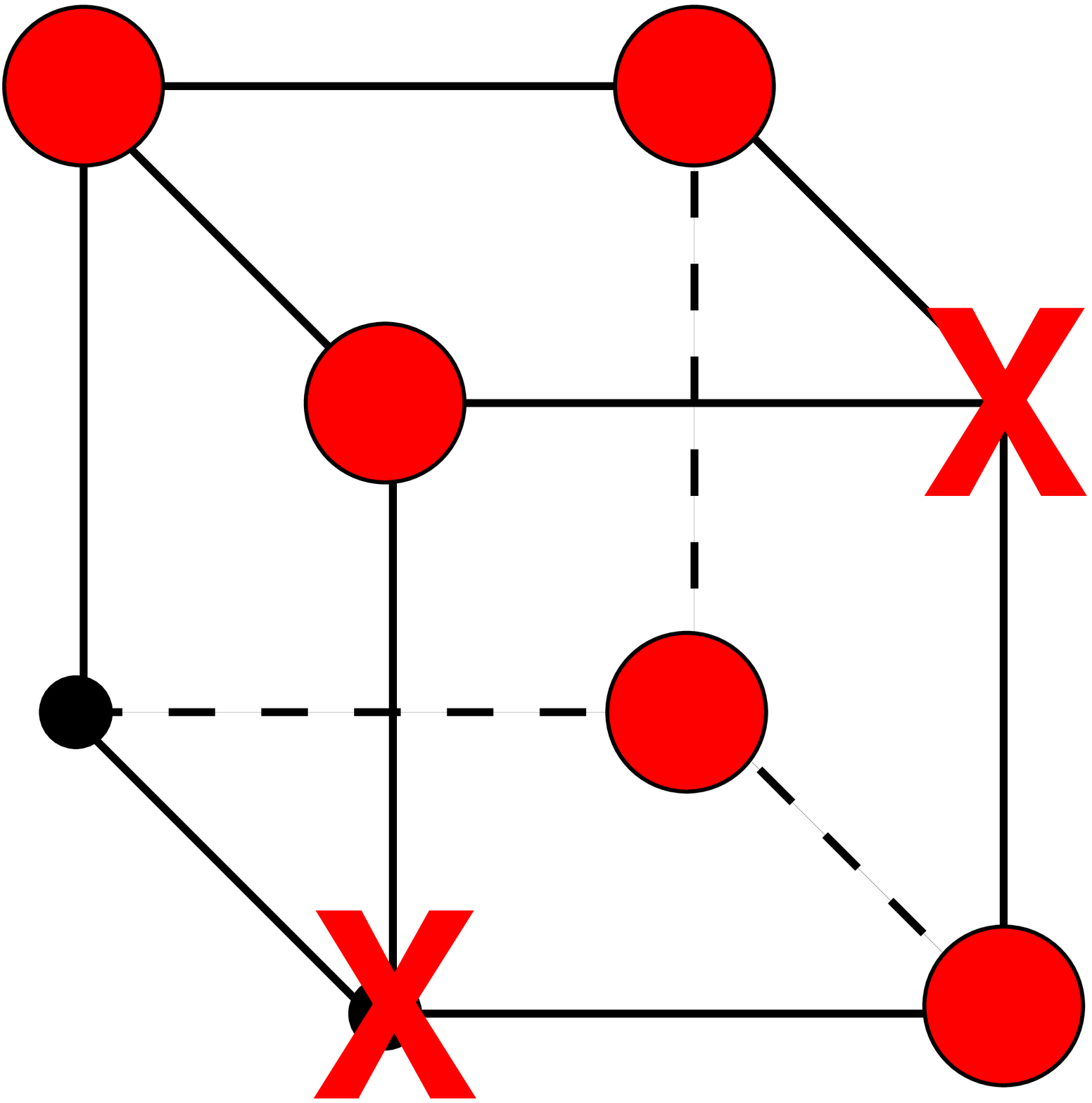}
\end{minipage}
& 6 & $0$ & 5 &
& & & \\
&&&&&&&&&&&\\
\hline
\end{tabular}
\caption{\label{tab:sus2} Configuration classes $[n_1,n_2]$ that contribute to $\chi_2$ on a $2^3$ lattice.  The degeneracy $g_2$, the fermion bag weight $\mathrm{Det}(W_1W_2)$ and the total monomer number $N_m$ are also given.}
\end{table*}

\end{document}